\documentclass[twocolumn,showpacs,aps,floatfix,prd,nofootinbib]{revtex4}

\usepackage{graphicx}
\usepackage{dcolumn}
\usepackage{array, longtable}
\usepackage{epsfig}
\usepackage{amsmath}
\usepackage{colordvi}
\usepackage{hhline}

\newcommand{\BaBarYear}{14}
\newcommand{\BaBarNumber}{001}
\newcommand{\SLACPubNumber}{15934}

 \newcommand{\BaBarType}      {PUB}  

\input babarsym

\def\Ecm       {\ensuremath {E_{\rm c.m.}}\xspace}

\def\mrec      {\ensuremath {m_{\rm rec}}\xspace}
\def\mrecc     {\ensuremath {m^c_{\rm rec}}\xspace}
\def\jetset   {\mbox{\tt Jetset \hspace{-0.5em}7.\hspace{-0.2em}4}\xspace}

\long\def\inst#1{\par\nobreak\kern 4pt\nobreak
    {\it #1}\par\vskip 10pt plus 3pt minus 3pt}

\begin{document}

\begin{flushleft}
\babar-\BaBarType-\BaBarYear/\BaBarNumber \\
SLAC-PUB-\SLACPubNumber \\
\end{flushleft}

\title{\large \bf
\boldmath
Cross sections for the
reactions $\epem\to \KS \KL$, $\KS \KL \pipi$,
 $\KS \KS\pipi$, and $\KS \KS K^+K^-$    
from events with initial-state radiation
} 

%
\author{J.~P.~Lees}
\author{V.~Poireau}
\author{V.~Tisserand}
\affiliation{Laboratoire d'Annecy-le-Vieux de Physique des Particules (LAPP), Universit\'e de Savoie, CNRS/IN2P3,  F-74941 Annecy-Le-Vieux, France}
\author{E.~Grauges}
\affiliation{Universitat de Barcelona, Facultat de Fisica, Departament ECM, E-08028 Barcelona, Spain }
\author{A.~Palano$^{ab}$ }
\affiliation{INFN Sezione di Bari$^{a}$; Dipartimento di Fisica, Universit\`a di Bari$^{b}$, I-70126 Bari, Italy }
\author{G.~Eigen}
\author{B.~Stugu}
\affiliation{University of Bergen, Institute of Physics, N-5007 Bergen, Norway }
\author{D.~N.~Brown}
\author{L.~T.~Kerth}
\author{Yu.~G.~Kolomensky}
\author{M.~J.~Lee}
\author{G.~Lynch}
\affiliation{Lawrence Berkeley National Laboratory and University of California, Berkeley, California 94720, USA }
% ...
\collaboration{The \babar\ Collaboration}
\noaffiliation


\begin{abstract}
We study the processes $\epem\to \KS \KL \gamma$, 
 $\KS \KL \pipi\gamma$, $\KS \KS \pipi\gamma$, 
 and $\KS \KS K^+K^-\gamma$,
where the photon is radiated from the initial state,
providing cross section measurements for the  hadronic states over a continuum  
 of center-of-mass energies.
The results are based on 469~\invfb of
  data collected with the \babar\ detector at SLAC.  We observe the 
  $\phi(1020)$ resonance in the $\KS \KL$ final state and measure the product
  of its electronic width and branching fraction with about 3\% uncertainty.
  We present a measurement of the $\epem\to \KS \KL $ cross section in the energy 
  range from 1.06 to 2.2~\gev and observe the production of a resonance at
  1.67~\gev.  We present the first measurements of the 
 $\epem\to \KS \KL \pipi$, $\KS \KS \pipi$, and $\KS \KS K^+K^-$
cross sections, and study the intermediate
  resonance structures.  We obtain the first observations of \jpsi decay 
  to the  $\KS \KL \pipi$, $\KS \KS \pipi$, and $\KS \KS K^+K^-$ final states.
  
\end{abstract}


\vfill
\maketitle

\setcounter{footnote}{0}

\section{Introduction}
\label{sec:Introduction}

The idea to use electron-positron annihilation events with initial-state
radiation (ISR) to study processes with energies below the
nominal \epem center-of-mass (\Ecm) energy was outlined in Ref.~\cite{baier}. 
The possibility of exploiting ISR
to measure low-energy cross sections at high-luminosity $\phi$ and $B$ factories 
is  discussed in Refs.~\cite{arbus, kuehn, ivanch}, 
and motivates the study described in this paper.  
This is of particular interest because of a three-standard-deviation
discrepancy between the current measured value of the muon anomalous magnetic moment ($g-2$) and that predicted by the Standard Model~\cite{dehz}, 
in which hadronic loop contributions are obtained from experimental
\epem annihilation cross sections at low \Ecm energies. 
The study of ISR events at $B$ factories 
provides independent results over a continuum of energy values
for hadronic
cross sections in this energy region and also contributes to the
investigation of low-mass resonance spectroscopy. 
                                
Studies of the ISR processes $e^+e^-\to\mumu\gamma$~\cite{Druzhinin1,isr2pi}
and $\epem \to X_h\gamma$, 
using data of the \babar\ experiment at SLAC,
where $X_h$ represents any of several exclusive multihadron final states,
have been reported previously.
The studied final states include:
charged hadron pairs $\pip\pim$~\cite{isr2pi}, $\Kp\Km$~\cite{isr2k}, and
$p\overline{p}$~\cite{isr2p};
four or six charged mesons~\cite{isr4pi,isr2k2pi,isr6pi};
charged mesons plus one or two \piz mesons~\cite{isr2k2pi,isr6pi,isr3pi,isr5pi};
and a \KS plus charged mesons~\cite{isrkkpi}.
Together, these demonstrate good detector efficiency for events of
this kind, 
and well understood tracking, particle identification, and \piz and
\KS reconstruction.

This paper reports measurements of the $\KS \KL$, $\KS \KL \pipi$, 
$\KS \KS \pipi$, and $\KS \KS K^+ K^-$ final states, 
produced in conjunction with a hard photon, that is 
assumed to result from ISR.
Candidate \KS decays are reconstructed in the $\pip\pim$ decay mode.
This is the first ISR measurement from \babar\ that includes \KL mesons,
which we detect via their nuclear interactions in the
electromagnetic calorimeter.
 We use the $\epem \to \gamma\phi \to \gamma\KS \KL$ reaction to
measure the $\KL$ detection efficiency directly from the data.
The $\epem \to \KS\KL$ cross section is measured from threshold to
2.2~\gev.
For the other final states, we measure cross sections from threshold to
4~\gev, study the internal structure of the events, 
and perform the first measurements of their \jpsi branching fractions.
Together with our previous measurements~\cite{isr2k,isr2k2pi}, these
results provide a much more complete understanding of the 
$K\Kbar$, $K\Kbar\pi\pi$, and
$K\Kbar K\Kbar$ final states in \epem annihilations.

\section{\boldmath The \babar\ detector and data set}
\label{sec:babar}

The data used in this analysis were collected with the \babar\ detector at
the \pep2\ asymmetric-energy \epem\ storage ring. 
The total integrated luminosity used is 468.6~\invfb~\cite{lumi}, 
which includes data collected at the $\Upsilon(4S)$
resonance (424.7~\invfb) and at a c.m.\ energy 40~\mev below this resonance (43.9~\invfb).

The \babar\ detector is described in detail elsewhere~\cite{babar}. 
Charged particles are
  reconstructed using the \babar\ tracking system,
which comprises the silicon vertex tracker (SVT) and the drift chamber (DCH)
inside the 1.5 T solenoid.
Separation of pions and kaons is accomplished by means of the detector of
internally reflected Cherenkov light  and energy-loss measurements in
the SVT and DCH. 
The hard ISR photon, photons from $\pi^0$ decays,
and \KL are detected in the electromagnetic calorimeter (EMC).  
Muon identification, provided by the instrumented flux return,
is used to select the $\mumu\gamma$ final state.

In order to study the detector acceptance and efficiency, 
we have developed a special package of simulation programs for
radiative processes based on 
the approach suggested by K\"uhn  and Czy\.z~\cite{kuehn2}.  
Multiple collinear soft-photon emission from the initial \epem state 
is implemented with the structure-function technique~\cite{kuraev,strfun}, 
while additional photon radiation from the final-state particles is
simulated using the PHOTOS package~\cite{PHOTOS}.  
The precision of the radiative simulation does not contribute more than 1\% 
uncertainty to the efficiency calculation.

The four-meson final states are generated according to a phase-space distribution.
We simulate the $\KS \KL \gamma$ channel using a model that includes
the $\phi(1020)$ and two additional resonances,
fitted to all available $\epem\to \KS \KL$ cross section
measurements~\cite{cmdphi,cmdksklff,kskldm1,ksklsnd,ksklolya},
which cover the range from threshold up to about 2.5~\gev.
Samples of roughly five times the number of expected events
are generated for each final state and
processed through the detector response simulation \cite{GEANT4}. 
These events are then reconstructed using the same software chain
as the data. Variations in detector
and background conditions are taken into account.

We also simulate a number of background processes.
Based on our experience with final states including kaons,
we consider the ISR processes
$\KS\KL (\phi)\eta$, $\KS K^{\pm}\pi^{\mp}$,
$\KS K^{\pm}\pi^{\mp}\piz$, $\KS\KL\pipi\piz$, and $\KS\KL\piz$,
with normalizations based on our previous measurements and
isospin relations.
In addition, 
we generate a large sample of the as-yet unobserved 
final state
$\KS\KL\ppz\gamma$, which is a potential background.
We also simulate several non-ISR backgrounds, including 
$\epem\to q \qbar$ $(q = u, d, s, c)$ events using the
\jetset~\cite{jetset} generator,
and $\epem\to\tau^+\tau^-$ events using the KORALB~\cite{koralb} generator. 

\begin{figure}[tbh]
\begin{center}
\includegraphics[width=0.9\linewidth]{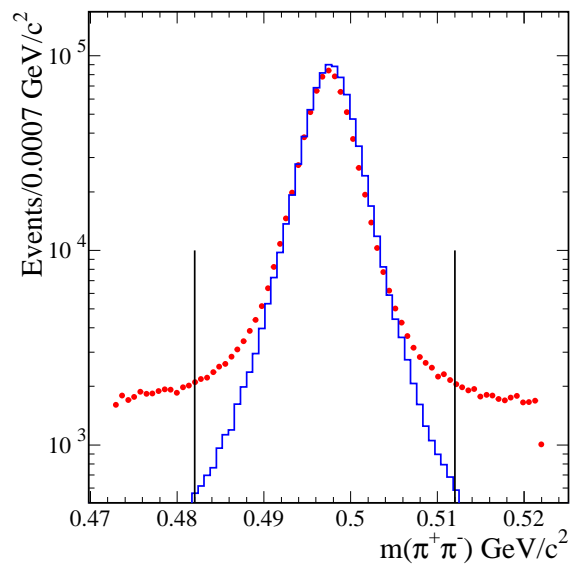}
\vspace{-0.4cm}
\caption{The $\pipi$ invariant mass distribution for the selected
  $\KS$ candidates for the data (points) and simulation (histogram).
  The vertical lines indicate the signal region.
}
\label{ksmass}
\end{center}
\end{figure}
\section{\boldmath The ISR photon and $\KS$ selection}
\label{sec:KS}
Photons are reconstructed as clusters of energy deposits in contiguous
crystals of the EMC.
We consider the cluster in the event with the highest energy in the
\epem c.m. frame, 
and require ISR event candidates to contain a cluster with
$E^\gamma_{\rm c.m.} > 3~\gev$, which we denote as the ISR photon.
The ISR photon detection efficiency has been studied using $\mu\mu\gamma$
events~\cite{isr2pi}, 
and we apply a polar-angle-dependent correction of typically
$-$1.5$\pm$0.5\% to the simulated efficiency.

In these events, 
we reconstruct \KS candidates decaying to two charged pions from pairs
of oppositely charged tracks not identified as electrons.
They must have a well reconstructed vertex between 0.2 and 40.0~cm in
radial distance from the beam axis,
and their total momentum must be consistent with originating from the 
interaction region. 
The $m(\pipi)$ invariant mass distribution for these \KS candidates 
is shown in Fig.~\ref{ksmass} for both data (points) and a simulation
(histogram) containing only genuine \KS.
The background level is relatively low and we select candidates in the
$482< m(\pipi)<512~\mevcc$ mass range (vertical lines on
Fig.~\ref{ksmass}), 
and use the sidebands 472--482 and 512--522~\mevcc 
to estimate the contribution from non-\KS backgrounds.

A few thousand events (about 1\% of the total number of events) have more than one selected \KS
candidate, 
and we use these to study the $\KS\KS\pipi$ and $\KS\KS\Kp\Km$
final states. 
Considering only the ``best'' \KS candidate, 
with $m(\pipi)$ closest to the nominal~\cite{PDG} \KS mass,
we also include these events in the $\KS \KL$ and $\KS \KL \pipi$
measurements. 
The \KS detection efficiency has been studied very carefully at \babar,
with data-MC differences in the efficiency determined as a function of
the \KS direction and momentum. 
We apply a correction event by event, 
which introduces an overall correction  $+$1.1$\pm$1.0\% to the number
of \KS.

We also require the event to contain exactly zero or two 
tracks that are consistent with originating from the interaction region, 
excluding those in the selected \KS candidate(s).
Any number of additional  tracks and EMC clusters is allowed.

\section{\boldmath $\KL$ detection and efficiency}
\label{sec:KL}
The decay length of the \KL meson is large, 
and the probability to detect a \KL decay in the DCH is low. 
Instead, we look for a cluster in the EMC resulting from the
interaction of a \KL with a nucleus in the EMC material.
Such clusters are indistinguishable from photon-induced clusters, 
and give poor resolution on the \KL energy.

In this section, we describe the use of a
clean sample of $\epem\to\phi\gamma$, $\phi\to\KS\KL$ events
to optimize our selection of \KL clusters and measure their detection
efficiency and angular resolution.
In sections~\ref{sec:phiparams} and~\ref{sec:xskskl},
we describe the use of the selected
\KL candidate clusters to study the $\phi$ resonance
and measure the $\epem\to\KS\KL$ cross section above the $\phi$
region, respectively.

\subsection{\boldmath The $\epem \to \phi\gamma \to \KS\KL\gamma$ process}
\label{sec:phigamma}
Using the four-momenta of the best selected \KS, the ISR photon, 
and the initial electron and positron,
we can calculate the recoil-mass-squared, 
\begin{equation}
m^2_{\rm rec} = (E_0 - E_{\gamma} - E_{\KS})^2 - 
(\vec p_0 - \vec p_{\gamma} - \vec p_{\KS})^2,
\label{mkl}
\end{equation}
where $E_0 = E^+ + E^-$ and $\vec p_0 = \vec p^+ + \vec p^-$ are the
energy and total momentum vector of the initial \epem system,
$E_{\gamma}$ and $\vec p_{\gamma}$ (with $E_{\gamma}\equiv |\vec p_{\gamma}|$) are the energy and momentum vector of the photon, and
$E_{\KS}$ and $\vec p_{\KS}$ are the energy and momentum vector of the
\KS candidate.
The presence of the reaction $\epem \to \KS \KL\gamma$ would be
evident as a peak in the \mrec distribution at the mass of the \KL.

\begin{figure}[tbh]
\begin{center}
\includegraphics[width=0.9\linewidth]{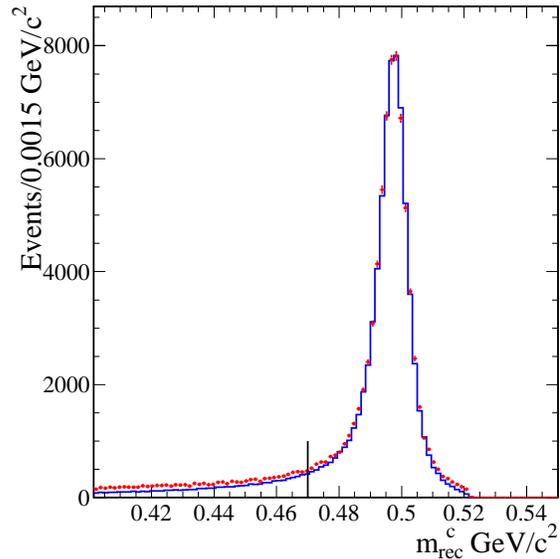}
\vspace{-0.4cm}
\caption{
The distribution of constrained recoil mass \mrecc,
obtained according to Eqs.~\ref{mkl} and~\ref{egam},
for selected $\gamma\KS$ candidates.
The points represent the data, 
and the histogram an MC simulation of
$\epem\to\gamma\phi\to\gamma\KS\KL$ events, 
normalized to the two most populated bins.
}
\label{klmass}
\end{center}
\end{figure}

Due to the large uncertainty of the measured ISR photon energy, 
the calculated value of \mrec also has a large uncertainty. 
However, 
if we assume the reaction $\epem \to \gamma\phi(1020) \to \gamma\KS\KL$, 
we can calculate the constrained ISR photon energy $E^c_{\gamma}$
according to:
\begin{equation}
E^c_{\gamma} = \frac{E^2_0 - p^2_0 - m^2_{\phi}}{2(E_0 - \vec p_0\cdot\vec
  n_{\gamma})},
\label{egam} 
\end{equation}
where $\vec n_{\gamma}$ is a unit vector along the ISR photon
direction and $m_{\phi}$ is the $\phi$ meson mass~\cite{PDG}.   
Using $E^c_{\gamma}$ instead of the measured $E_{\gamma}$ in
Eq.~\ref{mkl}, 
we obtain a much better resolution on the recoil mass \mrecc for
genuine events of that type.
The \mrecc distribution for our data is shown in Fig.~\ref{klmass}
as the points.
A simulated distribution for genuine 
$\epem \!\to\! \gamma\phi \!\to\! \gamma\KS\KL$ events is shown as the histogram.
Selecting events with $\mrecc > 0.4$~\gevcc 
(corresponding to $m(\KS \KL)<1.1$~\gevcc),
with the additional requirement
that there be no  other track within
a 0.2~cm radius of the interaction point, 
we obtain a very clean sample of $\KS \KL\gamma$ events, without any need to 
detect the \KL meson.

\begin{figure}[tbh]
\begin{center}
\includegraphics[width=0.9\linewidth]{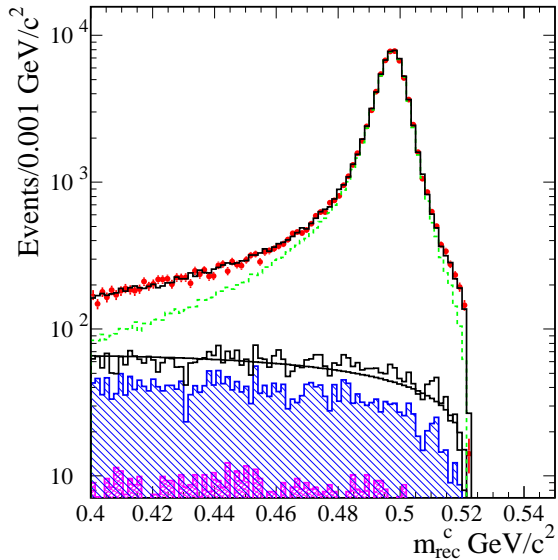}
\vspace{-0.4cm}
\caption{
The experimental \mrecc distribution (points) compared with our
estimated background contributions from (cumulatively):
$\KS \KL 2\piz\gamma$ (shaded area), 
$\KS \KL  \piz\gamma$ (hatched), and
$\phi(\KS \KL)\eta\gamma$ (open histogram). 
The simulated signal distribution is shown as the dashed histogram and
the sum of all simulated events as the solid histogram.
}
\label{klmass2}
\end{center}
\end{figure}

The non-$\KS$ background, 
estimated from the sidebands of the $m(\pip\pim)$ distribution in
Fig.~\ref{ksmass},
contributes 0.8\% of the events in Fig.~\ref{klmass}. 
This background arises from $\epem \to \gamma\gamma$ events in which
one photon converts to a misidentified electron-positron pair. 
We estimate backgrounds from other ISR final states containing a real
\KS using the simulation.
Normalized contributions to the \mrecc distribution for
$\KS\KL 2\piz$, $\KS \KL\piz (K^{*0}\overline{K})$, and  
$\phi(\KS \KL)\eta$ are shown in Fig.~\ref{klmass2}, 
cumulatively, as shaded, hatched, and open histograms. 
The simulated backgrounds from  $\epem\to q \qbar$ $(q = u, d, s, c)$,
and $\epem\to\tau\tau$ events are found to be negligible.


Fitting the simulated non-$\KS$ and ISR backgrounds with smooth
functions and summing them together with the signal simulation,
we obtain excellent agreement with the observed spectrum.
The total background is 6.9$\pm$0.5\% of the selected events.

The position of the $\KL$ peak in Fig.~\ref{klmass2} is very sensitive
to both the reconstructed $\KS$ candidate mass
and the assumed $\phi$-meson mass (see Eqs.~\ref{mkl},\ref{egam}).
There is a small 0.21$\pm$0.02~\mevcc data-MC difference in the \KS
peak position in Fig.~\ref{ksmass}.
As a cross check,
we correct the data for this difference and 
vary $m_\phi$ in Eq.~\ref{egam} for the data so that the experimental
\mrecc peak position matches that of the simulation.
This results in an estimate of $m_\phi = 1019.480\pm0.040\pm0.036$\mevcc, 
where the systematic uncertainty includes the effects of 
the nominal \Kz mass (0.024~\mevcc ~\cite{PDG}),
the $\KS$ momentum measurement in the DCH (0.020~\mevcc), and 
the DCH-EMC mis-alignment (0.018~\mevcc). 
This is consistent with the value, tabulated by the Particle Data Group (PDG)  
$m_\phi = 1019.455\pm0.020$\mevcc~\cite{PDG}.

Subtracting the non-$\KS$ and ISR-produced backgrounds,
we obtain $81\,012\pm285$ ($447\,434$ for the MC simulation) $\KS \KL\gamma$
events in the $\phi$ mass region without requiring $\KL$ detection. 

These events must satisfy our trigger and software filters,
which were designed for various classes of events.
We study efficiencies in data and simulation
using prescaled events not subject to these filters,
and obtain a correction of $(+3.9\pm2.3)$\%.
Furthermore, 
the pions from $\KS$ decays in this particular reaction have a
relatively large probability to overlap in the DCH, 
and the reconstruction efficiency for overlapping tracks is not well 
simulated.   
We introduce a $+1.5\pm0.6$\% correction for this effect. 

\subsection{\boldmath The $\KL$ detection efficiency}
\label{sec:KLeff}

We select events with $\mrecc > 0.47$\mevcc (vertical line in
Fig.~\ref{klmass}), 
reducing the background level from 6.9\% to 2.8\%.
Using the $\KS$ and ISR photon angles and momenta, 
we calculate the hypothetical \KL direction for each event, 
and look for an EMC cluster in that direction.

\begin{figure}[tbh]
\includegraphics[width=0.9\linewidth,height=0.8\linewidth]{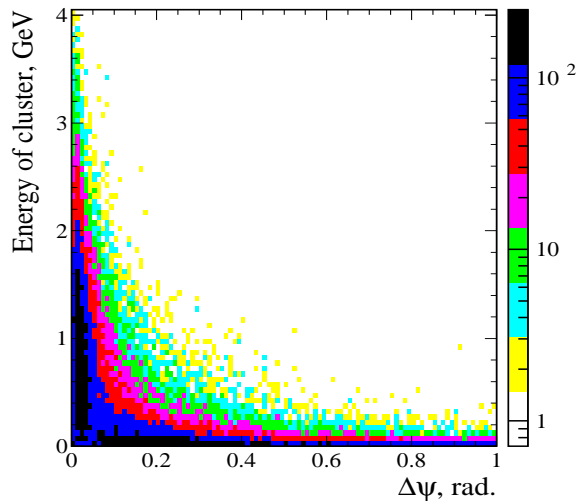}
\vspace{-0.4cm}
\caption{
The EMC cluster energy versus the opening angle between the measured
cluster direction and the predicted $\KL$ direction 
for all non-ISR clusters in the data.
}
\label{evspsi}
\end{figure}
\begin{figure}[tbh]
\includegraphics[width=0.9\linewidth,height=0.8\linewidth]{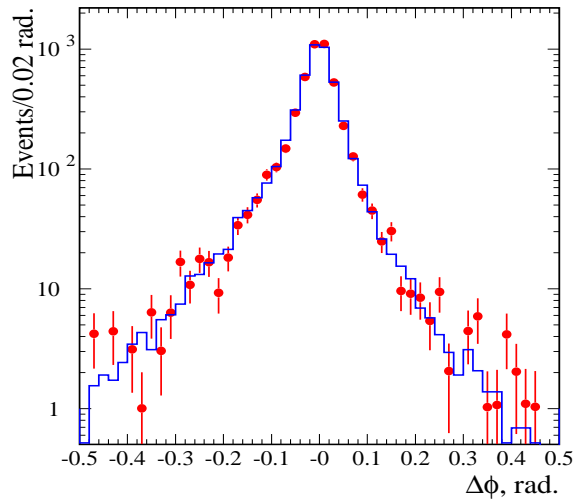}
\vspace{-0.4cm}
\caption{The difference in azimuthal angle between the EMC cluster
  direction and the predicted $\KL$ direction for selected clusters in
  the data (points) and MC-simulated $\KS\KL (\phi)\gamma$ events
  (histogram).
}
\label{dpsi}
\end{figure}
Figure~\ref{evspsi} shows a two-dimensional plot of the EMC cluster energy
versus the opening angle $\Delta\psi$ between the predicted $\KL$
direction and measured cluster direction 
for all clusters in the data except those assigned to the ISR photon.
A clean signal is observed at high cluster energies, but the background
from low-energy clusters is large.
We consider clusters with energy greater than 0.2~\gev,
and select the one closest to the predicted $\KL$ direction if it is
within 0.5 radians.
This yields $\KL$ detection probabilities of about 48\% in the data
and 51\% in the MC simulation.

We then study the resolution in polar ($\theta$) and azimuthal
($\phi$) angle of the selected $\KL$ clusters as a function of their
position in the detector and the predicted \KL energy.
The resolutions in the two angles are consistent, 
with no significant dependence on position or energy.
The overall $\Delta\phi$ distributions are shown for data and
simulation in Fig.~\ref{dpsi}.
Good agreement is seen,
with root-mean-square deviations of 0.035 radians.
We use this value in the kinematic fits.

\section{\boldmath The kinematic fit procedure}
\label{sec:Analysis}
Each candidate event selected in Sec.~\ref{sec:KS} is
subjected to a set of constrained kinematic fits
in which the four-momenta and covariance matrices of 
the initial \epem,
the ISR photon,
the best \KS candidate,
and the two tracks from the interaction region, if present,
are taken into account. 
The three-momentum vectors for each particle including the photon obtained from
these fits are determined with better
accuracy, and are used in further calculations.

First,
we consider each neutral cluster with $E>0.2$~\gev 
(excluding the ISR photon) 
as a \KL candidate,
and perform a three-constraint (3C) kinematic fit under the 
$\KS \KL\gamma$ or $\KS \KL\pipi\gamma$ hypothesis.
The angular resolutions for $\KL$ clusters discussed in the previous
section are used, and the \KL momentum is determined in the fit.
We retain the \KL candidate cluster giving the best \chisq value in
each event.

We then perform a kinematic fit under the $\KS
K^{\pm}\pi^{\mp}\pi^0\gamma$ hypothesis, 
where the cluster is assumed to be one photon from a \piz decay, 
rather than a \KL.
Such events can enter the sample if a charged kaon is misidentified
as a pion and only one photon from the \piz decay is considered.
Similarly, we perform fits under the hypotheses of the other
backgrounds discussed in Sec.~\ref{sec:babar},
giving us additional \chisq variables with which to suppress
these processes.

We perform additional fits to the events with more than one \KS
candidate under the $\KS \KS\pipi\gamma$ and $\KS\KS\Kp\Km\gamma$
hypotheses.
For each pair of $\KS$ candidates, 
a four-constraint (4C) kinematic fit is performed
using the four-momenta and covariance matrices of all initial- and
final-state particles.
The combination with the best \chisq for each hypothesis is retained.

\section{\boldmath The $\KS \KL $ final state ($m(\KS \KL)<1.08~\gevcc$)}
\label{sec:phiparams}
\subsection{Additional selection criteria and background subtraction}
\label{selcriphi}
\begin{figure}[tbh]
\begin{center}
\includegraphics[width=0.9\linewidth]{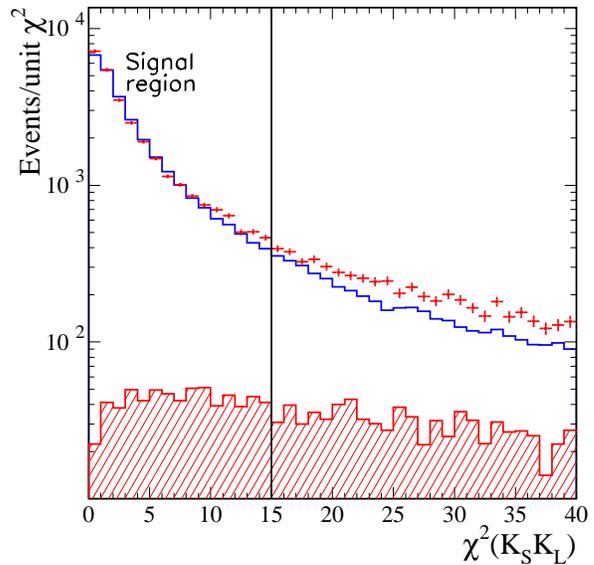}
\vspace{-0.4cm}
\caption{
The three-constraint \chisq distribution for $\KS$-cluster-$\gamma$
events in the data (points) fitted under the $\KS \KL\gamma$ hypothesis. 
The open histogram represents the same distribution for
the MC-simulated signal events,
normalized to the data in the region $\chisq < 10$,
and the shaded histogram represents the estimated background in the
data.
}
\label{kskl_chi2_phi}
\end{center}
\end{figure}

To study this mass region, we consider events selected as described in 
Sec.~\ref{sec:phigamma},
with $\mrecc > 0.4$~\gevcc (see Fig.~\ref{klmass}).
We select a \KL cluster where possible,
using the 3C fits described in Sec.~\ref{sec:Analysis}, 
and obtain the \chisq distribution for the best $\KS \KL\gamma$ candidate
shown in Fig.~\ref{kskl_chi2_phi} as the points.
The unshaded histogram is for the corresponding MC-simulated pure 
$\KS \KL\gamma$ events,
normalized to the data in the region $\chi^2<10$ where we expect very
low background.

The experimental and simulated distributions are broader than a
typical 3C \chisq distribution due to multiple soft-photon emission
from the initial state,
which is not taken into account in the fit, 
but is present in both the data and simulation. 
The observed difference at higher \chisq values is due to background
in the data and possibly a data-MC difference in the angular
uncertainty of the $\KL$ cluster. 

For further analysis we require $\chisq(\KS \KL)<15$
(vertical line in Fig.~\ref{kskl_chi2_phi}), 
and for these events we calculate the $\KL$ candidate mass according
to Eqs.~\ref{mkl},\ref{egam} and perform the background subtraction
described in Sec.~\ref{sec:phigamma}.
We obtain $27\,925\pm 176$ events for the data 
(871 background events are subtracted) 
and $164\,179$ events for the MC-simulation, 
representing samples with the $\KL$ detected.
Dividing by the corresponding numbers of events before $\KL$ cluster 
selection, 
we obtain $\KL$ detection efficiencies,
including the effects of the kinematic fit and \chisq selection, 
of 0.3447$\pm$0.0017 for the
data and 0.3724$\pm$0.0008 for the simulation.
The double ratio 0.9394$\pm$0.0052 is applied as a correction factor
to account for this data-MC difference.
This ratio is independent of momentum and polar angle of the $\KL$.
\begin{figure}[tbh]
\begin{center}
\includegraphics[width=0.9\linewidth]{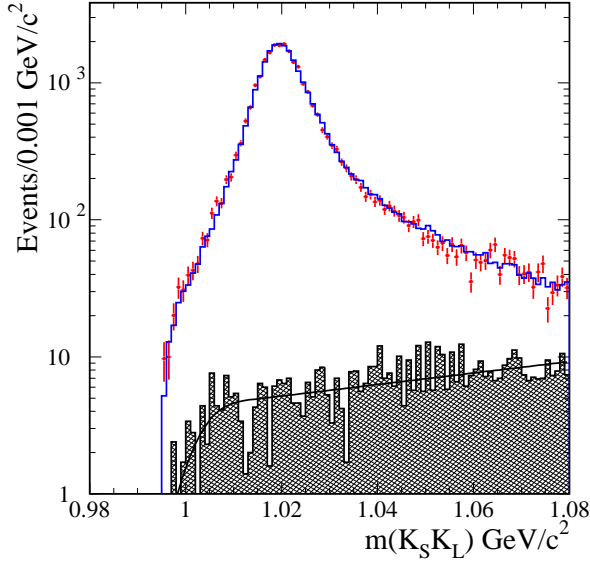}
\vspace{-0.4cm}
\caption{
The $\KS \KL$ invariant mass distribution in the data (points) and
signal-MC simulation (histogram) for candidate events in the
signal region of Fig.~\ref{kskl_chi2_phi}. 
The shaded histogram represents the estimated background, 
and the line is a smooth parametrization thereof.
}
\label{kskl_bkg}
\end{center}
\end{figure}

We use the four-vectors returned by the kinematic fit to calculate 
the $\KS \KL$ invariant mass,  the distribution 
of which is shown in Fig.~\ref{kskl_bkg}. 
The $\phi(1020)$ resonance is clearly visible, 
with a width of about 10~\mev,
much larger than the nominal width of the resonance~\cite{PDG}
due to the resolution of this final state.
The background, estimated as described above, 
is shown as the shaded histogram.
We fit it with a smooth, empirical function,
shown as the line,
and use the fit result in each bin for background subtraction.

\subsection{\boldmath Fit for the $\phi(1020)$ parameters}
\label{sec:xsphi}
To obtain the parameters of the $\phi(1020)$, 
we fit the background-subtracted distribution in Fig.~\ref{kskl_bkg}
with a cross section $\sigma(s)$ convolved with a resolution matrix
${\rm Res}(j,i)$. 
In each mass bin $i$:
\begin{eqnarray}
 N_{\KS \KL}(i)& = & \sum_{j=1}^{100}{N_{0}(j)\cdot
  {\rm Res}(j,i)}, \nonumber \\
N_{0}(j) & = & \int_{m_j}^{m_{j+1}} \sigma(s)\epsilon(s)\epsilon_{\KS \KL}^{\rm corr}L(s)ds,  
\label{nkskl}
\end{eqnarray}
where $s=m(\KS \KL)^2$, 
$\epsilon(s)$ is the simulated detection efficiency,
$\epsilon_{\KS \KL}^{\rm corr} = 0.939\cdot0.985\cdot0.961$ is the
data-MC efficiency correction factor for the \chisq cut, 
track overlap, and event filter,
$L(s)$ is the ISR luminosity,
calculated at leading-order~\cite{ivanch},
and $N_{0}(j)$ is the acceptance-corrected 
number of events expected for bin $j$.

\begin{figure}[tb]
\begin{center}
\includegraphics[width=0.9\linewidth]{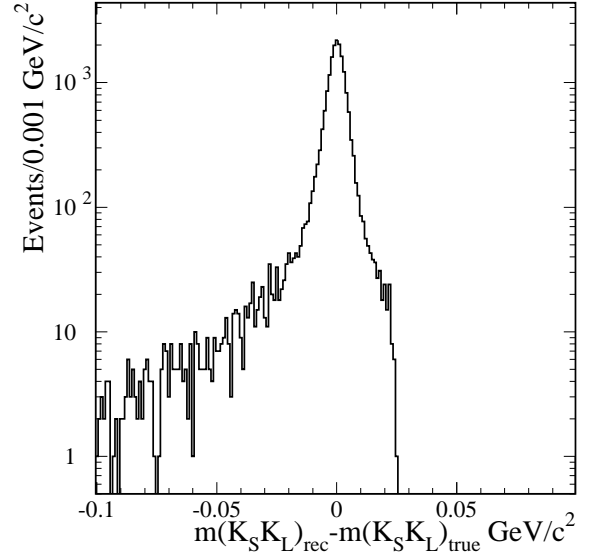}
\vspace{-0.4cm}
\caption{
The simulated distribution of differences between the reconstructed
and generated $\KS \KL$ invariant mass for the 1~\mev bin of
reconstructed mass at the $\phi$ peak. 
}
\label{mc_res}
\end{center}
\end{figure}

The 100$\times$100 resolution matrix is obtained from simulation
by binning the reconstructed versus simulated $\KS \KL$ invariant mass for
signal events in 1$\times$1~\mevcc intervals.
The distribution of differences between the reconstructed and simulated
masses  near 1.020~\gevcc,
corresponding to a row of this matrix,
is shown in Fig.~\ref{mc_res}.
The $\KS\KL$ threshold and a radiative tail are visible. 
We normalize each row to unit area,
and introduce an additional variable Gaussian smearing $\sigma_{\rm add}$, 
to account for any data-MC difference in the resolution.

We describe the cross section near the $\phi$ resonance
using formulae discussed in detail in Refs.~\cite{cmdphi, sndphi}:
\begin{eqnarray}
\sigma(s)&=&\frac{1}{s^{5/2}}\frac{q^3_{\KS \KL}(s)}{q^3_{\KS \KL}(m_{\phi}^2)}
\Bigg| \frac{\Gamma_{\phi}m^3_{\phi}\sqrt{m_{\phi}\sigma_{\phi\to \KS \KL}
    /C}}{D_{\phi}(s)} \nonumber \\
&-&\frac{\sqrt{\Gamma_{\phi}\Gamma_{\rho}m^3_{\rho}m^2_{\phi}6\pi\BR(\rho\to
  e^+ e^-)\BR(\phi\to \KS \KL)}}{D_{\rho}(s)} \nonumber\\
&+&\frac{\sqrt{\Gamma_{\phi}\Gamma_{\omega}m^3_{\omega}m^2_{\phi}6\pi\BR(\omega\to
  e^+ e^-)\BR(\phi\to \KS \KL)}}{D_{\omega}(s)} \nonumber\\
&+&A_{\KS \KL} \Bigg| ^2,
\label{sigmak}
\end{eqnarray}
where $q_{\KS \KL}(s) = \sqrt{s-4m^2_{K^0}}$ is a threshold term, 
$\sigma_{\phi\to \KS \KL}$ is the peak cross section value, 
$D_V(s) = s - m_V^2 + i\sqrt{s}\Gamma_V(s)$ is the propagator for a
vector resonance $V$,
 $C = 0.389\times 10^{12}$~nb\mev$^2/c^4$~\cite{PDG},
 and 
\begin{equation}
\Gamma_{V}(s) = \Gamma_{V}\cdot\sum_{V\to f}{\BR(V\to f)\frac{P_{V\to
      f}(s)}{P_{V\to f}(m_{V}^2)}}
\label{widths}
\end{equation}
describes the energy-dependent width, and for the $\phi$ we use the set of final states 
$f = \Kp\Km$, $\KS\KL$, $\pipi\piz$, and $\eta\gamma$,
with corresponding branching fractions $\BR(V\to f)$ and phase space
factors $P_{V\to f}(s)$.
We include the influence of the $\rho(770)$ and $\omega(782)$ 
resonances in the in the energy-dependent width according
to the ``ideal'' quark model,  
which assumes their decay rates to $\KS \KL$ are a factor of two lower
than that of the $\phi$. 
We use the relation 
\begin{equation}
\sigma(V\to f) = \frac{12\pi\BR(V\to e^+ e^-)\BR(V\to f)}{m^2_V}C
\label{sigbr}
\end{equation}
in Eq.~\ref{sigmak} for the corresponding cross sections.

We introduce a complex constant $A_{\KS \KL}$ to describe the
contributions of higher radial excitations of the $\rho$, $\omega$
and $\phi$ mesons to the cross section,
as well as any deviations from the ``ideal'' quark structure relations 
for the $\rho(770)$ and $\omega(782)$.
It can be written in terms of two free parameters,
a non-resonant cross section $\sigma_{\rm bkg}$, and a phase $\Psi$,
\begin{equation}
A_{\KS \KL} = m^2_{\phi}\sqrt{\sigma_{\rm bkg}m_{\phi}/C}\cdot e^{-i\Psi}.
\label{bkgamp}
\end{equation}
The fitted value of $\Psi$ is consistent with zero,
and we fix it to zero in the final fit, 
but propagate its fitted uncertainty as a systematic uncertainty
to account for model dependence.

\begin{figure}[tbh]
\begin{center}
\includegraphics[width=0.9\linewidth]{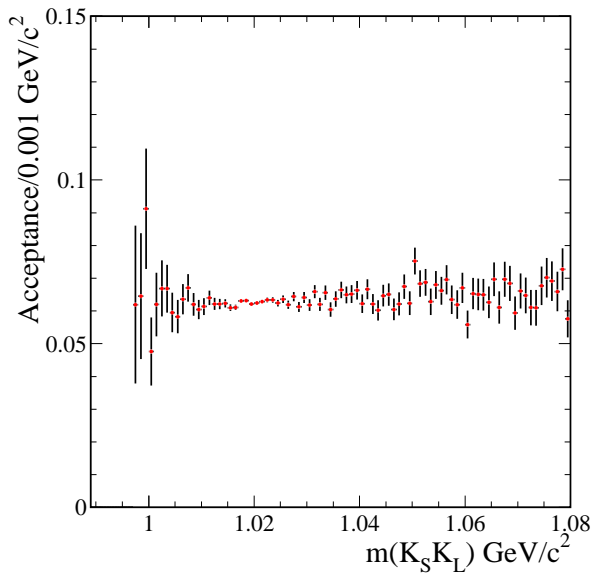}
\vspace{-0.4cm}
\caption{The simulated detection efficiency $\epsilon(s)$
  versus the generated $\KS\KL$ invariant mass,
  calculated by dividing the number of events in the signal region of
  Fig.~\ref{kskl_chi2_phi} by the number generated in each bin.
}
\label{mc_acc}
\end{center}
\end{figure}
\begin{figure}[tbh] 
\begin{center}
\includegraphics[width=0.9\linewidth]{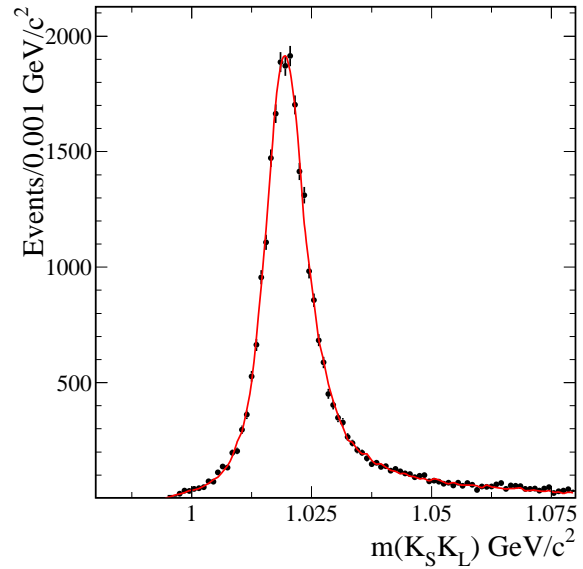}
\vspace{-0.4cm}
\caption{
The $\KS \KL$ invariant mass distribution in the $\phi(1020)$ region.
Only statistical uncertainties are shown. 
The curve represents the result of the fit described in the text.
}
\label{kskl_ee_babar}
\end{center}
\end{figure} 

The detection efficiency,
shown as a function of mass in Fig.~\ref{mc_acc},
is obtained by dividing the number of selected MC-simulated events
in each 0.001~\gevcc mass interval by the number generated in the same
interval.
The mass dependence is well described by a linear fit,
which we use in all calculations.
This efficiency includes the geometrical acceptance of the detector 
for the final-state photon and the charged pions from the $\KS$ decay, 
the inefficiency of the detector subsystems, 
and event losses due to additional soft-photon emission from the
initial state. 
It is not sensitive to the detector mass resolution.

The result of the fit is projected on the background-subtracted
invariant mass distribution in Fig.~\ref{kskl_ee_babar}.
We obtain the resonance parameters:
\begin{eqnarray*}
 \sigma_{\phi} & = &  1409\pm33\pm42\pm15~{\rm nb}, \\
      m_{\phi} & = & 1019.462\pm0.042\pm0.050\pm0.025~\mevcc, \\
 \Gamma_{\phi} & = & 4.205\pm0.103\pm0.050\pm0.045~\mev,\\
 \sigma_{\rm bkg} & = & 0.022 \pm 0.012~{\rm nb},
\end{eqnarray*}
where the first uncertainties are statistical, the second systematic, 
and the third due to model dependence, evaluated by varying
$\sigma_{\rm bkg}$ by its uncertainty.

We introduce an additional Gaussian
smearing to describe an uncertainty in the detector  resolution, and
obtain $\sigma_{\rm add}=0.6\pm0.2$~\mevcc, 
which improves the \chisq of the fit in the 1.0-1.05~\gevcc
region from 59 to 53, for 51 degrees of freedom.
We estimate systematic uncertainties of 0.05~\mevcc in mass and
0.05~\mev in width from the uncertainty of the $\sigma_{\rm add}$ value.
The other systematic uncertainties are summarized in
Table~\ref{error_tab}, 
along with the corrections applied to the measurements.
A total correction of $+$14.1$\pm$2.9\% is applied to the number of
events. 
The largest contribution to the uncertainty is from the software filter,
due to the limited number of available prescaled events.

\begin{table}[tbh]
\caption{
Summary of corrections and systematic uncertainties for the
measurement of the $\epem\to \KS \KL$ process in the $\phi$ resonance
region.
}
\label{error_tab}
\begin{tabular}{l c c} 
\hline\hline
Source & Correction & Uncertainty \\
\hline
Background filter efficiency & $+3.9\%$ & $2.3\%$ \\ 
Photon detection efficiency  & $+1.5\%$ & $0.5\%$ \\
$\KL$ detection efficiency   & $+6.1\%$ & $0.6\%$ \\ 
$\KS$ detection efficiency   & $+1.1\%$ & $1.0\%$ \\
Track overlap                & $+1.5\%$ & $0.6\%$ \\
ISR luminosity               & --       & $0.5\%$ \\
Backgrounds                  & --       & $0.5\%$ \\
Radiative corrections        & --       & $1.0\%$ \\
\hline
Total  (sum in quadrature)   & $+14.1\%$& $2.9\%$ \\
\hline\hline
\end{tabular}
\end{table}

Our parameter values are consistent with 
the most precise cross section measurement, 
$\sigma_{\phi} =1376\pm24~{\rm nb}$~\cite{cmdphi}, 
and with the PDG values $m_{\phi}=1019.455\pm0.020~\mevcc$
and $\Gamma_{\phi} = 4.26\pm0.04~\mev$ ~\cite{PDG}.
Since each row of the resolution matrix is normalized to unit area, 
the smearing procedure does not affect the total number of events,
which is proportional to the product $\Gamma_{\phi}\sigma_0$ of the
total width and peak cross section of the $\phi$.
Using this product as a free parameter in the fit, 
we obtain the product of the electronic width of the $\phi$ and its
branching fraction to $\KS\KL$,
\begin{equation*}
  \Gamma_{ee}\BR_{\KS \KL}  = 
0.4200\pm0.0033\pm0.0122\pm0.0019~\kev,
\end{equation*}
where the first uncertainty is statistical, the second systematic, 
and the third due to model dependence.
Using $\BR_{\KS \KL} = 0.342\pm0.004$ 
or $\Gamma_{\phi}=4.26 \pm 0.04$~\mev from Ref.~\cite{PDG}, 
we obtain 
$\Gamma_{ee} =  1.228  \pm 0.037 \pm 0.014~\kev$ or
$\BR_{ee}\BR_{\KS \KL} =  0.986\pm0.030\pm 0.009$, respectively,
where the first uncertainty is our total experimental uncertainty
and the second is from the PDG tables.
These values are consistent with the most recent measurement of
$\Gamma_{ee} =  1.235  \pm 0.022$~\kev~\cite{cmdphi2}, 
and with the PDG values of $\Gamma_{ee}=1.27  \pm 0.04~\kev$
and $\BR_{ee}\BR_{\KS \KL}=1.006\pm0.016$~\cite{PDG},
and have comparable precision.
\begin{figure}[tbh]
\begin{center}
\includegraphics[width=0.9\linewidth]{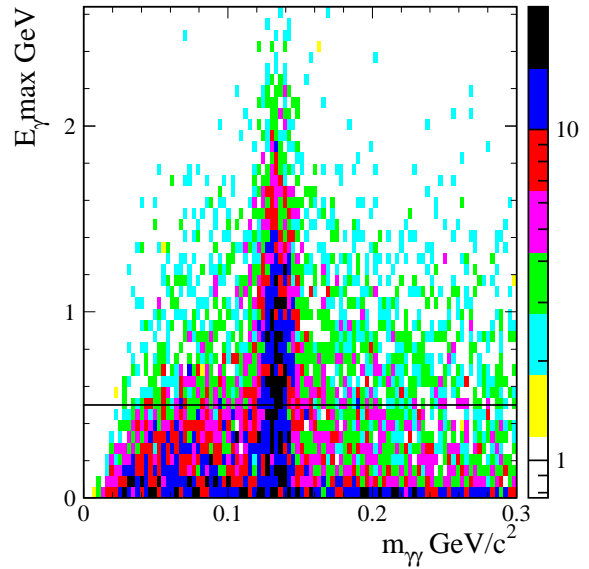}
\vspace{-0.4cm}
\caption{
Two-dimensional plot
of the higher cluster energy in a photon-candidate pair
versus the corresponding diphoton mass  $m_{\gamma\gamma}$ 
for all pairs of EMC clusters, containing neither the ISR photon nor the
$\KL$ candidate.
}
\label{egammax}
\end{center}
\end{figure}
\section{\boldmath The $\KS \KL$ final state ($m(\KS \KL)>1.06~\gevcc$)}
\label{sec:xskskl}
In this section we consider events with $m(\KS \KL)>1.06~\gevcc$.
Since the $\epem\to \KS \KL$ cross section drops much more
rapidly with increasing mass than the background,
we apply additional selection criteria compared to the criteria of Sec.~\ref{selcriphi}.
In all cases, we consider \KS candidates with 
$0.482< m(\pipi)<0.512~\mevcc$ (see Fig.~\ref{ksmass}) and use
sideband data to subtract non-$\KS$ background from all studied
quantities.

\subsection{Additional selection criteria}
\label{addkskl}
We consider all EMC clusters except those assigned to the ISR photon
and the $\KL$ as photon candidates, and combine each pair into a \piz
candidate.
Figure~\ref{egammax} shows 
a scatter plot of the higher of the energies $\rm E_{\gamma}max$
of the two photons assigned to the pair versus the corresponding
diphoton mass  $\rm m_{\gamma\gamma}$.
A large signal from events containing a $\piz$ is observed. 
To reduce this background, 
we require $\rm E_{\gamma}max < 0.5~\gev$ 
(horizontal line in Fig.~\ref{egammax}).
Since signal event may contain several background clusters,
this reduces the signal efficiency.
We measure this loss using events in the $\phi$ region, 
where no $\piz$ signal is observed, 
but background clusters are present in both data and simulation.  
We find losses of 10\% in the data and 7\% in the simulation,
and apply the 3\% difference as a correction.
\begin{figure}[tbh]
\includegraphics[width=0.9\linewidth]{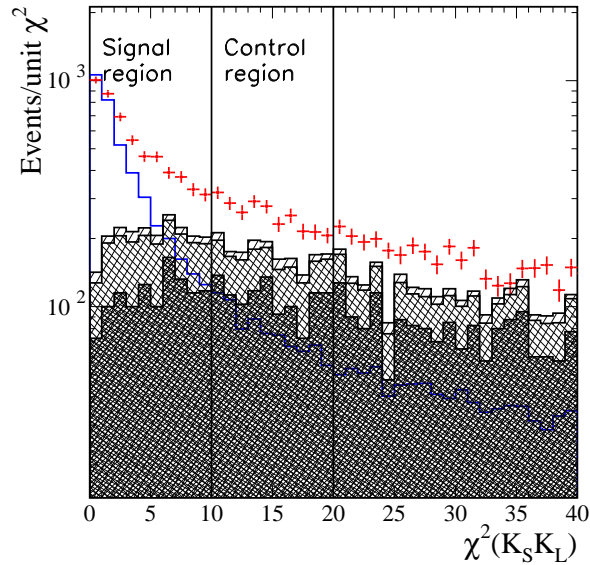}
\vspace{-0.4cm}
\caption{
 The 3C \chisq distributions for $\KS \KL\gamma$ candidate events in
 the data (points) and signal simulation (open histogram),
 fitted under the $\KS \KL$ hypothesis. 
 The shaded, cross-hatched, and hatched areas represent the simulated
 contributions from the ISR $\phi\eta$, $\KS \KL\piz$, and $\KS \KL\ppz$
 channels, respectively. 
}
\label{kskl_chi2_ff}
\end{figure}

The 3C \chisq distribution for the remaining candidate events with
$m(\KS \KL)>1.06~\gevcc$ is shown as the points in
Fig.~\ref{kskl_chi2_ff}.
The open histogram shows the corresponding simulated distribution
for genuine $\KS \KL\gamma$ events,
normalized to the data in the region $\chi^2<3$.
The shaded, cross-hatched and hatched areas represent the simulated
contributions from the ISR channels $\phi\eta$, $\KS \KL\piz$, and
$\KS \KL\ppz$, respectively.
These channels contribute significant background, and almost entirely
account for the difference between the data and
signal-MC \chisq distributions.
We find no significant contribution from simulated non-ISR backgrounds.

We select events with $\chisq(\KS \KL)<10$, 
and use events from the control region $10<\chisq(\KS \KL)<20$ 
(vertical lines in Fig.~\ref{kskl_chi2_ff})
to estimate the background in the signal region. 
The signal region contains 6264 data and $13\,292$ MC-simulated events, 
while the control region contains 2968 and 2670, respectively.
\begin{figure*}[tbh]
\begin{center}
\includegraphics[width=0.33\linewidth]{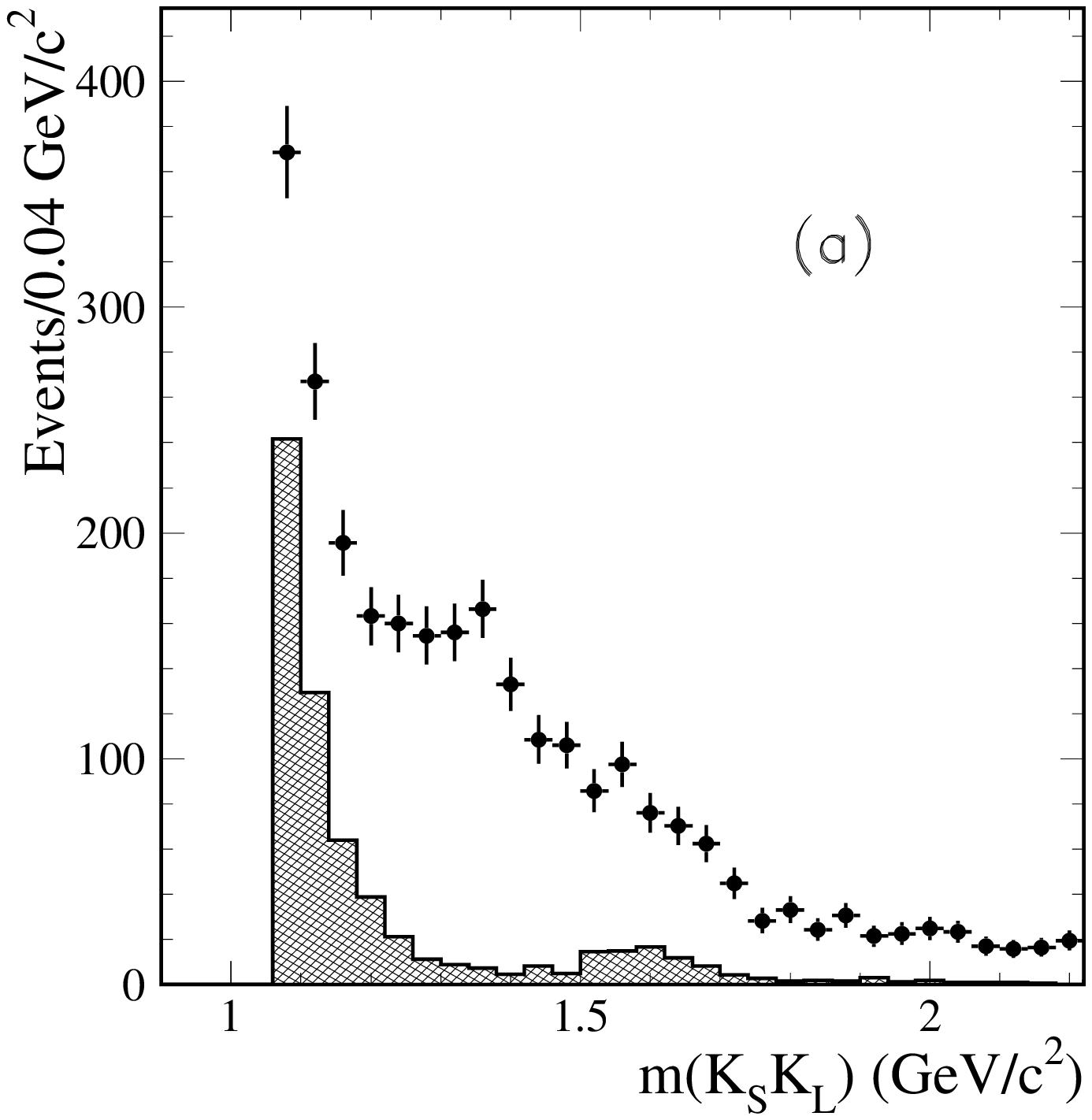}
\includegraphics[width=0.33\linewidth]{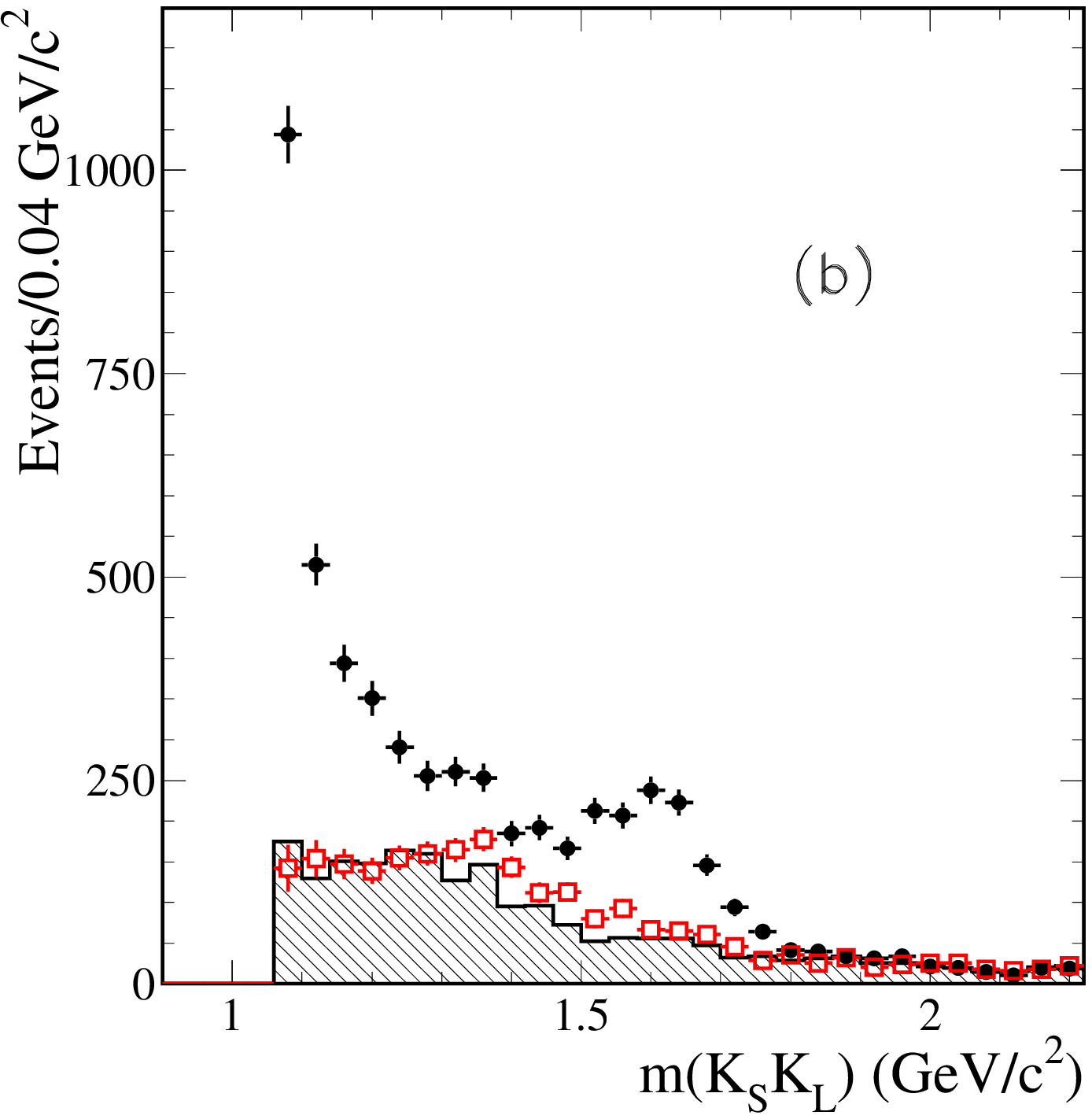}
\includegraphics[width=0.33\linewidth]{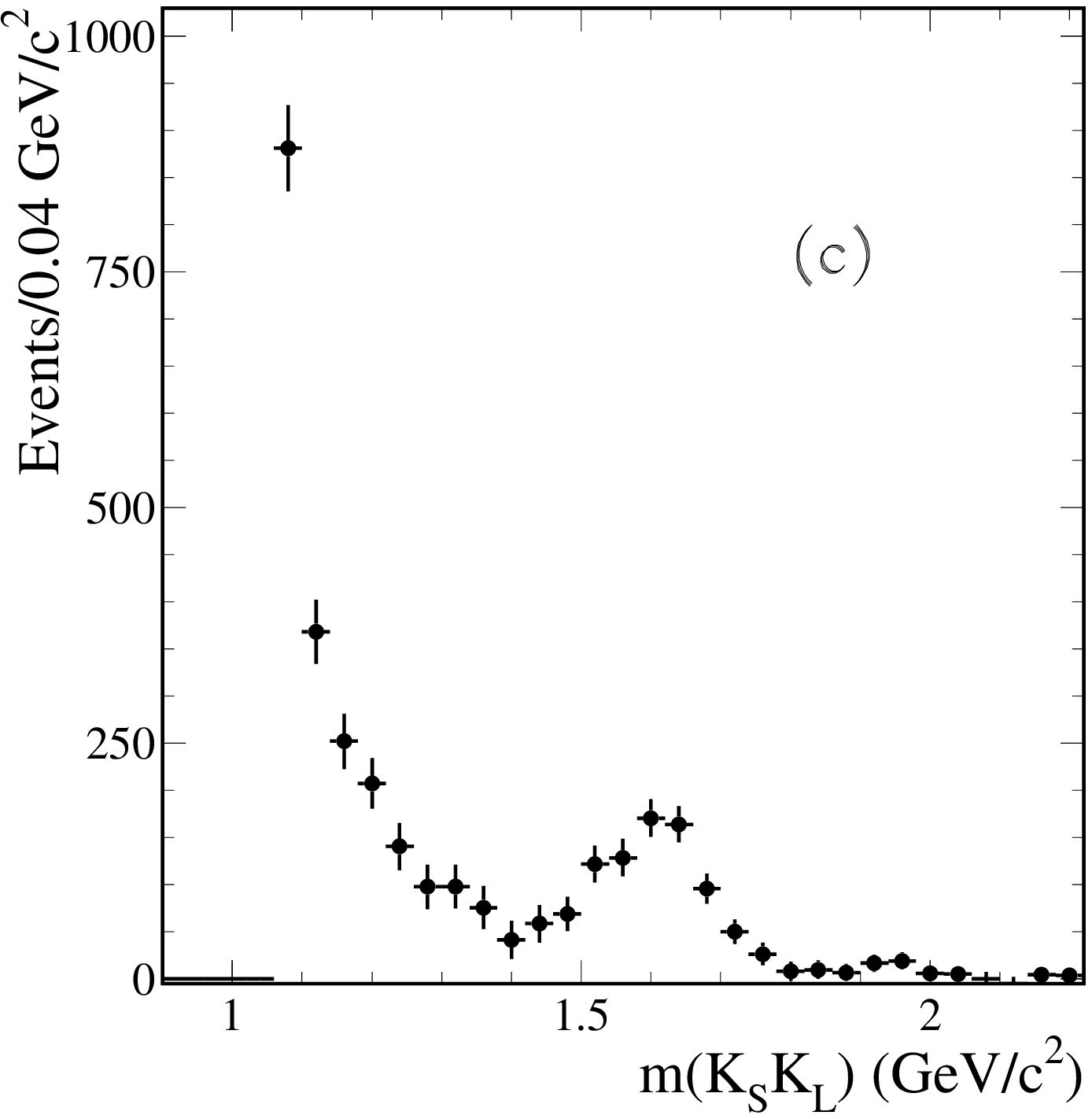}
\vspace{-0.4cm}
\caption{
The $\KS \KL$ invariant mass distribution for data (circles) in
the \chisq control (a) and signal (b) regions
(see Fig.~\ref{kskl_chi2_ff}). 
The histogram in (a) represents MC-simulated signal events in the
control region,
and that in (b) represents the simulated background from 
$\phi\eta$, $\KS\KL\pi^0$, and $\KS \KL 2\pi^0$ events;
The squares show the total estimated background, 
obtained as the difference between the data and simulated
distributions in (a),
normalized to the data in the signal region.
(c)
The $\KS \KL$ invariant mass distribution above 1.06~\gevcc for the
data after background subtraction.
}
\label{kskl_ff_control}
\end{center}
\end{figure*}

\subsection{Background subtraction}
\label{sec:ffbkg}
To obtain any distribution of the $\KS \KL$ signal events $N_0^d(m)$, 
we take the experimental events in the signal region of
Fig.~\ref{kskl_chi2_ff}, $N_s^d(m)$, 
and subtract the background 
events,  taken from the control region $N_c^d(m)$, 
corrected for the presence of signal events, 
estimated from MC-simulation $N_c^{MC}(m)$:
\begin{equation}
N_0^d(m) = N_s^d(m) - b\cdot (N_c^d(m) - a\cdot N_c^{MC}(m)),
\label{sig_contr}
\end{equation}
where $b=1.15$ is the simulated ratio of background events in the
signal and control regions, 
and $a =N_0^d(m)/N_s^{MC}(m)$ is a factor equalizing the number of
signal and simulated events. 

This procedure relies on good agreement between data and simulation in
both the \chisq and mass distributions.
As noted above, the MC simulation uses a ``world average'' cross
section,  
well measured below 1.4~\gev, 
but based on only the DM1 measurement~\cite{kskldm1},
which has large statistical uncertainties,
in the 1.4-2.4~\gev \Ecm region. 
We adopt an iterative procedure,
in which we reweight the simulated mass distribution to match our
measurement and repeat the subtraction until there is no change in
the results.

Figure~\ref{kskl_ff_control}a shows the $\KS \KL$ mass distribution for
data events in the \chisq control region of Fig.~\ref{kskl_chi2_ff} as
points,
with the shaded histogram showing the distribution for signal MC at
the final iteration.
The signal contribution is not large, 
and the difference between the data and weighted MC-simulated
distributions is scaled by $b=1.15$ to estimate the background in the
\chisq signal region.
The squares in Fig.~\ref{kskl_ff_control}b represent this background
estimate in each $\KS\KL$ invariant mass bin.
We also estimate the background directly from the MC simulation of the
ISR $\phi\eta$, $\KS\KL\piz$, and $\KS\KL 2\piz$ processes, 
shown as the histogram in Fig.~\ref{kskl_ff_control}b.
The two estimates agree relatively well, 
but the MC simulation does not incorporate the correct mass distributions for
these processes, 
and other unknown processes might contribute. 
The mass distribution after background subtraction is shown in
Fig.~\ref{kskl_ff_control}c.
In the 1.4-2.4~\gevcc mass region we select about 1000 events,
compared with only 58 events found by the DM1~\cite{kskldm1} experiment.

\begin{figure}[tbh]
\begin{center}
\includegraphics[width=0.8\linewidth]{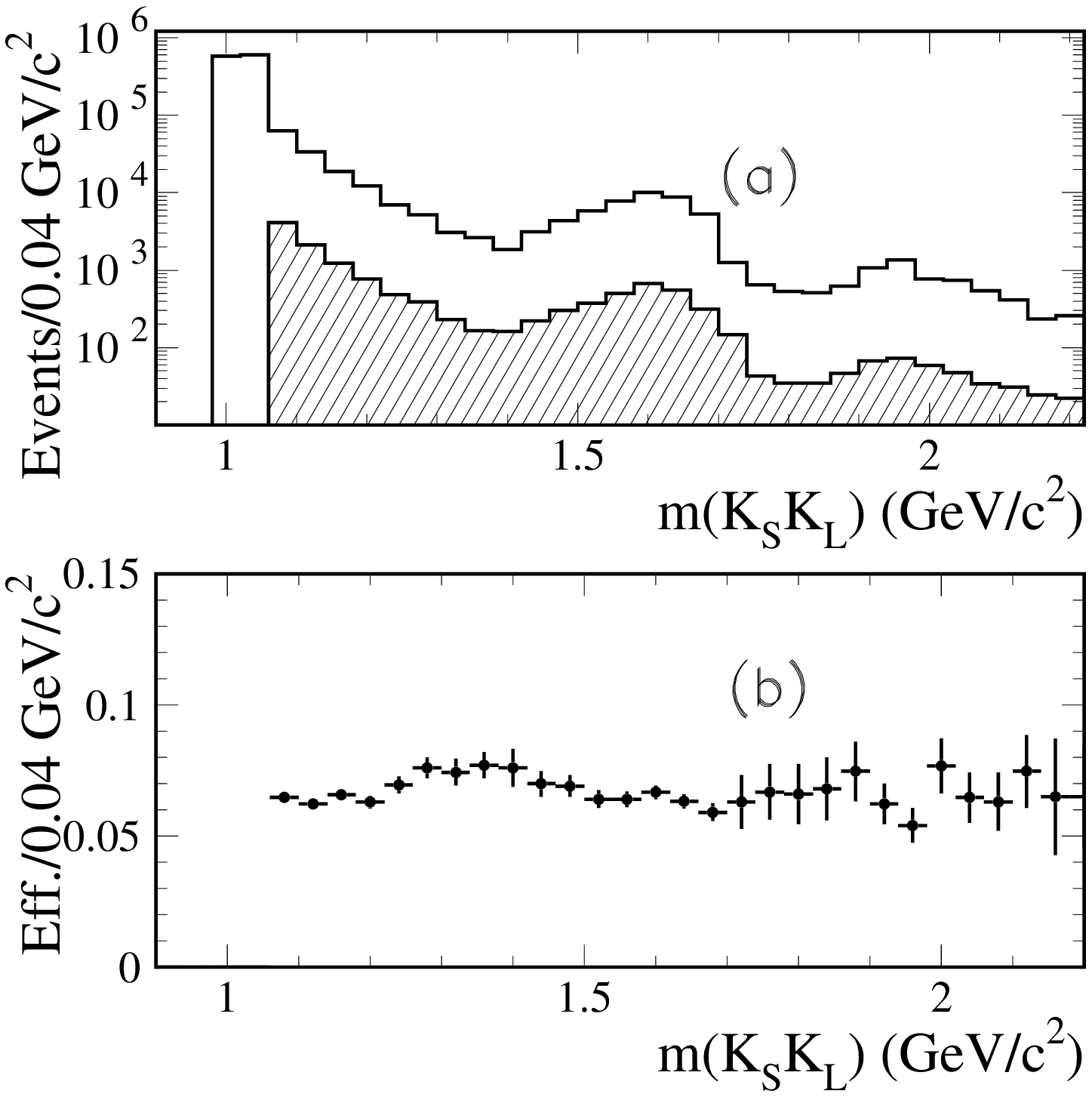}
\vspace{-0.4cm}
\caption{a)
The $\KS \KL$ mass distributions from the MC simulation for the
  signal (unshaded) and control (shaded) regions of
  Fig.~\ref{kskl_chi2_ff}. 
b) The mass dependence of the simulated net reconstruction and
  selection efficiency for the $\KS \KL$ final state in the
  $m(\KS \KL)>1.06~\gevcc$ region. 
}
\label{kskl_ff_eff}
\end{center}
\end{figure}
\begin{table*}
\caption{Summary of the $\epem\to K_S K_L$ 
cross section measurement. Uncertainties  are statistical only.}
\label{kskl_ff_tab}
\begin{ruledtabular}
\begin{tabular}{ c c c c c c c c }
$E_{\rm c.m.}$ (GeV) & $\sigma$ (nb)  
& $E_{\rm c.m.}$ (GeV) & $\sigma$ (nb) 
& $E_{\rm c.m.}$ (GeV) & $\sigma$ (nb) 
& $E_{\rm c.m.}$ (GeV) & $\sigma$ (nb)  
\\
\hline

   1.08 &  6.86 $\pm$  0.43 &   1.36 &  0.40 $\pm$  0.13 &   1.64 &  0.83 $\pm$  0.13 &   1.92 &  0.07 $\pm$  0.05 \\
   1.12 &  2.86 $\pm$  0.30 &   1.40 &  0.22 $\pm$  0.11 &   1.68 &  0.51 $\pm$  0.11 &   1.96 &  0.09 $\pm$  0.05 \\
   1.16 &  1.78 $\pm$  0.24 &   1.44 &  0.32 $\pm$  0.12 &   1.72 &  0.26 $\pm$  0.11 &   2.00 &  0.02 $\pm$  0.03 \\
   1.20 &  1.48 $\pm$  0.23 &   1.48 &  0.36 $\pm$  0.11 &   1.76 &  0.11 $\pm$  0.07 &   2.04 &  0.02 $\pm$  0.03 \\
   1.24 &  0.87 $\pm$  0.18 &   1.52 &  0.66 $\pm$  0.13 &   1.80 &  0.03 $\pm$  0.05 &   2.08 &  0.00 $\pm$  0.03 \\
   1.28 &  0.54 $\pm$  0.14 &   1.56 &  0.67 $\pm$  0.13 &   1.84 &  0.04 $\pm$  0.04 &   2.12 &  0.01 $\pm$  0.02 \\
   1.32 &  0.54 $\pm$  0.15 &   1.60 &  0.84 $\pm$  0.12 &   1.88 &  0.02 $\pm$  0.04 &   2.16 &  0.02 $\pm$  0.03 \\
 
\end{tabular}
\end{ruledtabular}
\end{table*}

\subsection{ Simulated detection efficiency }
\label{sec:ffeff}
The selection procedures applied to the data are also applied to the         
MC-simulated event sample. 
The resulting $\KS \KL$ invariant mass
distribution is shown in Fig.~\ref{kskl_ff_eff}(a) for the signal and
control (shaded histogram) regions.
The mass dependence of the detection
efficiency is obtained by dividing the number of reconstructed MC
events in each mass interval by the number generated in that interval. 
The results are shown in Fig.~\ref{kskl_ff_eff}(b). 
The 40~\mevcc mass intervals used are wider than the detector
resolution of 10~\mevcc, 
but a small effect of the resolution on the efficiency is visible, 
due to the very steep decrease in the cross section with increasing mass.

\begin{figure}[tbh]
\begin{center}
\includegraphics[width=0.9\linewidth]{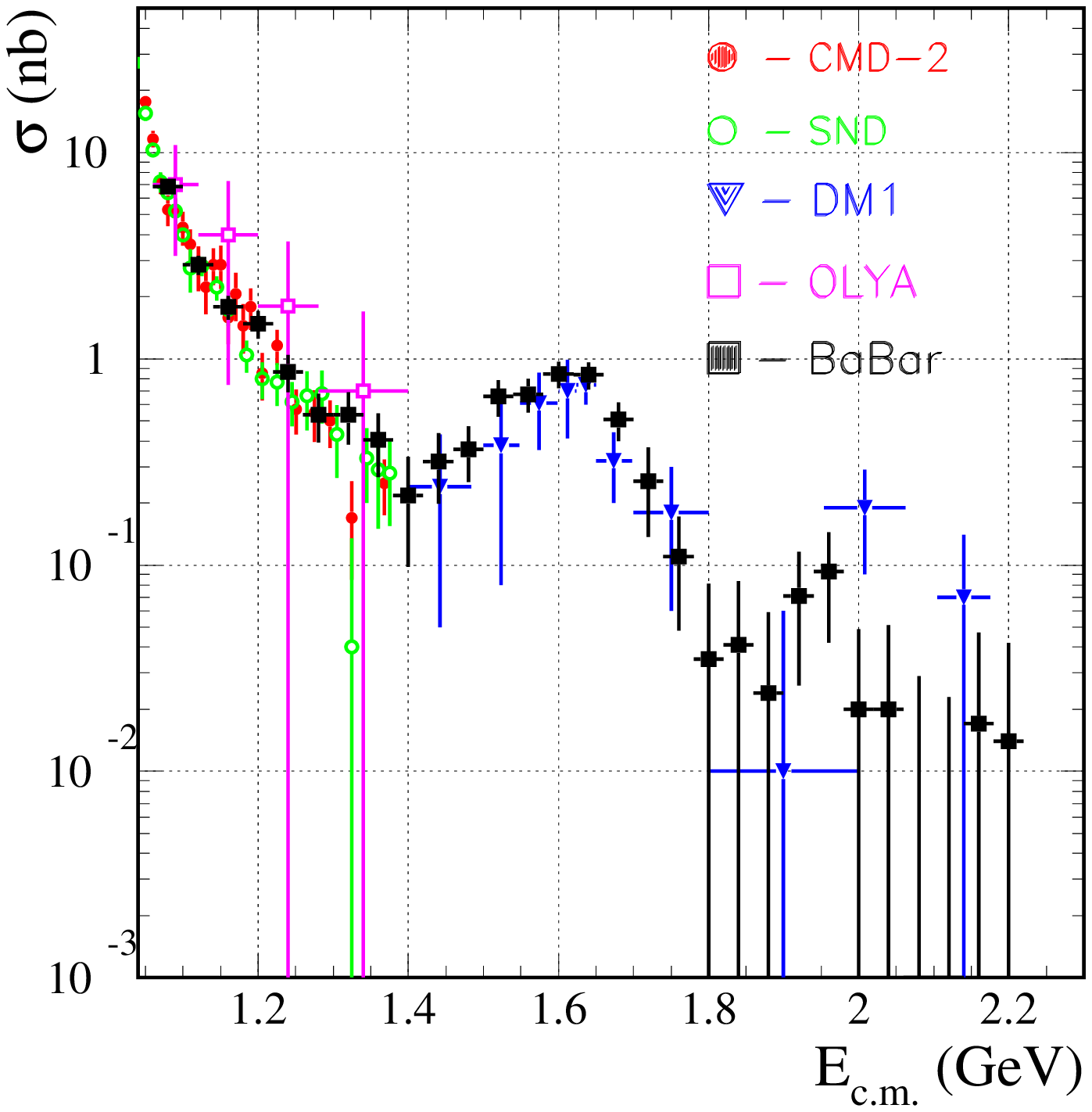}
\vspace{-0.4cm}
\caption{
The $\epem\to \KS \KL$ cross section
compared with all available data in this \Ecm region.
}
\label{kskl_ff_all}
\end{center}
\end{figure} 

\subsection{\boldmath The $\epem\to \KS \KL$ cross section for
  c.m.\ energies above 1.06~\gev}
\label{sec:xsff}
The cross section for \epem annihilation into $\KS \KL$ can be
calculated from 
\begin{equation}
  \sigma(\KS \KL)(E_{\rm c.m.})
  = \frac{dN_{\KS \KL\gamma}(E_{\rm c.m.})}
         {d{\cal L}(E_{\rm c.m.})
          \cdot\epsilon_{\KS \KL}^{\rm corr}
          \cdot\epsilon_{\KS \KL}^{\rm MC}(E_{\rm c.m.})
          \cdot R }  ,
\label{xsformular}
\end{equation}
where:
$E_{\rm c.m.}\equiv m(\KS \KL)$; 
$dN_{\KS \KL\gamma}$ is the number of selected $\KS \KL$ events after
background subtraction in the interval $dE_{\rm c.m.}$; 
$\epsilon_{\KS \KL}^{\rm MC}(E_{\rm c.m.})$ is the corresponding detection
efficiency, estimated from the MC simulation with corrections to the $\KS$
and ISR photon detection efficiencies;
and the factor $\epsilon_{\KS \KL}^{\rm corr}=0.939\cdot 0.961\cdot 0.97$ takes
into account the data-MC differences in $\KL$ detection and background
filter efficiencies, and the energy requirement on additional photon
clusters.
The radiative correction $R$ is within one percent of unity, 
with an estimated precision of about 1\%.
The differential luminosity $d{\cal L}(E_{\rm c.m.})$ associated with
the interval $dE_{\rm c.m.}$ centered at an effective collision
energy of $E_{\rm c.m.}$ is calculated using the leading order formula
(see for example Ref.~\cite{isr3pi}),
and the systematic uncertainty associated with the luminosity
determination is estimated to be 0.5\%.

The cross section for the reaction $\epem\to \KS \KL$ after all
corrections is shown as a function of energy in
Fig.~\ref{kskl_ff_all} and listed in Table~\ref{kskl_ff_tab}.
Only statistical uncertainties are shown.
The systematic uncertainty is dominated by the background subtraction
procedure and is strongly correlated across the entire energy range.
It is about 10\%  at 1.6~\gev, 
and increases with decreasing cross section to $\sim$30 (50)\% for values
below 0.5 (0.3) nb.
Also shown are all other available data, which are consistent with our results.
Below 1.4~\gev, our measurement has precision comparable to the
measurements by the CMD2~\cite{cmdksklff} and SND~\cite{ksklsnd}
experiments at VEPP-2M,
and is much more precise than the result from the OLYA
experiment~\cite{ksklolya}.
In the 1.4-2.4~\gev region,
our result is much more precise than the only other available measurement,
from the DM1~\cite{kskldm1} experiment.

\begin{figure}[tbh]
\begin{center}
\includegraphics[width=0.9\linewidth]{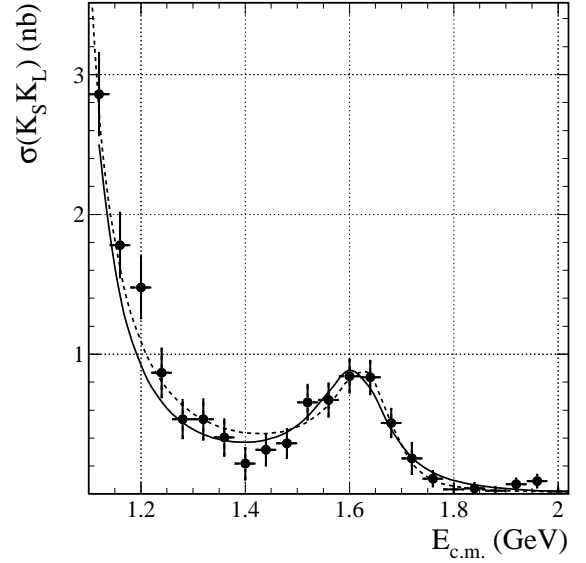}
\vspace{-0.4cm}
\caption{
The $\epem\to \KS \KL$ cross section (points)
compared with the results of the fits described in the text with
the non-resonant amplitude fixed to zero (solid lines) and floating (dashed
lines).
}
\label{kskl_ff_lin}
\end{center}
\end{figure} 
\begin{figure}[tbh]
\begin{center}
\includegraphics[width=0.9\linewidth]{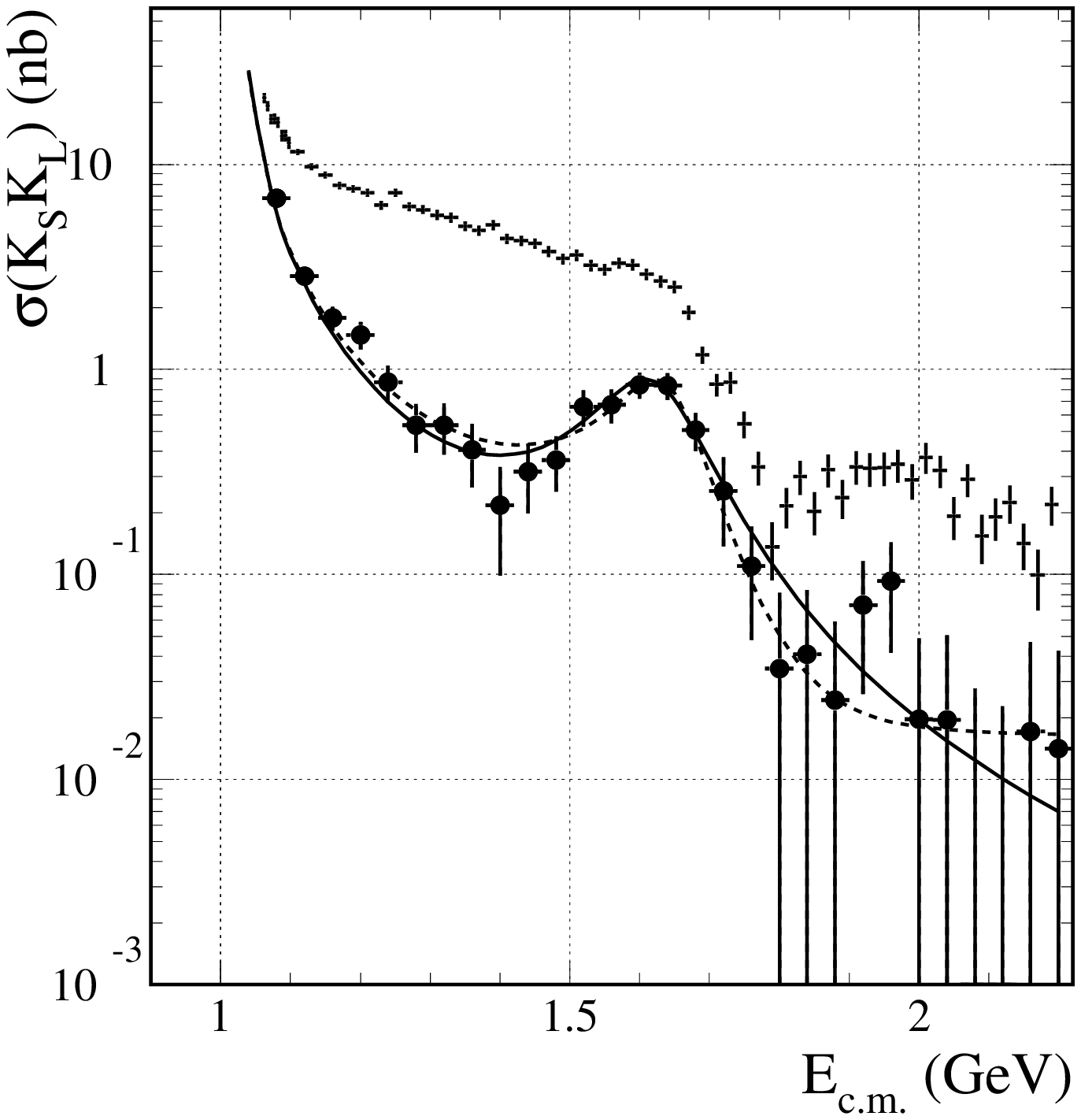}
\vspace{-0.2cm}
\caption{
Comparison of the $\epem \to \KS\KL$ cross section (points)
 with that for $\epem \to \Kp\Km$~\cite{isr2k} (crosses).
}
\label{kskl_ff_kk}
\end{center}
\end{figure} 
The measured cross section exhibits a distinctive structure around
1.6~\gev, 
indicating the presence of a vector resonance, perhaps the $\phi(1680)$.
Denoting it $\phi^\prime$,
we fit the cross section using Eq.~\ref{sigmak} with the additional
amplitude
\begin{equation}
-A_{\phi '}=\frac{\Gamma_{\phi '}m^3_{\phi '}\sqrt{ m_{\phi '}\sigma_{\phi '\to \KS \KL}
    /C}}{D_{\phi '}(s)}\frac{q^{3/2}_{\KS
    \KL}(m_{\phi}^2)}{q^{3/2}_{\KS \KL}(m_{\phi '}^2)}~.
\label{phiprimeamp}
\end{equation}
The energy-dependent width (see Eq.~\ref{widths}) assumes the
branching fractions and phase space factors of the major $\phi(1680)$
decay modes, 
$f = K^* K$, $\phi\eta$, $\phi\pi\pi$, and $\KS \KL$ taken from 
Refs.~\cite{isr2k2pi,isrkkpi}.
We fix the $\phi(1020)$ parameters to the values obtained in
Sec.~\ref{sec:xsphi}, and float the parameters of the $\phi '$.
Since the other vector states in this energy range, 
such as $\omega(1420,1650)$ and $\rho(1450,1700)$, 
are relatively wide and overlap considerably,
we again describe the sum of their contributions using the non-resonant
cross section $\sigma_{\rm bkg}$ and phase $\Psi$ of Eq.~\ref{bkgamp}.
First, we fix both to zero, 
and the fit yields a relatively good description of the data,
with \chisq=30 for (29-4) degrees of freedom.
The result of the fit (solid curve) is compared with the data in
Figs.~\ref{kskl_ff_lin} and~\ref{kskl_ff_kk}.

Next, we allow $\sigma_{bkg}$ and $\Psi$ to float in the fit, and obtain
$\Psi = 0.2\pm 0.6$ radians.
Since this is consistent with zero,
we fix it to zero and repeat the fit.
The result is shown as the dashed curves in
Figs.~\ref{kskl_ff_lin} and~\ref{kskl_ff_kk}.
We obtain an improved description of the cross section, 
with \chisq = 21 for (29-6) degrees of freedom and the fitted parameter
values:
\begin{eqnarray*}
 \sigma_{\phi '} & = &  0.46  \pm 0.10 \pm 0.05~{\rm nb}, \\
      m_{\phi '} & = &  1674 \pm 12 \pm 6~\mevcc, \\
 \Gamma_{\phi '} & = &  165 \pm 38 \pm 70~\mev,\\ 
\sigma_{\rm bkg} & = & 0.36 \pm 0.16~{\rm nb},
\end{eqnarray*}
where the first uncertainties are statistical
and the second systematic,
dominated by the difference between fixed and floated $\Psi$.
The relative phase between the non-resonant background and the $\phi$ 
resonance is consistent with that between the $\phi '$ and $\phi$
resonances, but the uncertainty is very large.

Our parameter values for this resonance are consistent with those 
of the PDG for the $\phi(1680)$, 
and with the results of similar fits performed in
Refs.~\cite{isr2k2pi,isrkkpi} for the $K^* K$, $\phi\eta$, and
$\phi\pi\pi$ decay modes of the $\phi(1680)$. 
However, as shown in Fig.~\ref{kskl_ff_kk}, 
the cross section for $\epem \to \Kp\Km$ is quite different from that
for $\KS\KL$,
indicating substantial interference between the iso-scalar and
iso-vector amplitudes in this energy range.
The fitting function used above is not able to reproduce the $\Kp\Km$
data~\cite{isr2k}, 
and therefore the results should be taken with caution.
A simultaneous fit to
the cross sections for $\epem\to\pipi$ (pure iso-vector), 
$\pipi\piz$ (pure iso-scalar),
$K^+ K^-$, $\KS \KL$, 
and perhaps other multihadron final states
is needed to extract the iso-scalar and iso-vector components,
together with reliable resonance parameter values.

The product $\Gamma_{\phi '}\sigma_{\phi '}$ is proportional to the total
number of events and does not depend on the experimental resolution. 
Using this product as a free parameter in the fit and Eq.~\ref{sigbr}
we obtain for the $\phi(1680)$ candidate
\begin{equation}
  \Gamma_{ee}\BR_{\KS \KL}  = 
(14.3  \pm 2.4 \pm 1.5 \pm 6.0)~\ev\ , 
\label{geephiprim}
\end{equation}
where the first uncertainty is statistical,
the second systematic,
and the third due to model dependence.
There is no independent measurement of the $\phi(1680)\to\KS\KL$ 
branching fraction that could be used to calculate $\Gamma_{ee}$. 
However, we have also measured $\Gamma_{ee}\BR_{K^* K} = 369\pm53$~\ev,
$\Gamma_{ee}\BR_{\phi\eta}=138\pm43$~\ev~\cite{isrkkpi}, and
$\Gamma_{ee}\BR_{\phi\pi\pi}=42\pm5$~\ev~\cite{isr2k2pi}. 
We assume these are the four dominant decay modes,
estimate their rates, and use them in our $\Gamma(s)$ calculation. 

\section{\boldmath The $\KS \KL\pipi$ final state}
\label{sec:kskl2pi}
We now consider the events with exactly two  tracks not
assigned to the \KS candidate,
but consistent with originating from the same event vertex.
This final state has four charged particles, and therefore large
backgrounds from ISR and non-ISR multihadron events.
We make additional requirements on the two tracks and the rest of the
event in order to suppress these backgrounds.

\begin{figure*}[tb]
\begin{center}
\includegraphics[width=0.33\linewidth]{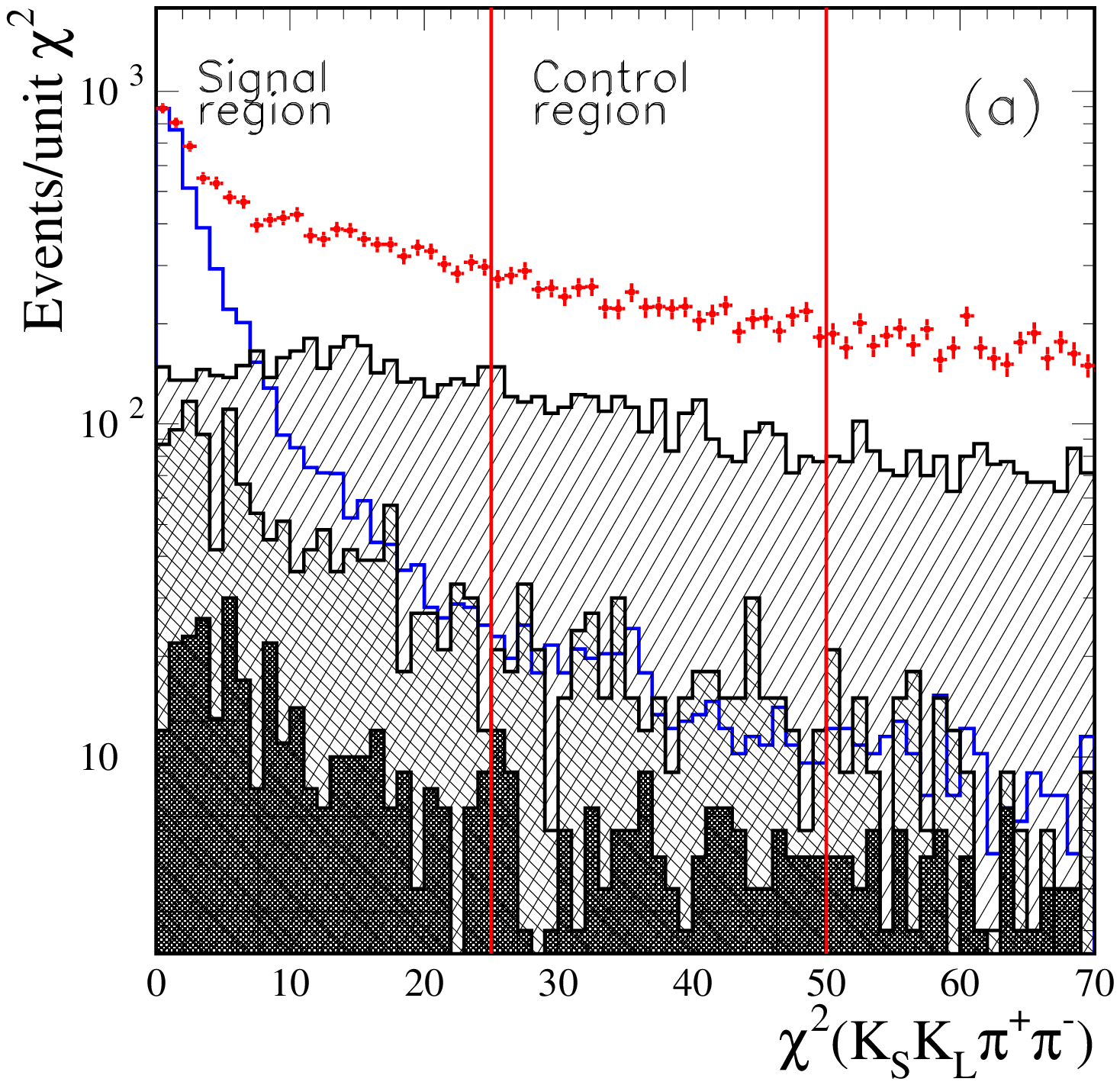}
\includegraphics[width=0.33\linewidth]{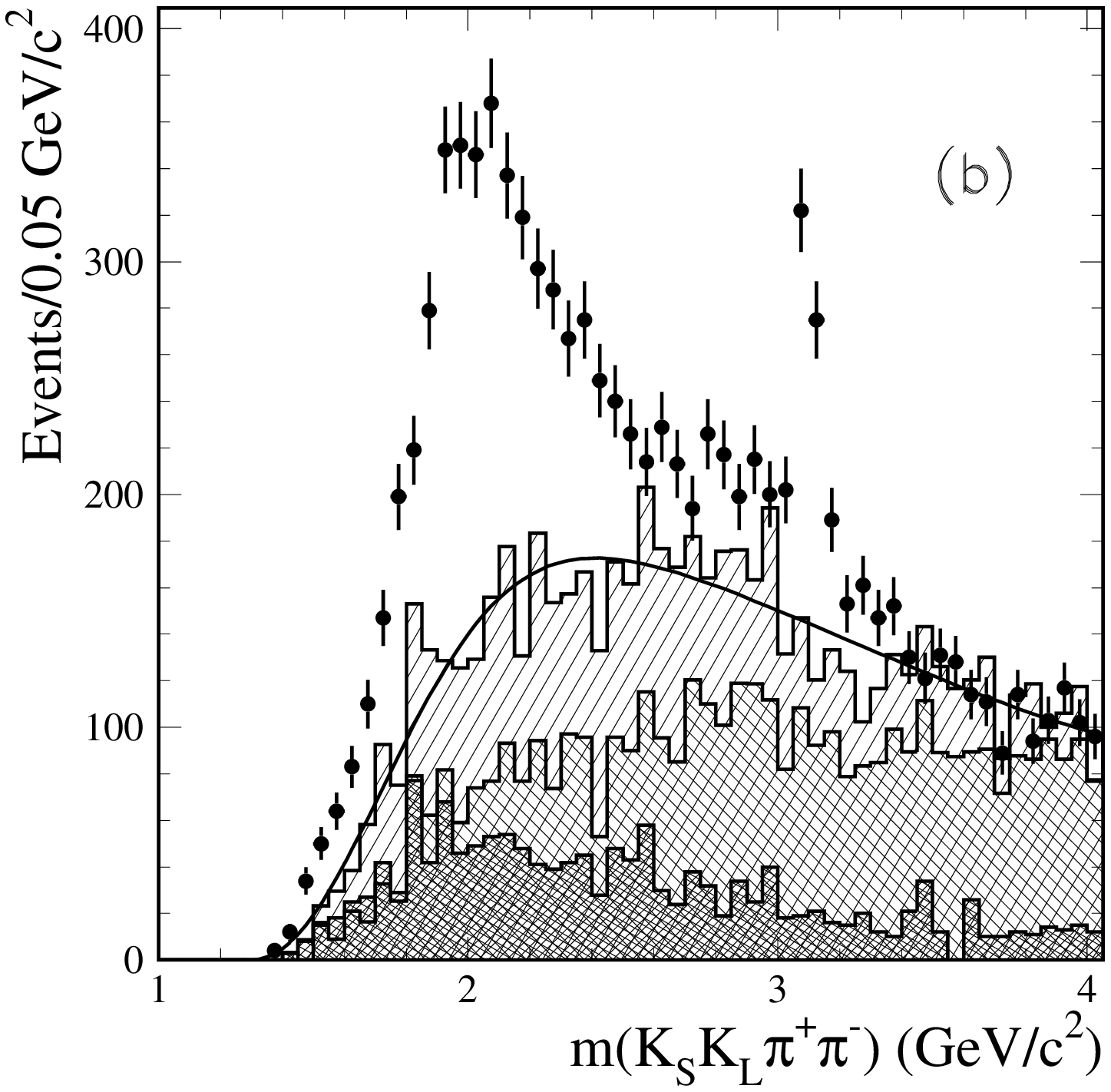}
\includegraphics[width=0.33\linewidth]{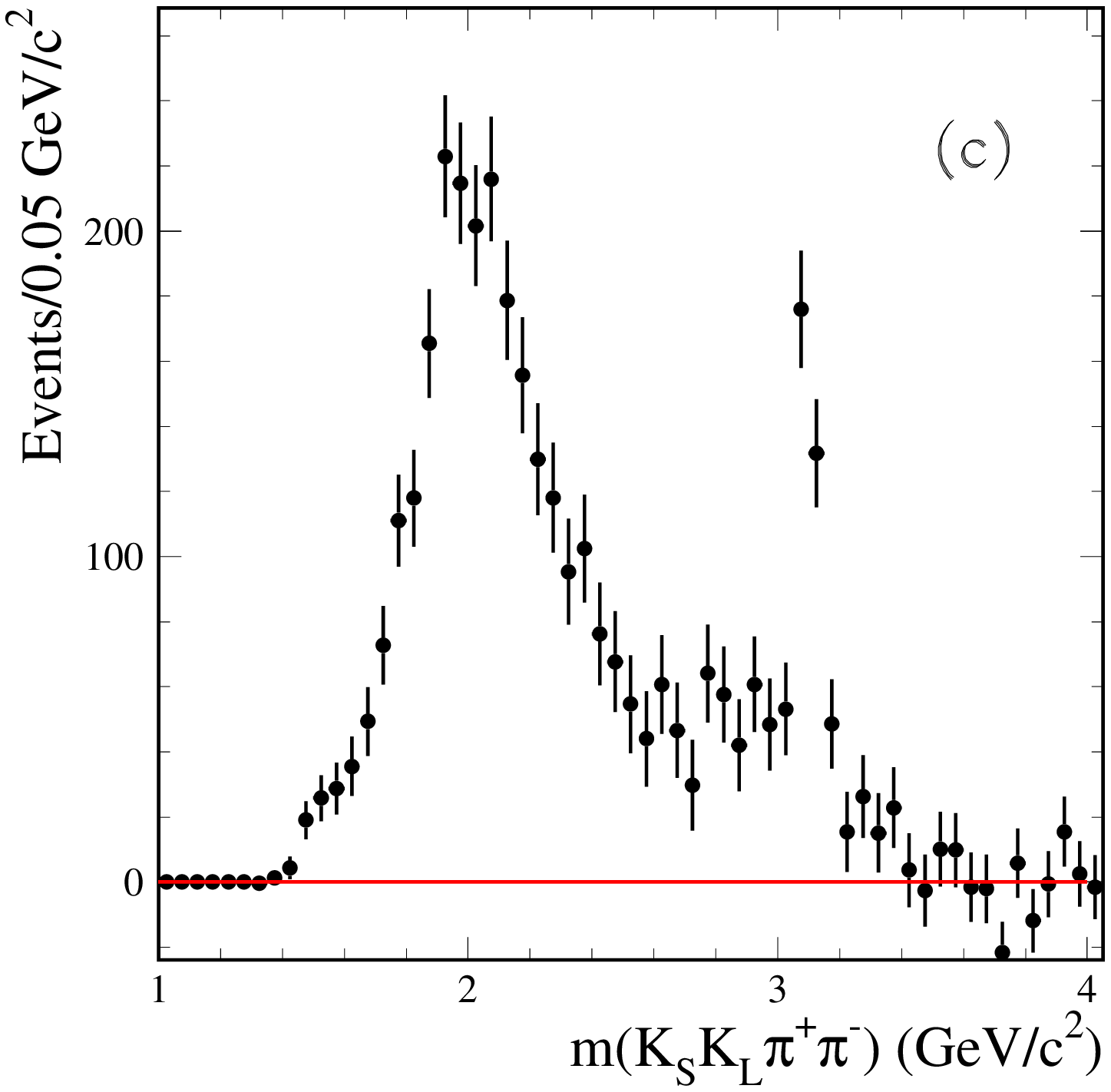}
\vspace{-0.5cm}
\caption{
  (a) The three-constraint \chisq distributions for data (points) and
     MC simulated $\KS \KL\pipi\gamma$ events (open histogram).
     The shaded, cross-hatched, and hatched histograms represent the
     simulated contributions from ISR $\phi\eta$, ISR $\KS K\pi\piz$,
     and non-ISR $q\bar q$ events, respectively. 
  (b) The $\KS \KL\pipi$ invariant mass distribution for data events in the
     signal region of (a) (points).
     The shaded and cross-hatched histograms represent the simulated
     contributions from ISR $\phi\eta$+$\KS K\pi\pi^0$ and non-ISR
     $q\bar q$ events, respectively,
     and the hatched area represents that estimated from the control
     region. 
     The curve shows the empirical fit used for background subtraction.
  (c) The $\KS \KL\pipi$ invariant mass distribution after background
     subtraction.
}
\label{kskl2pi_chi2}
\end{center}
\end{figure*}

\subsection{Additional selection criteria}
\label{addkskl2pi}
The two additional tracks must not be identified as \Kpm,
and are required to extrapolate to within $\pm$3~cm of the collision
point in 
the direction along the beam axis and 0.25 cm in the perpendicular direction.
The event must contain no other 
tracks that extrapolate to within 1~cm of the axis,
which is also the lower limit on the radial position of the $\KS\to\pipi$
vertex.
Considering all pairs of EMC clusters 
except those assigned to the ISR photon and \KL candidates,
we observe a large signal from $\piz$,
similar to that shown in Fig.~\ref{egammax}. 
As in that case,
we require $\rm E_{\gamma}max < 0.5~\gev$, 
reducing backgrounds from several sources
with a loss of 3\% in signal efficiency, 
as shown in Sec.~\ref{addkskl}. 

ISR $\KS K^{\pm}\pi^{\mp}\pi^0$ events with the charged kaon
mis-identified as a pion and a cluster from a $\pi^0$ photon taken as
the $\KL$ candidate are indistinguishable from signal events. 
To reduce this background, 
we pair the $\KL$ cluster with all other EMC clusters.
For every such pair with $M(\gamma\gamma )$ within 0.03~\gevcc of the
$\piz$ mass,
we perform a kinematic fit to the 
$\KS K^{\pm}\pi^{\mp}\pi^0\gamma$ hypothesis and require 
$\chisq(\KS K\pi\pi^0)>100$.

The 3C \chisq distribution for the remaining candidate events under the 
$\KS\KL\pipi\gamma$ hypothesis is shown as the points in  
Fig.~\ref{kskl2pi_chi2}a, 
with the corresponding MC-simulated pure $\KS \KL\pipi\gamma$ events
shown as the open histogram.
The simulated distribution is normalized to the data in the region
$\chi^2<1$, where the contribution of higher-order
ISR is small and the background contamination is lowest, 
but still amounts to about 15\% of the signal.
The shaded, cross-hatched, and hatched areas represent the simulated
contributions from the ISR $\phi\eta$, ISR $\KS K^{\pm}\pi^{\mp}\piz$, and 
non-ISR $q\bar q$ channels, respectively. 
These backgrounds account for only half of the observed data-MC
difference in the distribution at large \chisq values.

We define a signal region $\chisq(\KS \KL\pipi)<25$ 
and a control region $25<\chisq(\KS \KL\pipi)<50$ 
(vertical lines in Fig.~\ref{kskl2pi_chi2}),
from which we estimate backgrounds in the signal region. 
The signal region contains $10\,788$ data and 6825 MC events, 
while the control region contains 5756 and 633 events, respectively.

\subsection{Background subtraction}
\label{sec:kskl2pibkg}
The background to the  $\KS \KL\pipi$ mass distribution is subtracted
in two stages.
The \chisq distributions for the $\KS K^{\pm}\pi^{\mp}\pi^0$ and
non-ISR $q\bar q$ events peak at low values, 
since their kinematics are similar to those of signal events.
We therefore subtract their MC-simulated contribution from both the
signal and control regions of Fig.~\ref{kskl2pi_chi2}(a).
There are large uncertainties in their normalizations,
but this has little effect on the total uncertainty.
The mass distribution for the data in the signal region before
background subtraction is shown in 
Fig.~\ref{kskl2pi_chi2}(b)  
as the points,
with the simulated $\KS K^{\pm}\pi^{\mp}\pi^0$ and $q\bar q$ 
events shown as the shaded and cross-hatched histograms, respectively.

We estimate the remaining background using the mass distributions for
the remaining events in the signal and control regions, 
according to Eq.~\ref{sig_contr} of Sec.~\ref{sec:ffbkg}.
The contribution is shown as the hatched area in 
Fig.~\ref{kskl2pi_chi2}(b).   
We fit the sum of all backgrounds with a polynomial function to reduce
the statistical fluctuations (curve in 
Fig.~\ref{kskl2pi_chi2}(b))   
and use this fit for the background subtraction.
The resulting mass distribution for $\epem \to \KS\KL\pipi$ events is
shown in 
Fig.~\ref{kskl2pi_chi2}(c).   
We observe 3320 events in the mass range from threshold to 4.0~\gevcc.
In addition to a main peak around 2~\gevcc, a \jpsi signal and
a possible structure just below 3~\gevcc are visible.

We estimate the systematic uncertainty due to the background subtraction
to be about 10\% for $m(\KS \KL\pipi)<2.5$~\gevcc
(i.e.,\ a 30\% uncertainty on a 30\% total background),
increasing to about 30\% in the 2.5-3.0~\gevcc region and 
reaching 100\% above 3.4~\gevcc, where background dominates.

\begin{figure}[tbh]
\begin{center}
\includegraphics[width=0.9\linewidth]{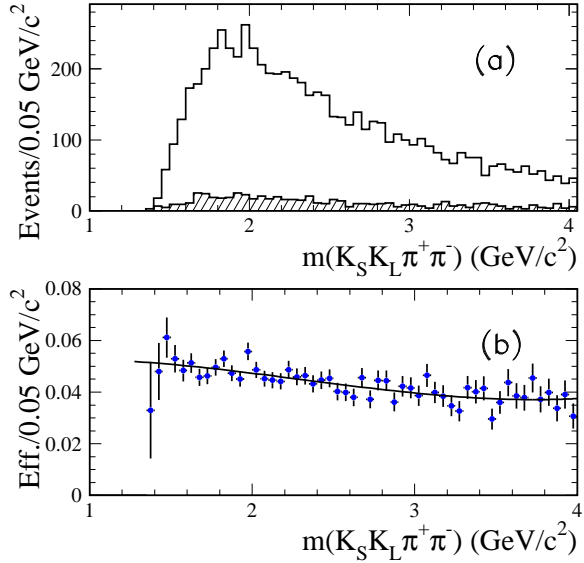}
\vspace{-0.4cm}
\caption{
(a) The $\KS \KL\pipi$ invariant mass distribution for MC-simulated
    signal events in the signal (open histogram) 
    and control region (shaded) of Fig.~\ref{kskl2pi_chi2}.
(b) The net reconstruction efficiency from the simulation.
}
\label{kskl2pi_eff}
\end{center}
\end{figure}

\subsection{Simulated detection efficiency }
\label{sec:kskl2pieff}
The selection procedures applied to the data are also applied to the         
MC-simulated event sample. 
The resulting $\KS \KL\pipi$ invariant mass
distribution is shown in Fig.~\ref{kskl2pi_eff}(a) for the signal and
control (shaded histogram) regions.
The detection efficiency as a function of mass is obtained by dividing
the number of reconstructed MC events in each 0.05~\gevcc mass
interval by the number generated in that interval,
and is shown in Fig.~\ref{kskl2pi_eff}(b). 
The 50~\mevcc mass interval used is wider than the detector resolution
of about 25~\mevcc.
Since the cross section has no sharp structures 
(except for the $J/\psi$ signal, which is discussed below), 
we apply no corrections for the resolution.
We apply all the corrections discussed above for data-MC differences
in the tracking, photon, and $\KL$ detection efficiencies.
\begin{table*}
\caption{Summary of the $\epem\to K_S K_L\pipi$ 
cross section measurement. Uncertainties are statistical only.}
\label{kskl2pi_tab}
\begin{ruledtabular}
\begin{tabular}{ c c c c c c c c }
$E_{\rm c.m.}$ (GeV) & $\sigma$ (nb)  
& $E_{\rm c.m.}$ (GeV) & $\sigma$ (nb) 
& $E_{\rm c.m.}$ (GeV) & $\sigma$ (nb) 
& $E_{\rm c.m.}$ (GeV) & $\sigma$ (nb)  
\\
\hline

  1.425 &  0.03 $\pm$  0.02 &  2.075 &  0.99 $\pm$  0.09 &  2.725 &  0.11 $\pm$  0.05 &  3.375 &  0.07 $\pm$  0.04 \\
  1.475 &  0.12 $\pm$  0.04 &  2.125 &  0.81 $\pm$  0.08 &  2.775 &  0.24 $\pm$  0.05 &  3.425 &  0.01 $\pm$  0.04 \\
  1.525 &  0.15 $\pm$  0.04 &  2.175 &  0.69 $\pm$  0.08 &  2.825 &  0.21 $\pm$  0.05 &  3.475 & -0.01 $\pm$  0.03 \\
  1.575 &  0.17 $\pm$  0.05 &  2.225 &  0.56 $\pm$  0.08 &  2.875 &  0.15 $\pm$  0.05 &  3.525 &  0.03 $\pm$  0.03 \\
  1.625 &  0.20 $\pm$  0.05 &  2.275 &  0.51 $\pm$  0.07 &  2.925 &  0.22 $\pm$  0.05 &  3.575 &  0.03 $\pm$  0.03 \\
  1.675 &  0.27 $\pm$  0.06 &  2.325 &  0.40 $\pm$  0.07 &  2.975 &  0.17 $\pm$  0.05 &  3.625 &  0.00 $\pm$  0.03 \\
  1.725 &  0.39 $\pm$  0.06 &  2.375 &  0.43 $\pm$  0.07 &  3.025 &  0.18 $\pm$  0.05 &  3.675 & -0.01 $\pm$  0.03 \\
  1.775 &  0.58 $\pm$  0.07 &  2.425 &  0.31 $\pm$  0.06 &  3.075 &  0.60 $\pm$  0.06 &  3.725 & -0.06 $\pm$  0.03 \\
  1.825 &  0.60 $\pm$  0.08 &  2.475 &  0.27 $\pm$  0.06 &  3.125 &  0.44 $\pm$  0.05 &  3.775 &  0.02 $\pm$  0.03 \\
  1.875 &  0.83 $\pm$  0.08 &  2.525 &  0.22 $\pm$  0.06 &  3.175 &  0.16 $\pm$  0.05 &  3.825 & -0.03 $\pm$  0.03 \\
  1.925 &  1.09 $\pm$  0.09 &  2.575 &  0.17 $\pm$  0.06 &  3.225 &  0.05 $\pm$  0.04 &  3.875 &  0.00 $\pm$  0.03 \\
  1.975 &  1.03 $\pm$  0.09 &  2.625 &  0.23 $\pm$  0.06 &  3.275 &  0.08 $\pm$  0.04 &  3.925 &  0.04 $\pm$  0.03 \\
  2.025 &  0.94 $\pm$  0.09 &  2.675 &  0.18 $\pm$  0.05 &  3.325 &  0.05 $\pm$  0.04 &  3.975 &  0.01 $\pm$  0.03 \\

\end{tabular}
\end{ruledtabular}
\end{table*}

\begin{figure}[tbh]
\begin{center}
\includegraphics[width=0.9\linewidth]{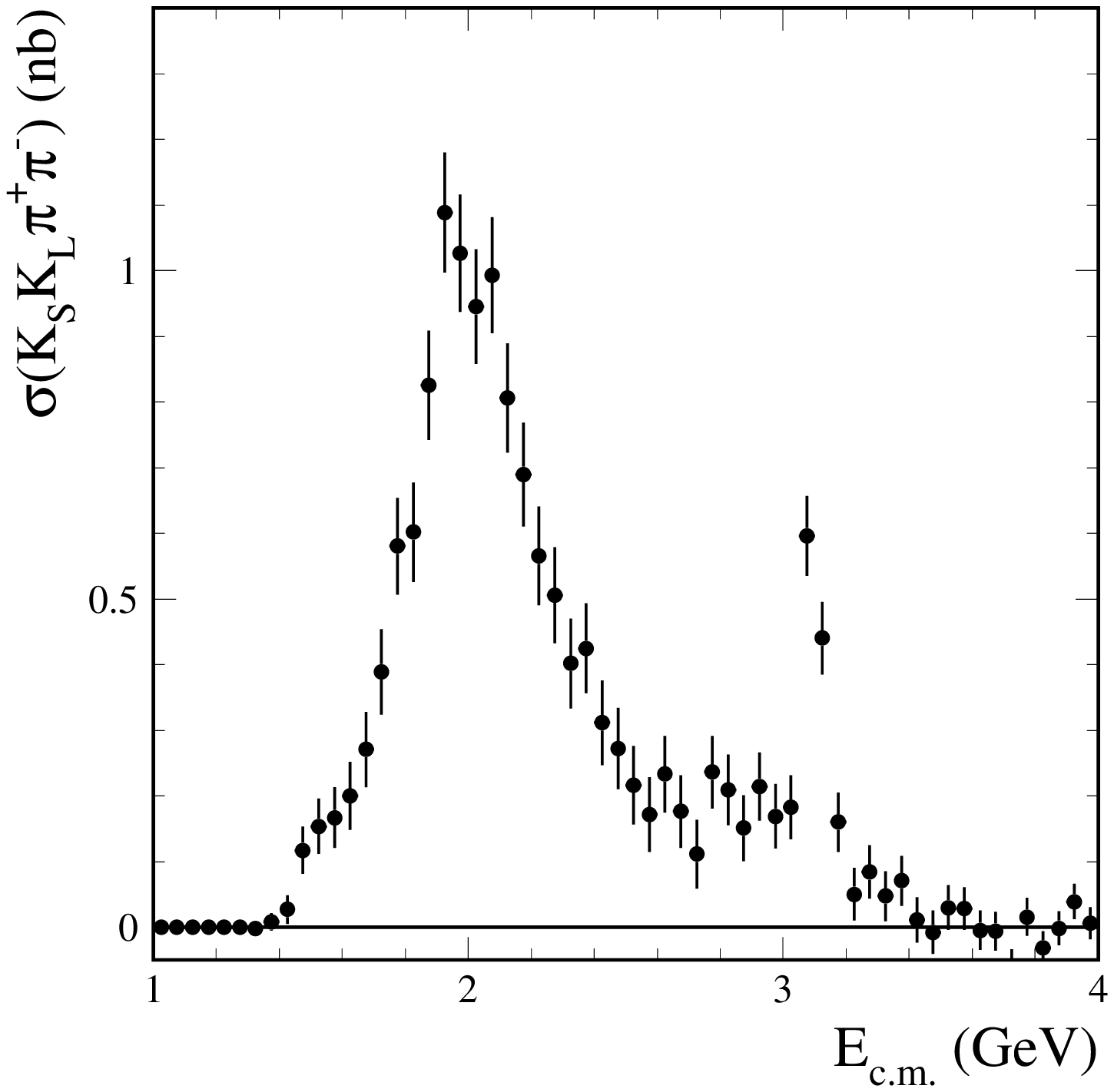}
\vspace{-0.4cm}
\caption{
The $\epem\to \KS \KL\pipi$ cross section.
The error bars are statistical only.
}
\label{kskl2pi_xs}
\end{center}
\end{figure}
\subsection{\boldmath The $\epem\to \KS \KL\pipi$ cross section }
\label{sec:xskskl2pi}
The cross section for the reaction $\epem\to \KS \KL\pipi$ is
calculated using Eq.~\ref{xsformular} with the corrections described
above,
plus an additional 3\% correction for the requirement on the maximum
energy of extra EMC clusters.
The cross section is shown as a function of energy in
Fig.~\ref{kskl2pi_xs}, 
and listed in Table~\ref{kskl2pi_tab}. 
There are no previous measurements for this final state. 
The cross section shows a threshold rise at 1.5~\gev,
a maximum value of about 1~nb near 2~\gev, 
and a slow decrease toward higher energies, 
perturbed by the \jpsi signal.

Only statistical uncertainties are shown.
The total systematic uncertainty is dominated by the background
subtraction procedure.
It amounts to about 10\% at 2~\gev, 
where the cross section peaks, 
and increases with decreasing cross section to $\sim$30\% near 1.5 and
3~\gev, and to 100\% well above 3~\gev.

\begin{figure}[tbh]
\begin{center}
\includegraphics[width=0.9\linewidth]{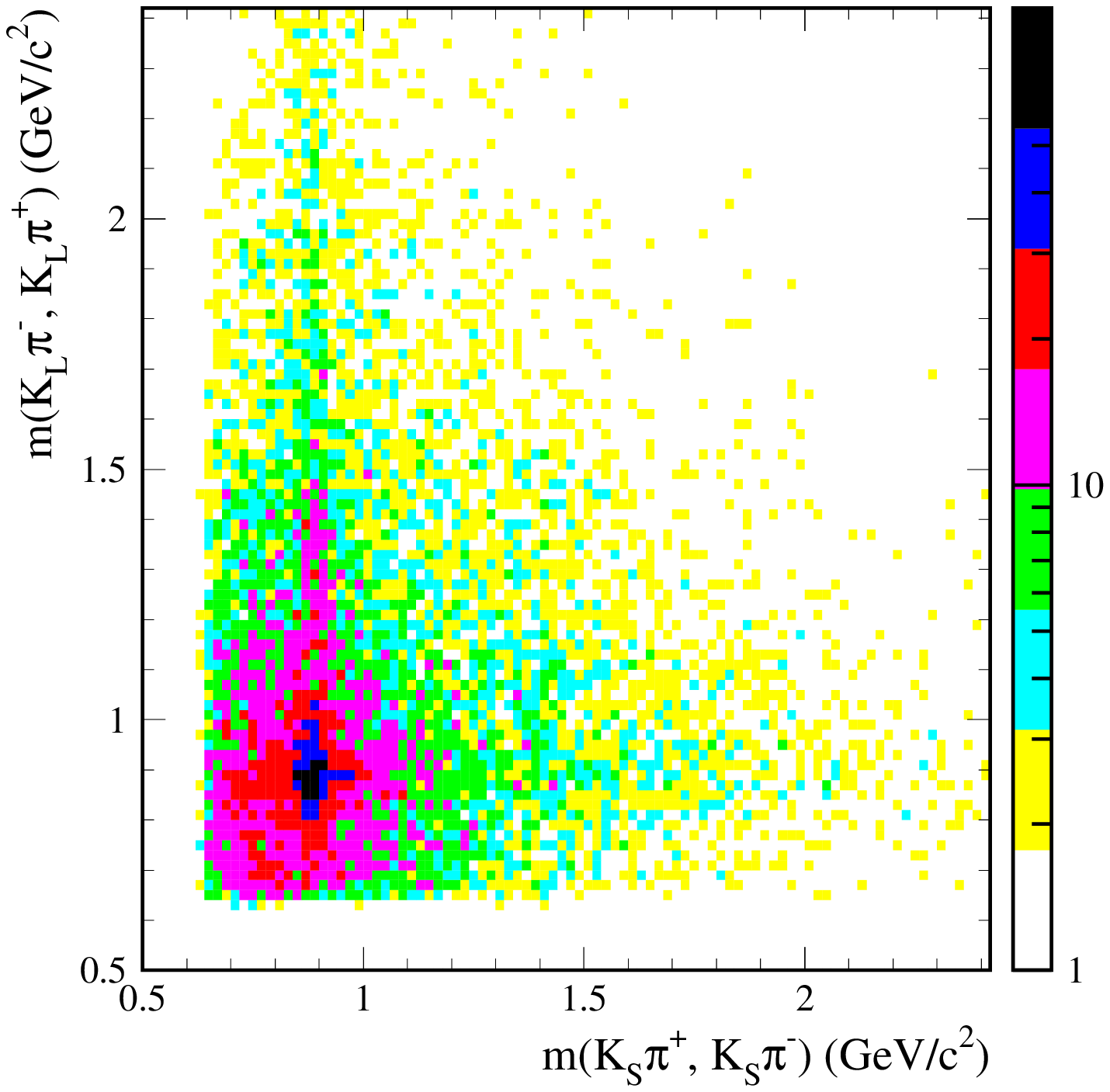}
\vspace{-0.4cm}
\caption{The $\KL\pi^{\pm}$ invariant mass versus the
             $\KS\pi^{\mp}$ invariant mass (two entries per event).  
}
\label{kstarpi2d}
\end{center}
\end{figure}

\subsection{\boldmath The $K^{*}(892)^{\pm}$ and $K_2^{*}(1430)^{\pm}$
  contributions } 
\label{sec:kstar}

\begin{figure}[tbh]
\begin{center}
\includegraphics[width=0.98\linewidth]{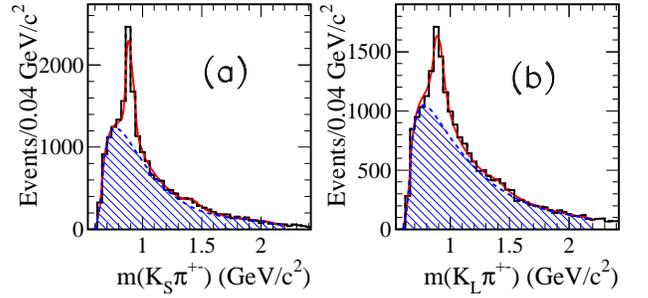}
\vspace{-0.4cm}
\caption{
The (a) $\KS\pipm$ and (b) $\KL\pipm$ mass projections of 
Fig.~\ref{kstarpi2d}. 
The curves represent the results of the fits described in the text,
with the hatched areas representing the non-resonant components.
}
\label{kstarpi}
\end{center}
\end{figure}
\begin{figure}[tbh]
\begin{center}
\includegraphics[width=0.9\linewidth]{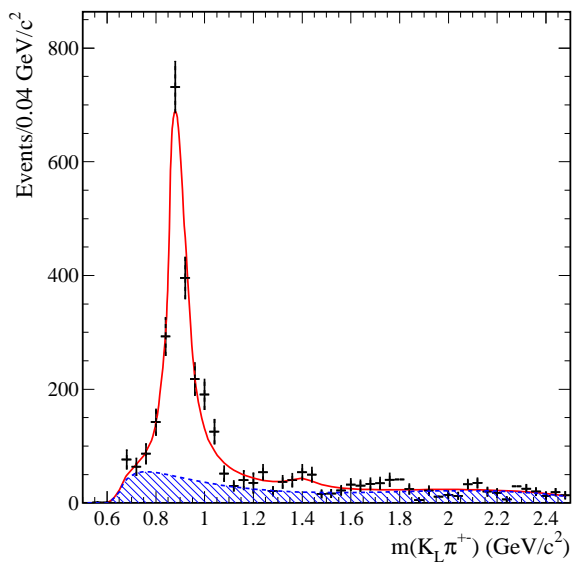}
\vspace{-0.2cm}
\caption{ 
The number of $K^{*}(892)^{\pm}$ events obtained from fits to the
$\KS \pi^{\pm}$ invariant mass distribution in each 0.04~\gevcc
interval of $\KL\pi^{\mp}$ mass.   
The curve represents the result of the fit described in the text,
with the hatched areas representing the non-resonant component.
  }
\label{k2stars}
\end{center}
\end{figure}

Figure~\ref{kstarpi2d} shows a scatter plot of the $\KL\pi^{\pm}$
invariant mass versus the $\KS\pi^{\mp}$ invariant mass,
with two entries per event. 
Clear bands corresponding to the $K^{*}(892)^{\pm}$ resonances are
visible. 
Indications of $K_2^{*}(1430)^{\pm}$ production are also seen in 
the projections shown in Fig.~\ref{kstarpi}.

We fit these projections with a sum of two Breit-Wigner functions
and a function describing the non-resonant contribution,
yielding $3335\pm115$ $K^{*}(892)^{\pm}\to \KS\pi^{\pm}$ decays,
$3200\pm151$ $K^{*}(892)^{\pm}\to \KL\pi^{\pm}$ decays,
and a total of $286\pm99$ $K_2^{*}(1430)^{\pm}$ decays.
The total number of $K^{*}(892)^{\pm}$ decays is larger than the
number of $\KS \KL\pipi$ events, 
indicating correlated production of $K^{*}(892)^+ K^{*}(892)^-$
pairs.
In each 0.04~\gevcc bin of $\KL\pi^{\mp}$ mass
we fit the $\KS\pi^{\pm}$ mass distribution with the same function,
and the resulting numbers of $K^{*}(892)^{\pm}$ decays are shown in 
Fig.~\ref{k2stars}.

A strong signal of $2098\pm61\pm200$ $K^{*}(892)^{\pm}$ is observed,
where the second uncertainty is due to variations of the fitting
procedure. 
This corresponds to the production of $K^{*}(892)^+ K^{*}(892)^-$
pairs in about 63\% of all observed $\KS \KL\pipi$ events.
We also find $105\pm23\pm50$ events at the $K_2^{*}(1430)^{\pm}$ mass, 
corresponding to $K^{*}(892)^\pm K_2^{*}(1430)^\mp$ correlated
production.
We have observed such correlated production previously in the 
$K^+K^-\ppz$ channel~\cite{isr2k2pi};
these results are compared and discussed below.

\begin{figure}[tbh]
\begin{center}
\includegraphics[width=\linewidth]{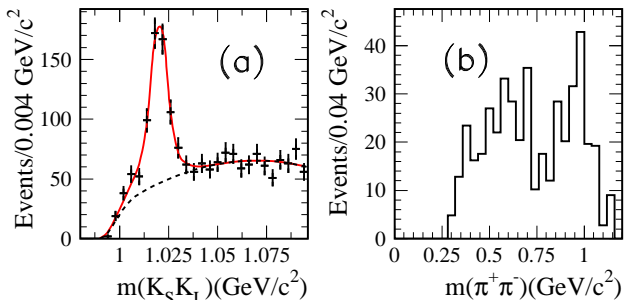}
\vspace{-0.4cm}
\caption{
(a) The $\KS \KL$ invariant mass distribution for the selected 
    $\KS \KL\pipi$ events. 
    The solid and dashed lines represent the result of the fit
    described in the text and its non-$\phi$ component, respectively.
(b) The $\pipi$ invariant mass distribution for events in the $\phi$
    peak (see text).
}
\label{phipipi}
\end{center}
\end{figure}
\subsection{\boldmath The $\phi(1020)\pipi$ contribution } 
\label{sec:phi2pi}
Figure~\ref{phipipi}(a) shows the $\KS \KL$ invariant mass
distribution for the selected $\KS \KL\pipi$ events.
A clear $\phi(1020)$ signal is visible.
Fitting with a Gaussian plus polynomial function yields
$424\pm 30$ $\phi \to \KS\KL$ decays,
corresponding to about 13\% of the events.

We calculate the $\pipi$ invariant mass for events in the $\phi$ region,
1.01$<m(\KS\KL)<$1.04~\gevcc, 
and subtract the non-resonant contribution using events in the
sideband 1.04$<m(\KS\KL)<$1.07~\gevcc.
We show the resulting $m(\pipi)$ distribution in Fig.~\ref{phipipi}(b).
It is consistent with those observed in the $\phi\pipi$ and $\phi\ppz$
final states~\cite{isr2k2pi},
where $f_0(980)$ signals were clearly seen.
Fitting the $m(\KS \KL)$ distribution in bins of the $\KS\KL\pipi$
mass, 
we obtain a $\phi\pipi$ invariant mass spectrum consistent with those
observed in the $K^+ K^-\pipi$ and $K^+ K^-\ppz$ final
states~\cite{isr2k2pi}.
However, 
the statistical uncertainties are quite large,
and so we do not present the distribution or calculate
a cross section for this intermediate state.
\section{\boldmath The $\KS \KS\pipi$ final state}
\label{sec:ksks2pi}
\subsection{Final Selection and Backgrounds}
\label{addksks2pi}
This final state contains six charged pions and no neutral particles
other than the ISR photon.
We consider the events from Sec.~\ref{sec:KS} with at least two \KS
candidates,
and the combination of two \KS candidates and two charged tracks in
each event giving the best \chisq for a 4C fit under the $\KS\KS\pipi$ 
hypothesis (see Sec.~\ref{sec:Analysis}).
To reduce the background from multihadronic \qqbar events, 
we reject events in which both of the charged tracks not in a
\KS candidate are identified as kaons. 

\begin{figure}[tbh]
\includegraphics[width=0.9\linewidth]{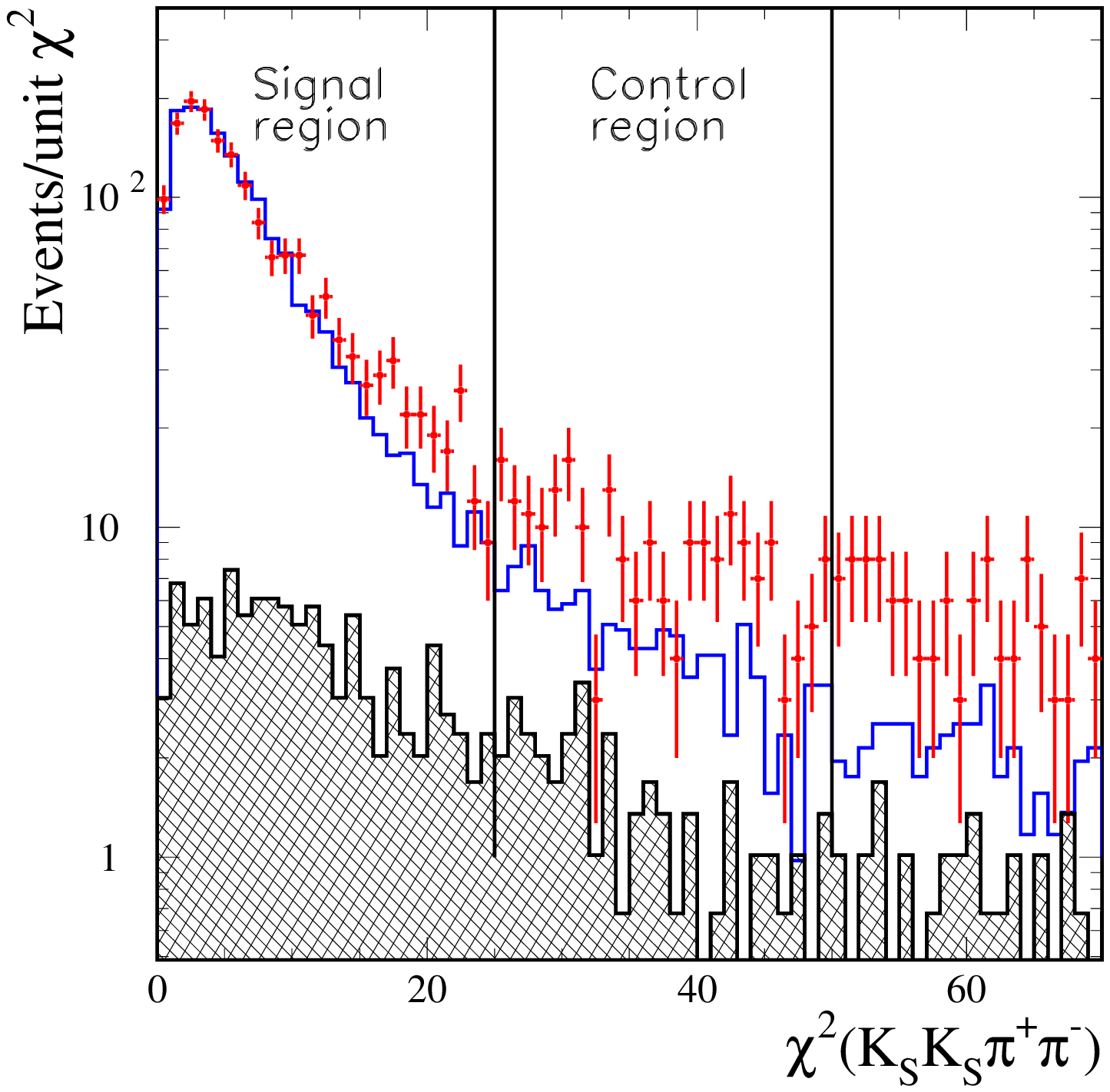}
\vspace{-0.4cm}
\caption{
The four-constraint \chisq distributions for $\KS \KS\pipi\gamma$
candidate events selected in the data (points) and
signal-MC simulation (open histogram) 
fitted under the $ \KS \KS\pipi$ hypothesis. 
The cross-hatched 
histogram represents the simulated
background contribution from 
non-ISR \qqbar events. 
}
\label{ksks2pi_chi2}
\end{figure}

The $\chisq(\KS \KS\pipi)$ distribution for the selected events in the
data is shown in Fig.~\ref{ksks2pi_chi2} (points),
along with that for selected simulated ISR $\KS\KS\pipi$ events 
(open histogram),
which is normalized to the data in the region 
$\chisq(\KS \KS\pipi)\! <\! 10$ where the backgrounds and radiative
corrections do not exceed 5\%.
Both distributions are broader than those for a typical 4C \chisq
distribution due to higher-order ISR, 
and the data include contributions from background processes.

The cross-hatched histogram in Fig.~\ref{ksks2pi_chi2} represents
the background from non-ISR $\epem \!\!\to\! \qqbar$ events.
These predominantly contain a hard $\piz$, giving a false ISR photon,
and have kinematics similar to the signal,   
giving a peak at low values of $\chisq(\KS \KS\pipi)$. 
We evaluate this background in a number of \Ecm ranges using the
selected data and \qqbar events simulated with JETSET.
Combining each ISR photon candidate with all other EMC clusters in the
same event, 
we compare the \piz signals in the resulting data and simulated
$\gamma\gamma$ invariant mass distributions. 
The simulation gives an \Ecm dependence consistent with the data,
so we normalize its prediction using the overall data-over-MC ratio of
\piz signals, and subtract that from the data.

All remaining background sources are either negligible or yield a
 $\chisq(\KS \KS\pipi)$ distribution that is nearly uniform over the range
shown in Fig.~\ref{ksks2pi_chi2}.
We define signal and control regions, 
$\chisq(\KS \KS\pipi)\! <\! 25$ and 
$25\! <\! \chisq(\KS \KS\pipi)\!\! <\!50$, respectively 
(see Fig.~\ref{ksks2pi_chi2}),
and use them to estimate and subtract the sum of the remaining
backgrounds as described in Sec.~\ref{sec:ffbkg}.
The signal region of Fig.~\ref{ksks2pi_chi2} contains 1704
data and 8309 MC-simulated events; the control region contains
219 data and 580 simulated events.

\begin{figure}[tbh]
\begin{center}
\includegraphics[width=0.9\linewidth]{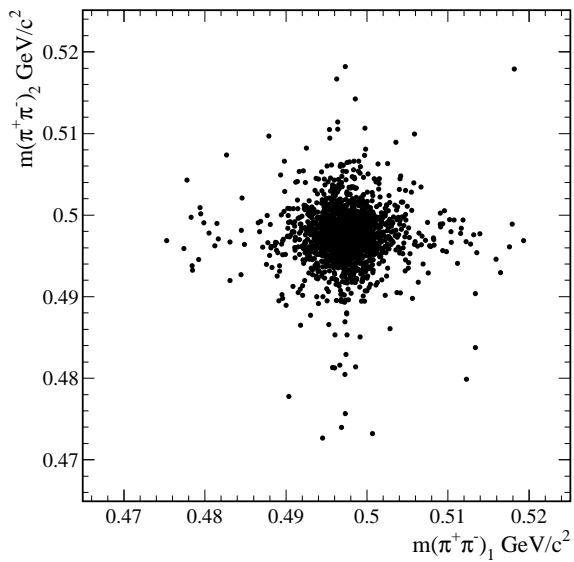}
\vspace{-0.4cm}
\caption{
Scatter plot of the $\pipi$ invariant mass of one $\KS$ candidate
versus 
that of the other $\KS$ candidate 
calculated using the results of the constrained fit.
}
\label{mkaons}
\end{center}
\end{figure}

We recalculate the masses of the two \KS candidates using the results of
the kinematic fit.
Figure~\ref{mkaons} shows a scatter plot of the invariant mass of
one $\KS$ candidate versus that of the other for events in the signal
region. 
Any background from events not containing two $\KS$ mesons is very low. 

\begin{figure}[tbh]
\begin{center}
\includegraphics[width=0.9\linewidth]{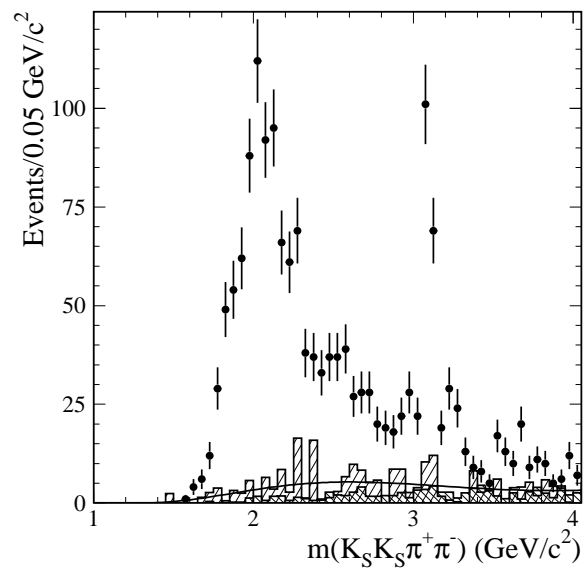}
\vspace{-0.4cm}
\caption{
The $\KS \KS\pipi$ invariant mass distribution (points) for events in
the signal region of Fig.~\ref{ksks2pi_chi2}.
The cross-hatched and hatched histograms represent the backgrounds
from non-ISR \qqbar events and others estimated from the \chisq control
region of Fig.~\ref{ksks2pi_chi2}, respectively.
The curve represents the smooth empirical fit to the total background
used for subtraction.
}
\label{ksks2pi_raw}
\end{center}
\end{figure}
The $m(\KS\KS\pipi)$ distribution for the events in the signal region
of Fig.~\ref{ksks2pi_chi2} is shown in Fig.~\ref{ksks2pi_raw} as the
points. 
The contributions from non-ISR events and the background estimated
from the control region are shown as cross-hatched and hatched
histograms, respectively. 
We fit the sum of all backgrounds with a second order polynomial to
reduce fluctuations,
and use the result (curve in Fig.~\ref{ksks2pi_raw}) for the
background subtraction.
This gives 1479 signal events with masses between threshold and
4.0~\gevcc.
We estimate the systematic uncertainty due to background subtraction
to be about 5\% of the signal for $m(\KS \KS\pipi)<2.5$~\gevcc, 
increasing to about 20\% in the 2.5-3.0~\mevcc region and 50-70\%
above 3.0~\gevcc, where background dominates.  
\subsection{ Simulated detection efficiency }
\label{sec:ksks2pieff}
\begin{figure}[tbh]
\begin{center}
\includegraphics[width=0.9\linewidth]{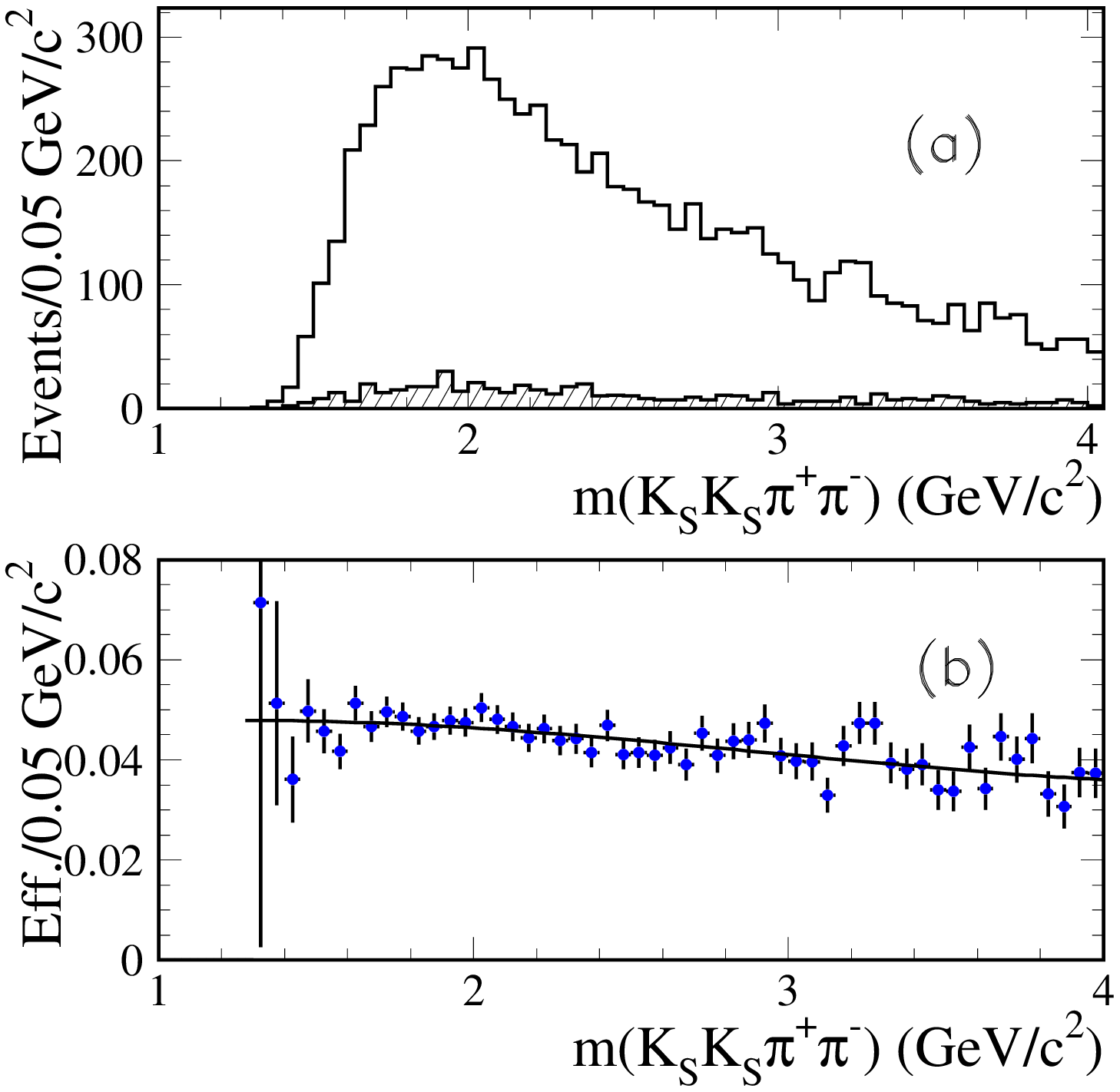}
\vspace{-0.4cm}
\caption{
(a) The $\KS\KS\pipi$ invariant mass distribution for the MC-simulated
    signal events in the signal and control (shaded) regions of
    Fig.~\ref{ksks2pi_chi2}.
(b) The net reconstruction and selection efficiency from the simulation.
}
\label{ksks2pi_eff}
\end{center}
\end{figure}

The MC-simulated $\KS\KS\pipi$ invariant-mass distribution is shown in
Fig.~\ref{ksks2pi_eff}(a) for events in the signal and control (shaded
histogram) regions.
The mass dependence of the detection efficiency
is shown in Fig.~\ref{ksks2pi_eff}(b). 
The mass interval used, 50~\mevcc per bin, is wider than the 10~\mevcc
detector resolution, 
and the cross section has no sharp structure 
(except the $J/\psi$ signal, discussed below), 
so we apply no corrections for the resolution.
We apply all the corrections discussed above for data-MC differences
in track, 
\KS, and photon 
detection efficiency.

\subsection{\boldmath Cross section for $\epem\to \KS \KS\pipi$ }
\label{sec:xsksks2pi}
\begin{figure}[tbh]
\begin{center}
\includegraphics[width=0.9\linewidth]{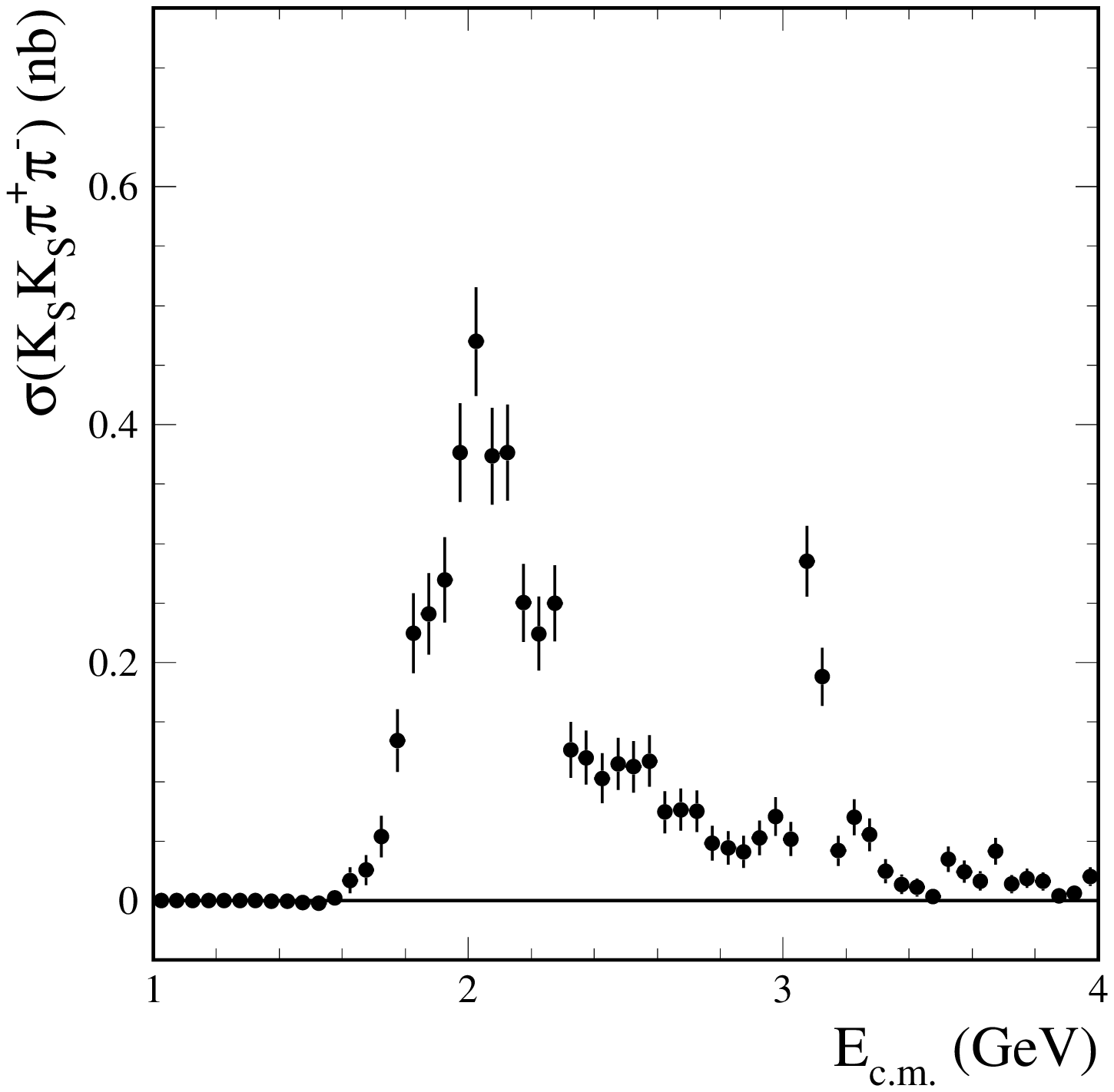}
\vspace{-0.4cm}
\caption{
The $\epem\to \KS \KS\pipi$ cross section.
}
\label{ksks2pi_xs}
\end{center}
\end{figure}

We calculate the $\epem \!\!\to\!\KS \KS\pipi $  cross section as a
function of the effective c.m.\ energy using
Eq.~\ref{xsformular} shown in Sec.~\ref{sec:xsff}.
The fully corrected cross section is shown in Fig.~\ref{ksks2pi_xs}
and listed in Table~\ref{ksks2pi_tab},
with statistical uncertainties only.
There are no other measurements for this final state. 
The cross section shows a slow rise from threshold at 1.5~\gev,
a maximum value of about 0.5 nb near 2~\gev,
and a slow decrease with increasing energy, 
punctuated by a clear \jpsi signal.
The systematic uncertainty is dominated by the uncertainty of the
backgrounds,  
and totals 5\% relative at the peak of the cross section, 
increasing to ~20\% at 3~\gev, and 50-70\% at higher energies.

\begin{table*}
\caption{Summary of the $\epem\to K_S K_S\pipi$ 
cross section measurement. Uncertainties are statistical only.}
\label{ksks2pi_tab}
\begin{ruledtabular}
\begin{tabular}{ c c c c c c c c }
$E_{\rm c.m.}$ (GeV) & $\sigma$ (nb)  
& $E_{\rm c.m.}$ (GeV) & $\sigma$ (nb) 
& $E_{\rm c.m.}$ (GeV) & $\sigma$ (nb) 
& $E_{\rm c.m.}$ (GeV) & $\sigma$ (nb)  
\\
\hline

   1.63 &  0.02 $\pm$  0.01 &   2.22 &  0.22 $\pm$  0.03 &   2.83 &  0.04 $\pm$  0.01 &   3.42 &  0.01 $\pm$  0.01 \\
   1.67 &  0.03 $\pm$  0.01 &   2.28 &  0.25 $\pm$  0.03 &   2.88 &  0.04 $\pm$  0.01 &   3.47 &  0.00 $\pm$  0.01 \\
   1.73 &  0.05 $\pm$  0.02 &   2.33 &  0.13 $\pm$  0.02 &   2.92 &  0.05 $\pm$  0.01 &   3.53 &  0.04 $\pm$  0.01 \\
   1.77 &  0.13 $\pm$  0.03 &   2.38 &  0.12 $\pm$  0.02 &   2.97 &  0.07 $\pm$  0.02 &   3.58 &  0.02 $\pm$  0.01 \\
   1.83 &  0.22 $\pm$  0.03 &   2.42 &  0.10 $\pm$  0.02 &   3.03 &  0.05 $\pm$  0.01 &   3.63 &  0.02 $\pm$  0.01 \\
   1.88 &  0.24 $\pm$  0.03 &   2.47 &  0.12 $\pm$  0.02 &   3.08 &  0.28 $\pm$  0.03 &   3.67 &  0.04 $\pm$  0.01 \\
   1.92 &  0.27 $\pm$  0.04 &   2.53 &  0.11 $\pm$  0.02 &   3.13 &  0.19 $\pm$  0.02 &   3.72 &  0.01 $\pm$  0.01 \\
   1.98 &  0.38 $\pm$  0.04 &   2.58 &  0.12 $\pm$  0.02 &   3.17 &  0.04 $\pm$  0.01 &   3.78 &  0.02 $\pm$  0.01 \\
   2.03 &  0.47 $\pm$  0.05 &   2.63 &  0.07 $\pm$  0.02 &   3.22 &  0.07 $\pm$  0.01 &   3.83 &  0.02 $\pm$  0.01 \\
   2.08 &  0.37 $\pm$  0.04 &   2.67 &  0.08 $\pm$  0.02 &   3.28 &  0.05 $\pm$  0.01 &   3.88 &  0.00 $\pm$  0.01 \\
   2.13 &  0.38 $\pm$  0.04 &   2.72 &  0.08 $\pm$  0.02 &   3.33 &  0.03 $\pm$  0.01 &   3.92 &  0.01 $\pm$  0.01 \\
   2.17 &  0.25 $\pm$  0.03 &   2.78 &  0.05 $\pm$  0.01 &   3.38 &  0.01 $\pm$  0.01 &   3.97 &  0.02 $\pm$  0.01 \\

\end{tabular}
\end{ruledtabular}
\end{table*}

\begin{figure}[tbh]
\begin{center}
\vspace{-0.2cm}
\includegraphics[width=0.9\linewidth]{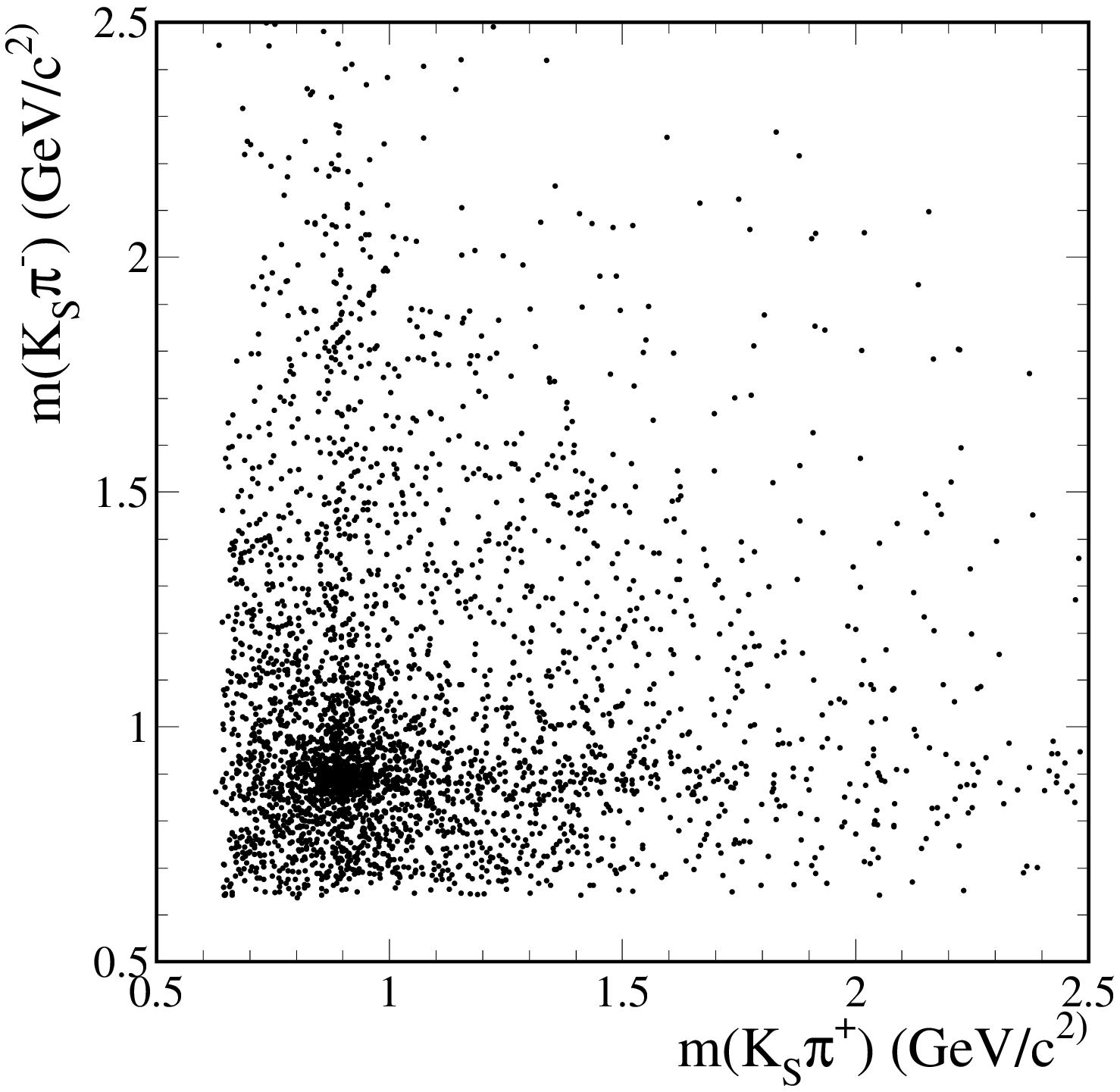}
\vspace{-0.4cm}
\caption{The $\KS\pi^-$ invariant mass versus the $\KS\pi^+$  
  invariant mass (two entries per event). 
  }
\label{kstarpi2dks}
\end{center}
\end{figure}
\begin{figure}[tbh]
\begin{center}
\vspace{-0.2cm}
\includegraphics[width=1.0\linewidth]{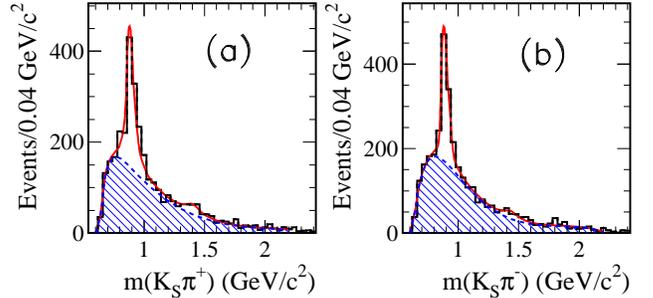}
\vspace{-0.4cm}
\caption{
The (a) $m(\KS\pi^+)$ and (b) $m(\KS\pi^-)$ projections of
Fig.~\ref{kstarpi2dks}.
The lines and hatched areas represent the resuls of the fits described
in the text and their non-$K^*$ components, respectively.
}
\label{kstarpiks}
\end{center}
\end{figure}
\begin{figure}[tbh]
\vspace{-0.2cm}
\begin{center}
\includegraphics[width=0.9\linewidth]{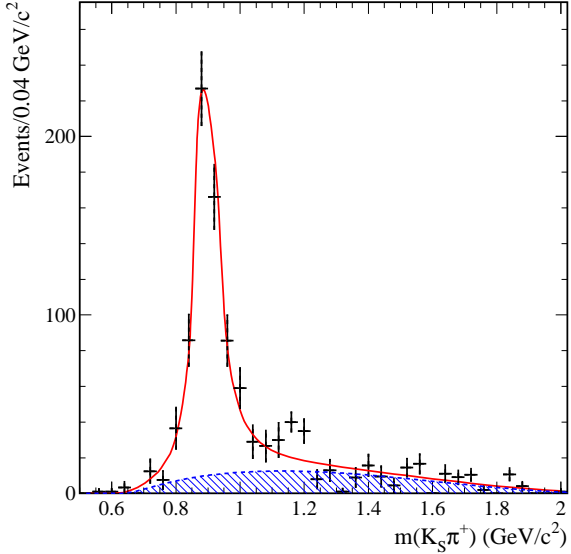}
\vspace{-0.2cm}
\caption{
The fitted number of $K^{*}(892)^+$ events in each 0.04~\gevcc
interval of the $\KS\pim$ mass.   
The curve represents the result of the fit described in the text, 
with the hatched area representing the non-resonant component.
  }
\label{k2starsks}
\end{center}
\end{figure}
\begin{figure}[tbh]
\vspace{-0.2cm}
\begin{center}
\includegraphics[width=0.9\linewidth]{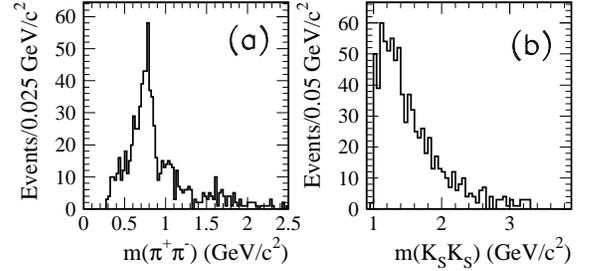}
\vspace{-0.2cm}
\caption{ 
The (a) $\pipi$ and (b) $\KS \KS$ invariant mass distributions for the  
selected $\KS\KS\pipi$ events
with $K^{*}(892)^+ K^{*}(892)^-$ events excluded (see text).
  }
\label{rhoksks}
\end{center}
\end{figure}
\subsection{\boldmath The $K^{*}(892)^{\pm}$ and $K_2^{*}(1430)^{\pm}$
  contributions }
\label{sec:kstar2ks}
Figure~\ref{kstarpi2dks} shows a scatter plot of the $\KS\pim$
invariant mass versus the $\KS\pip$ invariant mass, 
with two entries per event. 
Clear bands associated with the $K^{*}(892)^{\pm}$ are visible here,
as are peaks in the projections shown in Fig.~\ref{kstarpiks}.
The projections also show indications of $K_2^{*}(1430)^{\pm}$
production.

Fitting the projections with a sum of two Breit-Wigner functions
and a threshold function yields
827$\pm$29 $K^{*}(892)^+ \!\!\to\! \KS\pi^+$ and 
856$\pm$50 $K^{*}(892)^- \!\!\to\! \KS\pi^-$ decays, as well as
116$\pm$40 $K_2^{*}(1430)^+$ and
 70$\pm$34 $K_2^{*}(1430)^-$ decays.
The total number of $K^{*}(892)^{\pm}$ decays is larger than the
number of $\KS\KS\pipi$ events, 
indicating correlated production of $K^{*}(892)^+ K^{*}(892)^-$ pairs.
We fit the $\KS\pip$ invariant mass distributon in 0.04~\gevcc bins of
the $\KS\pim$ mass,
and show the number of $K^{*}(892)^+$ decays in each bin in
Fig.~\ref{k2starsks}.
A clear $K^{*}(892)^+$ signal is observed;
a fit yields $742\pm30\pm100$ pair production events,
$\epem \to K^{*}(892)^+ K^{*}(892)^- \to \KS\KS\pipi$,
where the second uncertainty is due to variation of the starting
values of the fit parameters.
This accounts for 50\% of the selected events and 88\% of the
$K^*(892)^\pm$ production.
We find no significant signal at the $K_2^{*}(1430)^+$ mass, 
and hence no evidence for 
$\epem \to K^{*}(892)^\pm K_2^{*}(1430)^\mp$ events.

The number of correlated $K^{*}(892)^+ K^{*}(892)^- $ production
events in this channel  
($742\pm104$  events with 4.5\% efficiency)
can be compared with the corresponding numbers in the
$\KS\KL\pipi$ channel 
($2098\pm209$ events with 5\% efficiency), presented above,
 and in the $\Kp\Km\ppz$ final state 
($1750\pm60$ events with 8\% efficiency), 
from our previous measurement~\cite{isr2k2pi} using the same
integrated luminosity. 
Normalizing these to the same 5\% efficiency, 
we obtain the ratios $(824\pm116)$:$(2098\pm209)$:$(1094\pm38)$. 
These are consistent with the 1:2:1 ratios expected assuming equal
production of \KS and \KL in $K^{*}(892)^{\pm}$ decays.

The size of the data sample is not large enough to apply
this procedure to every
$m(\KS\KS\pipi)$ bin and extract the 
$\epem \!\to\! K^{*}(892)^+ K^{*}(892)^-$ cross section. 
However,
considering events with both $m(\KS\pip)$ and $m(\KS\pim)$ within
$\pm$0.15\gevcc of the nominal $K^{*}(892)^\pm$ mass~\cite{PDG}, 
we conclude that the $ K^{*}(892)^+ K^{*}(892)^-$
 contribution almost completely dominates for
$m(\KS\KS\pipi)$ below 2.5~\gev.
For the events outside this box,
we show the $\pipi$ and $\KS\KS$ invariant mass distributions in
Fig.~\ref{rhoksks}.
The $\rho(770)$ resonance is prominent in the \pipi spectrum, 
whereas the $\KS\KS$ spectrum shows no significant structure.
The three resonant channels $K^{*}(892)^+ K^{*}(892)^-$,
$K^{*}(892)^{\pm} \KS\pi^{\mp}$ (see Fig.~\ref{kstarpi2dks}), and 
$\rho(770)\KS \KS$ dominate the $\KS\KS\pipi$ cross section within our 
measured range,
and there is a small contribution from
$K^{*}(1430)^{\pm} \KS\pi^{\mp}$.

\section{\boldmath The $\KS \KS \Kp \Km$ final state}
\label{sec:ksks2k}
\subsection{Final selection and background}
\label{addksks2k}
We consider the events from Sec.~\ref{sec:KS} with at least two \KS
candidates,
and the combination of two \KS candidates and two charged tracks in
each event giving the best \chisq for a 4C fit under the $\KS\KS\Kp\Km$ 
hypothesis (see Sec.~\ref{sec:Analysis}).
To reduce the background from multi-pionic events, 
we require that both of the charged tracks not in the
\KS candidates be identified as kaons. 

\begin{figure}[tbh]
\begin{center}
\includegraphics[width=0.9\linewidth]{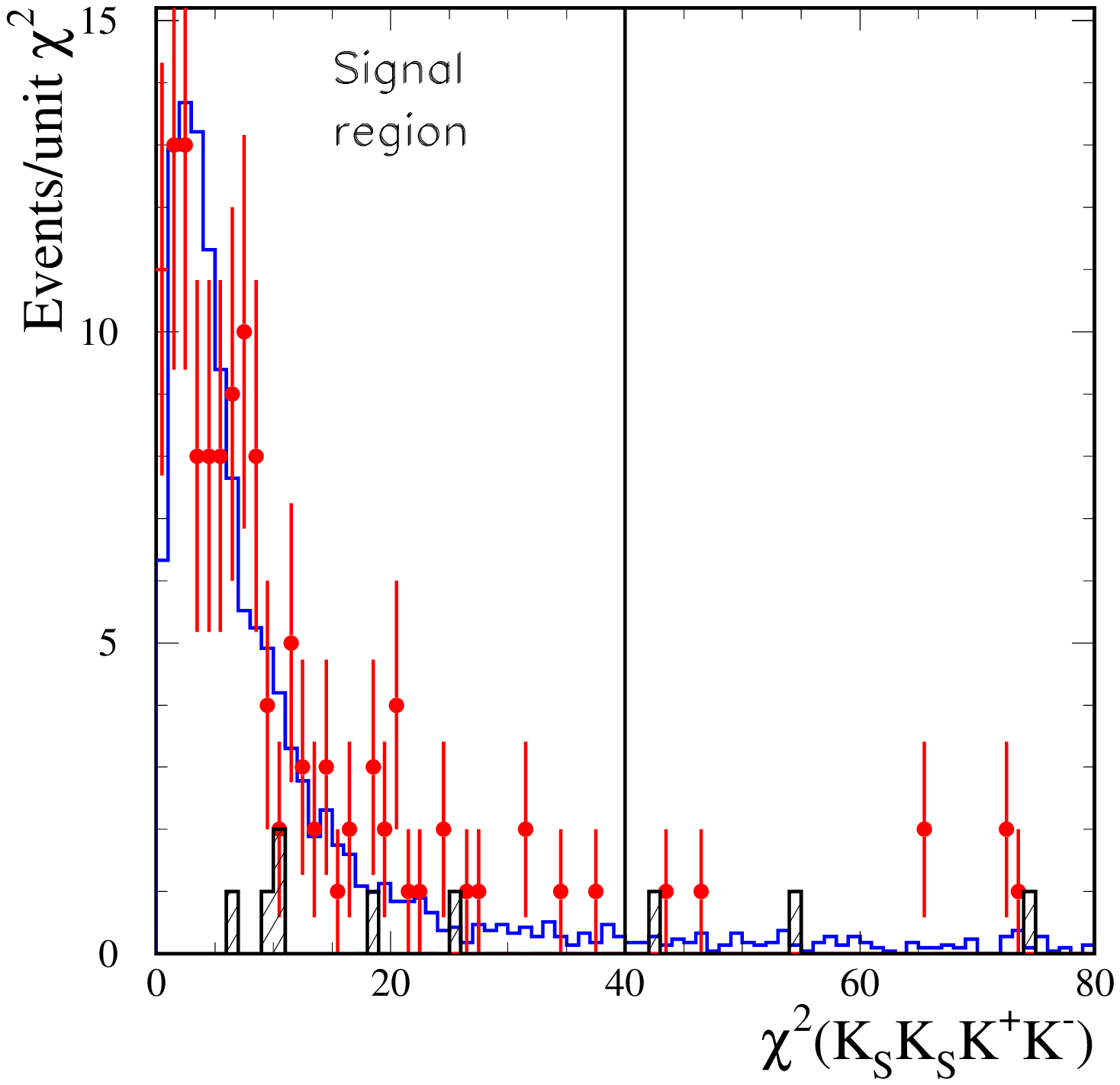}
\vspace{-0.4cm}
\caption{
The four-constraint \chisq distributions for $\KS\KS\Kp\Km\gamma$
candidate events in the data (points) and signal MC simulation (open
histogram) fitted under the $\KS\KS\Kp\Km$ hypothesis. 
The cross-hatched histrogram represents the simulated contribution
from non-ISR \qqbar events.
}
\label{ksks2k_chi2}
\end{center}
\end{figure}

The $\chisq(\KS\KS\Kp\Km)$ distribution for the selected events 
is shown in Fig.~\ref{ksks2k_chi2} (points)
along with that for simulated ISR $\KS\KS\Kp\Km$ events 
(open histogram),
where the latter distribution is normalized to the data
in the region $\chisq(\KS \KS\pipi)\! <\! 8$.
There is very little background:
simulated ISR events in other channels do not satisfy the selection;
there is no significant \piz peak in the data;
and the signal MC describes the data well, even at high \chisq values.
The simulation predicts only a few $\epem \to \qqbar \to
\KS\KS\Kp\Km\piz$ events, 
which are shown as the hatched histogram in Fig.~\ref{ksks2k_chi2}.

\begin{figure}[tbh]
\begin{center}
\includegraphics[width=0.9\linewidth]{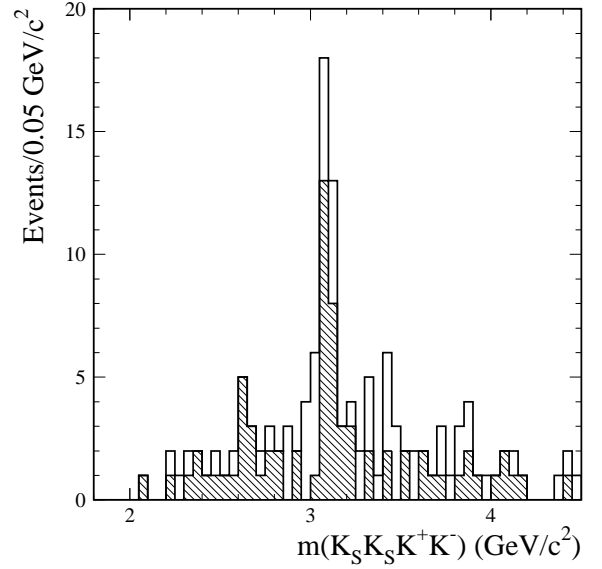}
\vspace{-0.4cm}
\caption{
The $\KS\KS\Kp\Km$ invariant mass distribution for data events in the
signal region, $\chisq(\KS \KS K^+ K^-)<40$ (open histogram). 
The subset of events with $m(\Kp\Km)<1.04~\gevcc$,
predominantly $\KS\KS\phi(1020)$ events,
is shown as the shaded histogram.
}
\label{ksks2k_raw}
\end{center}
\end{figure}
\begin{figure}[tbh]
\begin{center}
\includegraphics[width=0.9\linewidth]{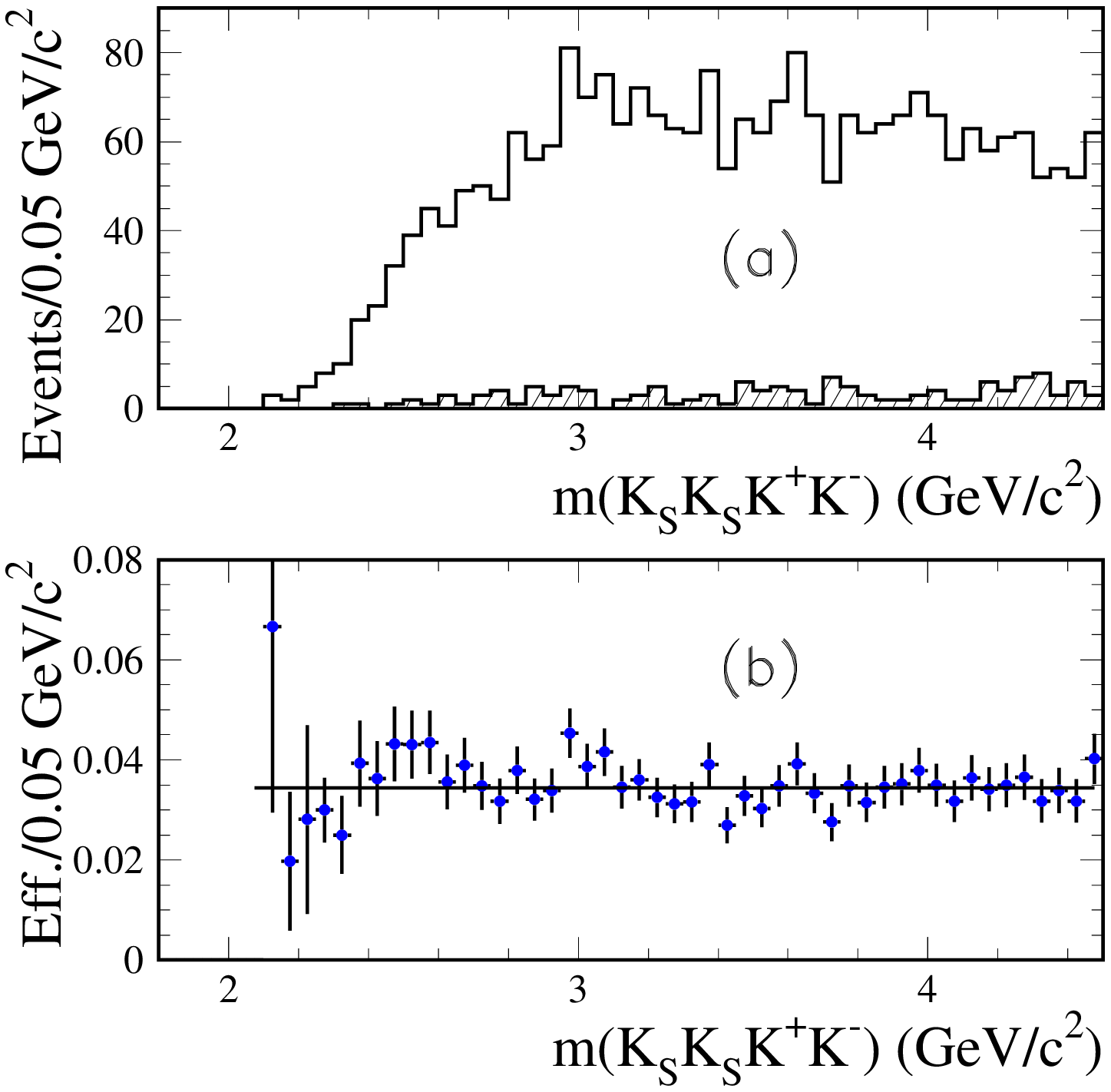}
\vspace{-0.4cm}
\caption{
(a) The $\KS\KS\Kp\Km$ invariant mass distribution for the MC-simulated
    signal events in the signal and control (shaded) regions of
    Fig.~\ref{ksks2k_chi2}.
(b) The net reconstruction and selection efficiency from the simulation.
}
\label{ksks2k_eff}
\end{center}
\end{figure}

We select events with $\chisq(\KS \KS K^+ K^-)<40$, 
obtaining 129 events in the data with masses between threshold and
4.5~\gevcc,
and 2544 events in the signal MC simulation.
The $\KS \KS K^+ K^-$ invariant mass distribution is shown as the open
histogram in Fig.~\ref{ksks2k_raw}.
We do not subtract any background,
nor do we assign any systematic uncertainty to account for possible background
contributions.

\subsection{ Simulated detection efficiency}
\label{sec:ksks2keff}
\begin{table*}
\caption{Summary of the $\epem\to K_S K_S K^+ K^-$ 
cross section measurement. Uncertainties are statistical only.}
\label{ksks2k_tab}
\begin{ruledtabular}
\begin{tabular}{ c c c c c c c c }
$E_{\rm c.m.}$ (GeV) & $\sigma$ (nb)  
& $E_{\rm c.m.}$ (GeV) & $\sigma$ (nb) 
& $E_{\rm c.m.}$ (GeV) & $\sigma$ (nb) 
& $E_{\rm c.m.}$ (GeV) & $\sigma$ (nb)  
\\
\hline
   2.05 & 0.003 $\pm$ 0.003 &   2.75 & 0.010 $\pm$ 0.004 &   3.45 & 0.013 $\pm$ 0.004 &   4.15 & 0.003 $\pm$ 0.002 \\
   2.15 & 0.000 $\pm$ 0.003 &   2.85 & 0.010 $\pm$ 0.004 &   3.55 & 0.006 $\pm$ 0.003 &   4.25 & 0.000 $\pm$ 0.003 \\
   2.25 & 0.008 $\pm$ 0.004 &   2.95 & 0.011 $\pm$ 0.005 &   3.65 & 0.004 $\pm$ 0.002 &   4.35 & 0.001 $\pm$ 0.001 \\
   2.35 & 0.010 $\pm$ 0.005 &   3.05 & 0.012 $\pm$ 0.005 &   3.75 & 0.005 $\pm$ 0.003 &   4.45 & 0.003 $\pm$ 0.002 \\
   2.45 & 0.007 $\pm$ 0.004 &   3.15 & 0.005 $\pm$ 0.003 &   3.85 & 0.009 $\pm$ 0.003 &        &  \\
   2.55 & 0.007 $\pm$ 0.004 &   3.25 & 0.010 $\pm$ 0.004 &   3.95 & 0.002 $\pm$ 0.002 &  &  \\
   2.65 & 0.017 $\pm$ 0.006 &   3.35 & 0.009 $\pm$ 0.004 &   4.05 & 0.004 $\pm$ 0.002 &  & \\
\end{tabular}
\end{ruledtabular}
\end{table*}

The MC-simulated $\KS\KS\Kp\Km$ invariant-mass distribution is shown in
Fig.~\ref{ksks2k_eff}(a) for events in the signal and control (shaded
histogram) regions.
The mass dependence of the detection efficiency
is shown in Fig.~\ref{ksks2k_eff}(b). 
The mass interval used, 50~\mevcc per bin, is wider than the 10~\mevcc
detector resolution, 
and the cross section has no sharp structure 
(except the $J/\psi$ signal, discussed below), 
so we apply no corrections for the resolution.
We apply all the corrections discussed above for data-MC differences
in track, 
\KS, and photon 
detection efficiency.

\begin{figure}[tbh]
\begin{center}
\includegraphics[width=0.9\linewidth]{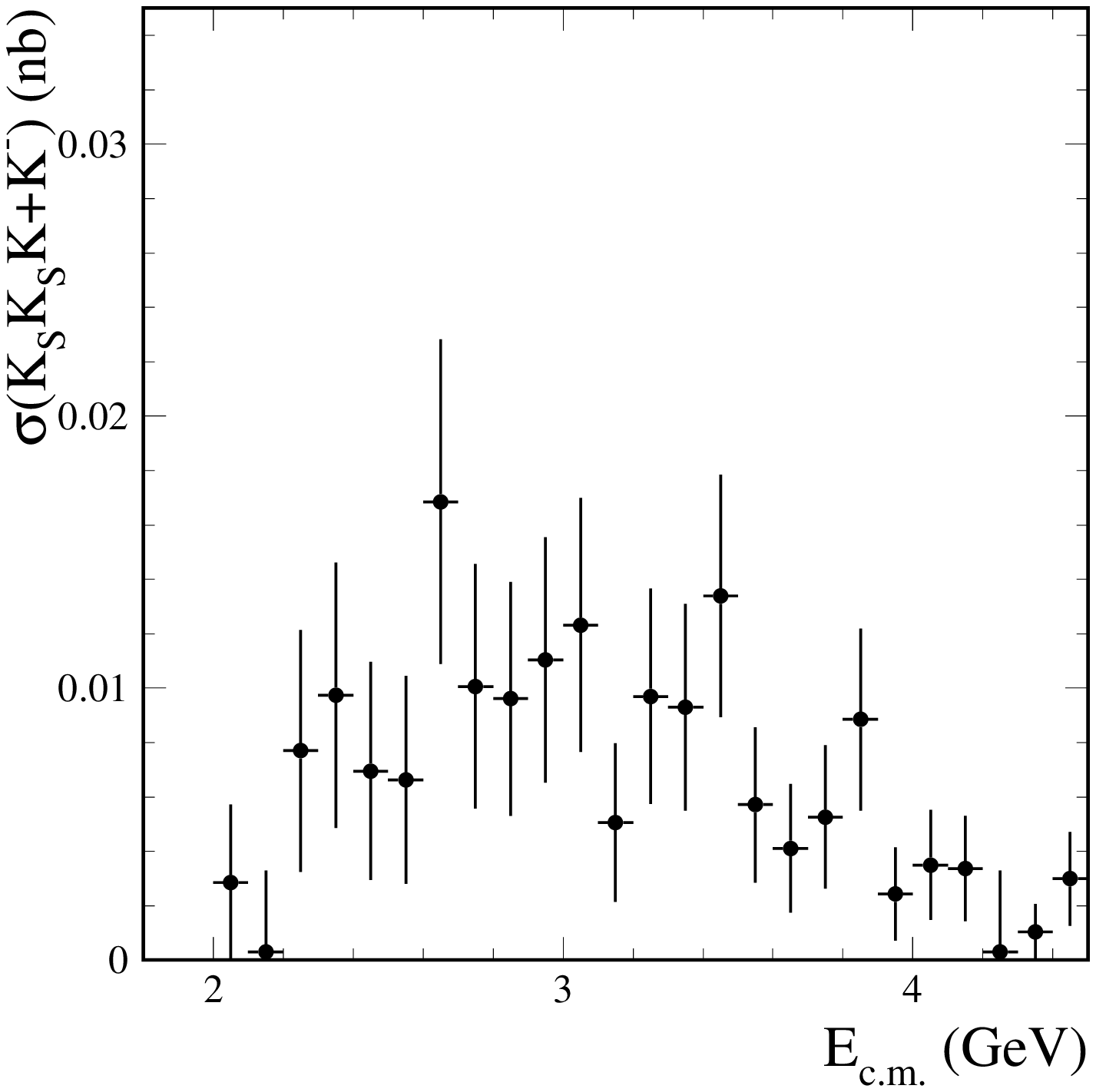}
\vspace{-0.4cm}
\caption{
 The $\epem\to \KS\KS\Kp\Km$ cross section.
 Events with invariant mass within 0.05~\gevcc of the \jpsi mass are
 excluded.
}
\label{ksks2k_xs}
\end{center}
\end{figure}
\begin{figure*}[tbh]
\begin{center}
\includegraphics[width=0.33\linewidth]{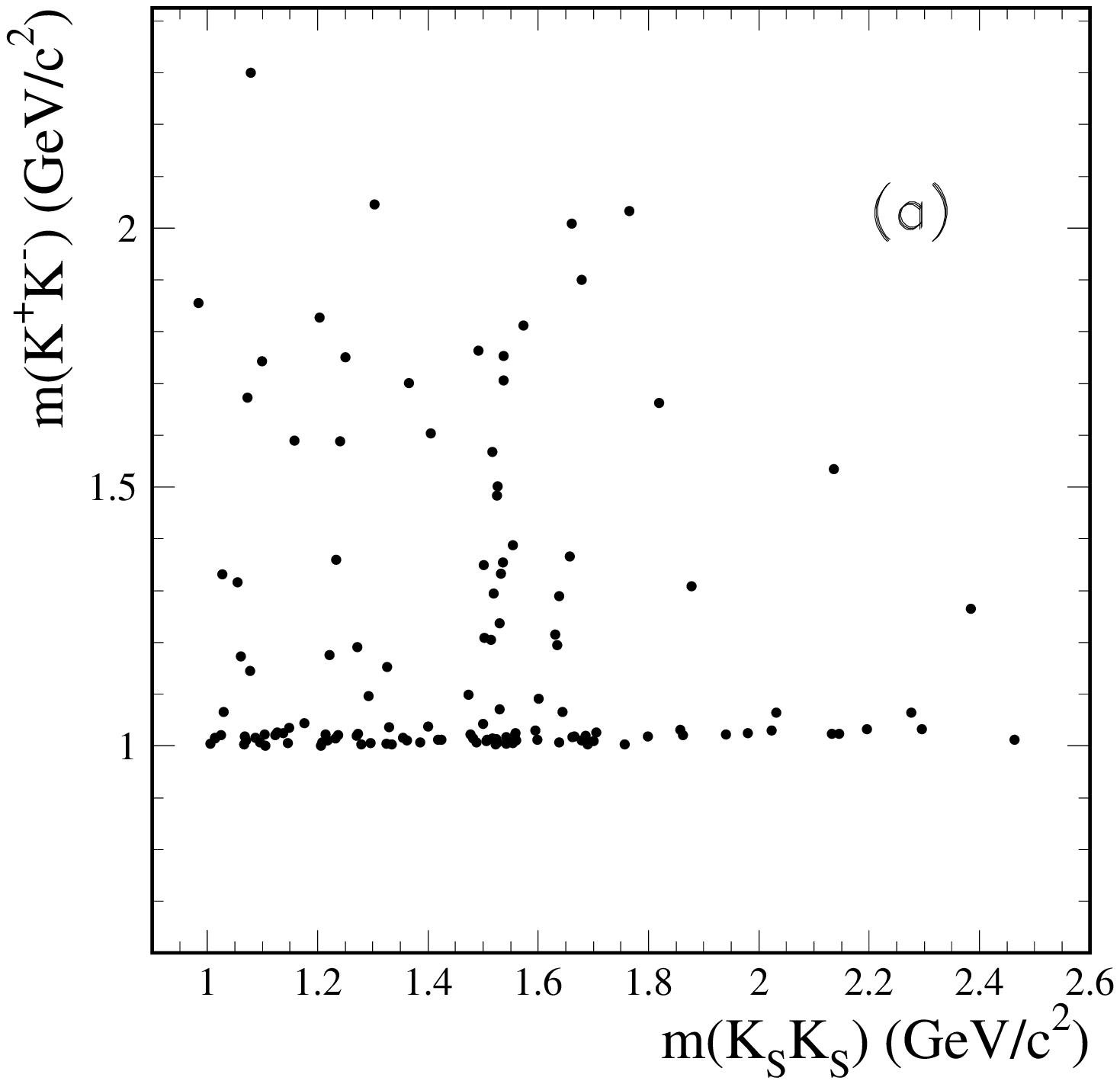}
\includegraphics[width=0.33\linewidth]{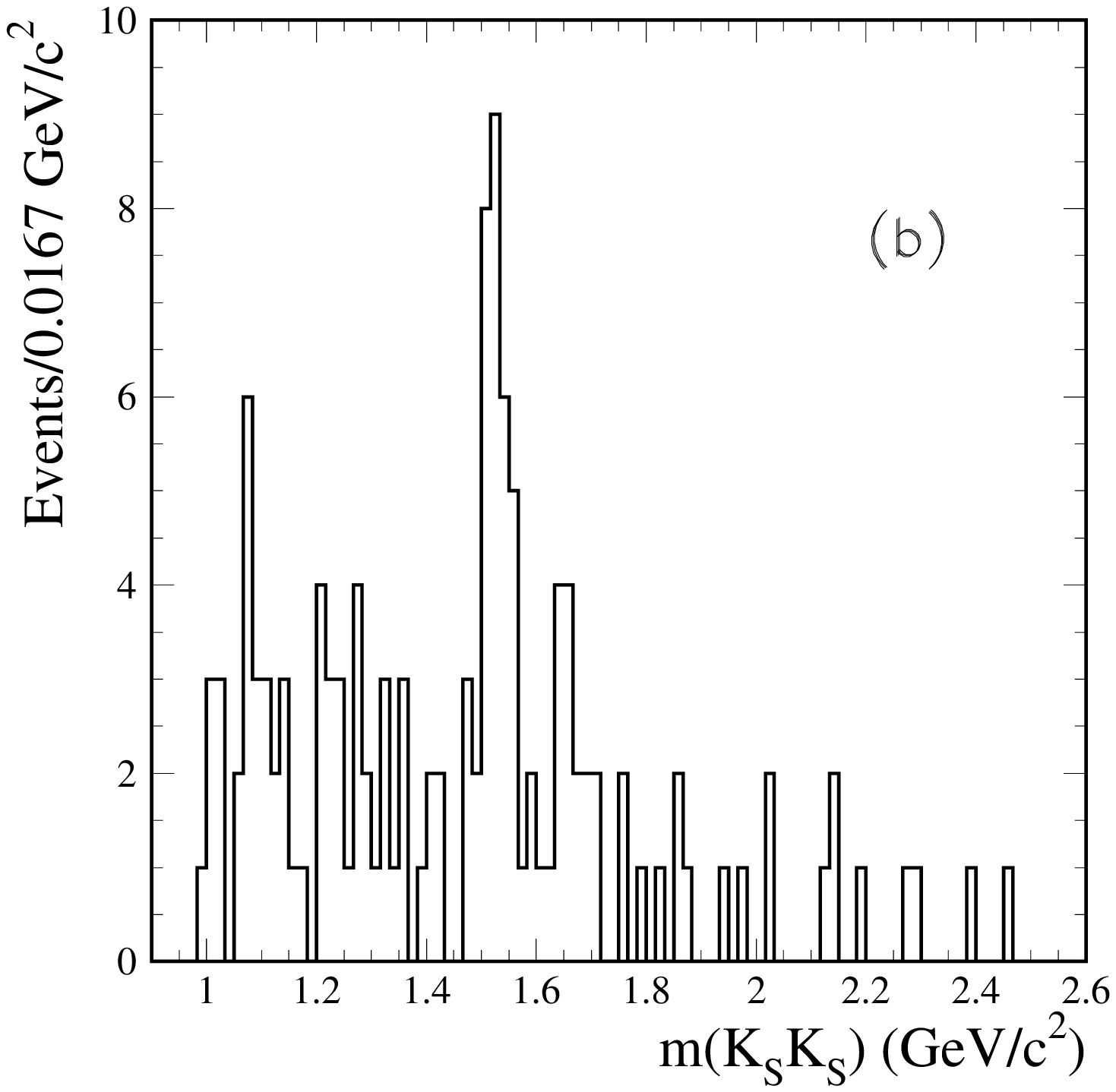}
\includegraphics[width=0.33\linewidth]{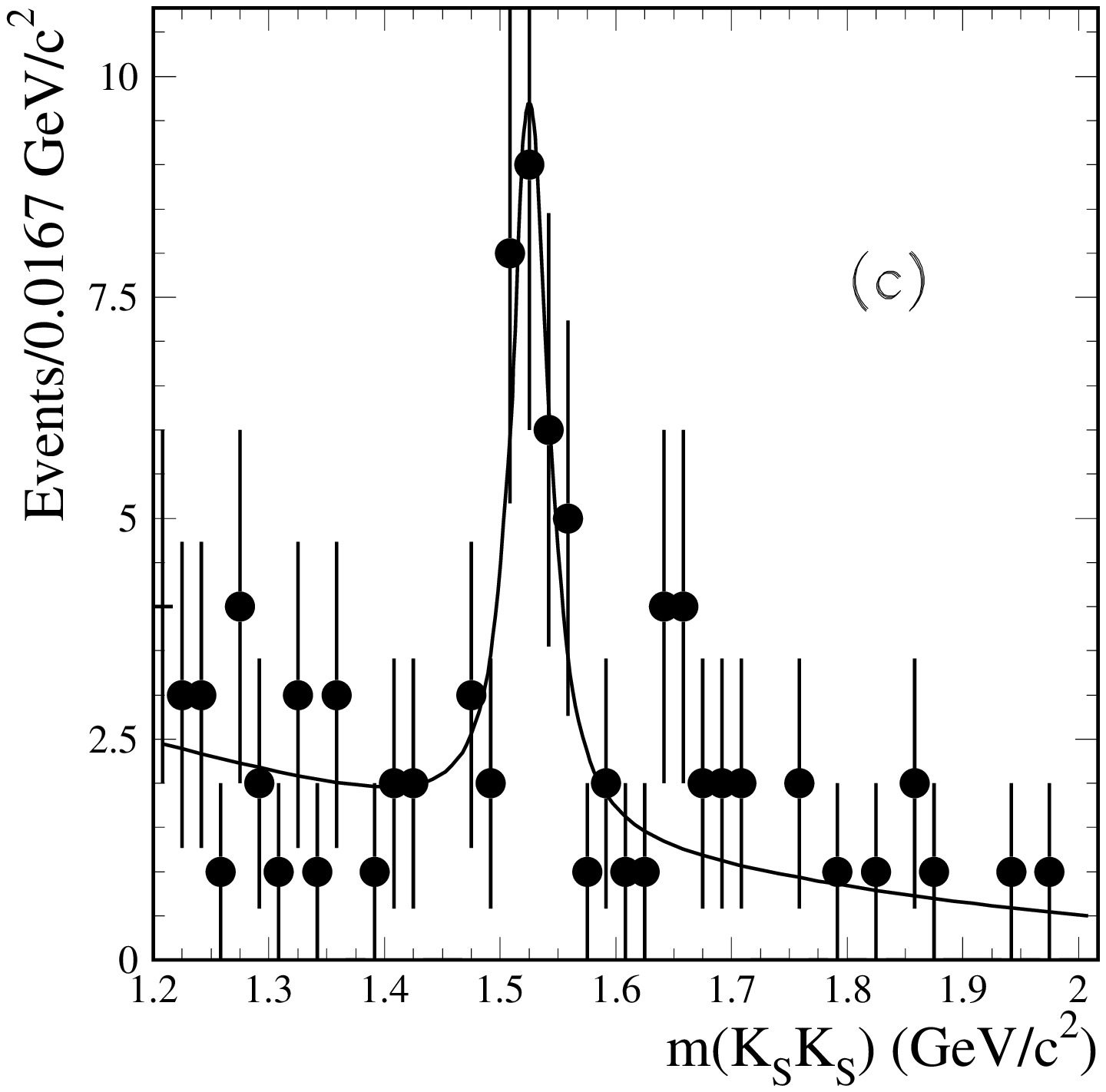}
\vspace{-0.4cm}
\caption{
(a) The $\Kp\Km$ versus $\KS\KS$ invariant mass for
    all selected $\KS\KS\Kp\Km$ events in the data.
(b) The $m(\KS\KS)$ projection of (a).
(c) An expanded view of (b) in which the line represents the result of
    the fit described in the text. 
}
\label{ksksvs2k}
\end{center}
\end{figure*}
\subsection{\boldmath Cross section for $\epem\to \KS \KS K^+ K^-$}
\label{sec:xsksks2k}
We remove the events within $\pm$0.05~\gevcc of the \jpsi signal 
(which is discussed below),
and calculate the $\epem \to \KS\KS\Kp\Km$ cross section using 
Eq.~\ref{xsformular}.
The fully corrected cross section is shown in as a function of energy
in Fig.~\ref{ksks2k_xs} and listed in Table~\ref{ksks2k_tab}.
There are no previous measurements of this final state. 
The systematic uncertainties are smaller than the statistical terms
and do not exceed 5\%.

\subsection{\boldmath Internal structure of the $\KS\KS\Kp\Km$ system}
\label{sec:phifprime}
Figure~\ref{ksksvs2k}(a) shows a scatter plot of the $\Kp\Km$
invariant mass versus the $\KS\KS$ invariant mass for all selected
events. 
A strong $\phi(1020)$ band is evident. 
Requiring $m(K^+ K^-)<1.04$~\gevcc, 
we obtain the contribution from $\phi\KS\KS$ events shown in
Fig~\ref{ksks2k_raw} as the shaded histogram.
This mode dominates at all masses.

There is also structure for $m(\KS\KS)$ near 1.5~\gevcc,
which is more visible as a peak in the $m(\KS\KS)$ projection of 
Fig.~\ref{ksksvs2k}(b).
We fit this mass region with a Breit-Wigner plus a second-order
polynomial function.
An expanded view is shown in Fig.~\ref{ksksvs2k}(c),
along with the result of the fit.
We obtain 29$\pm$7 events with Breit-Wigner mass and width
\begin{eqnarray}
   m   & = & 1.526 \pm 0.007~\gevcc \nonumber \\
\Gamma & = & 0.037 \pm 0.012 ~\gev. \nonumber
\end{eqnarray}
These parameters may be compared with the averages~\cite{PDG} 
for the $f_2^{'}(1525)$ resonance, $m(f_2^{'}) = 1.525 \pm 0.005~\gevcc$ and
$\Gamma(f_2^{'}) = 0.073^ {+0.006}_ {-0.005} ~\gev$; the mass is
consistent but the width is about 3 standard deviations lower.

\begin{figure*}[tbh]
\begin{center}
\includegraphics[width=0.33\linewidth]{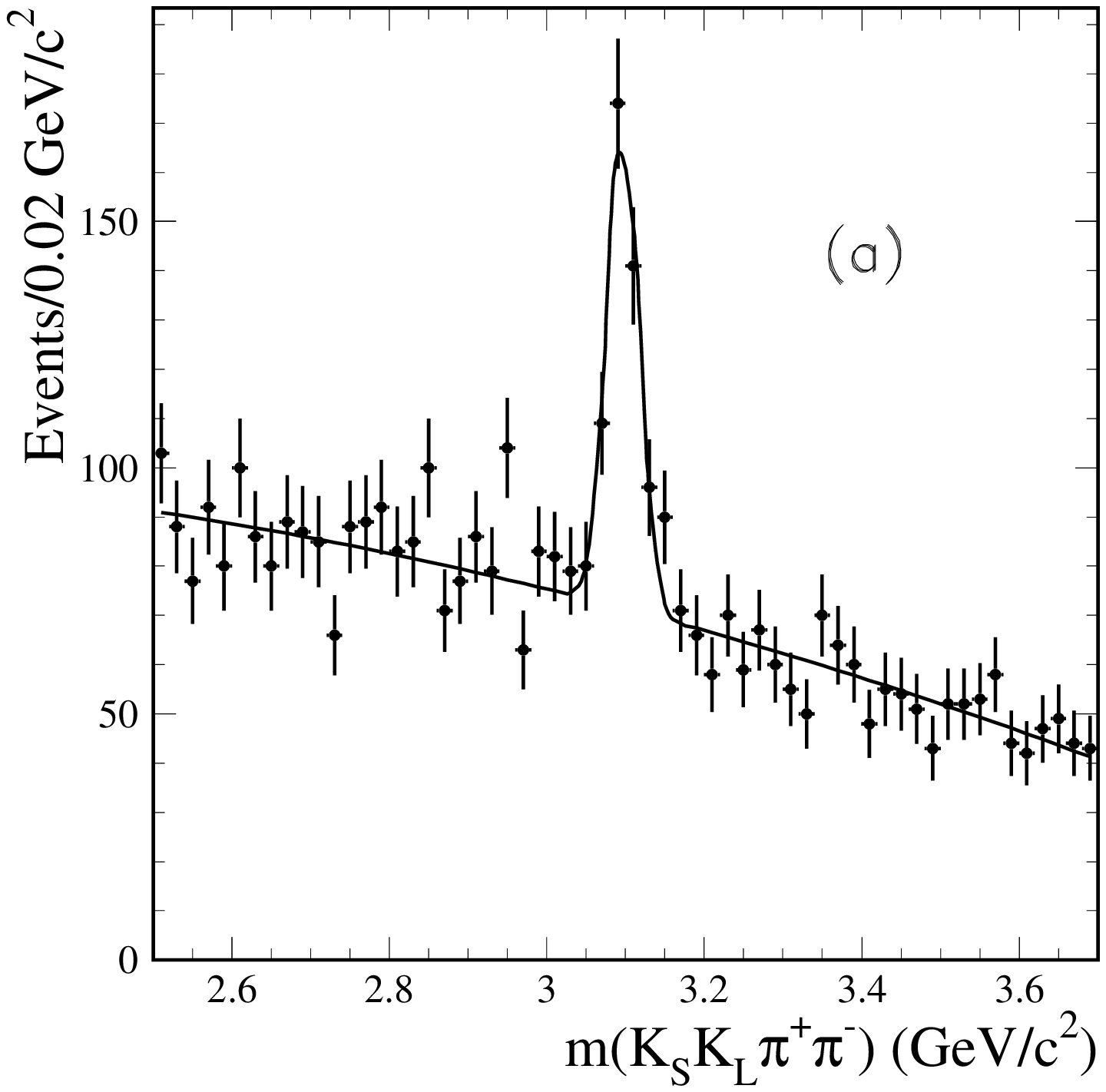}
\includegraphics[width=0.33\linewidth]{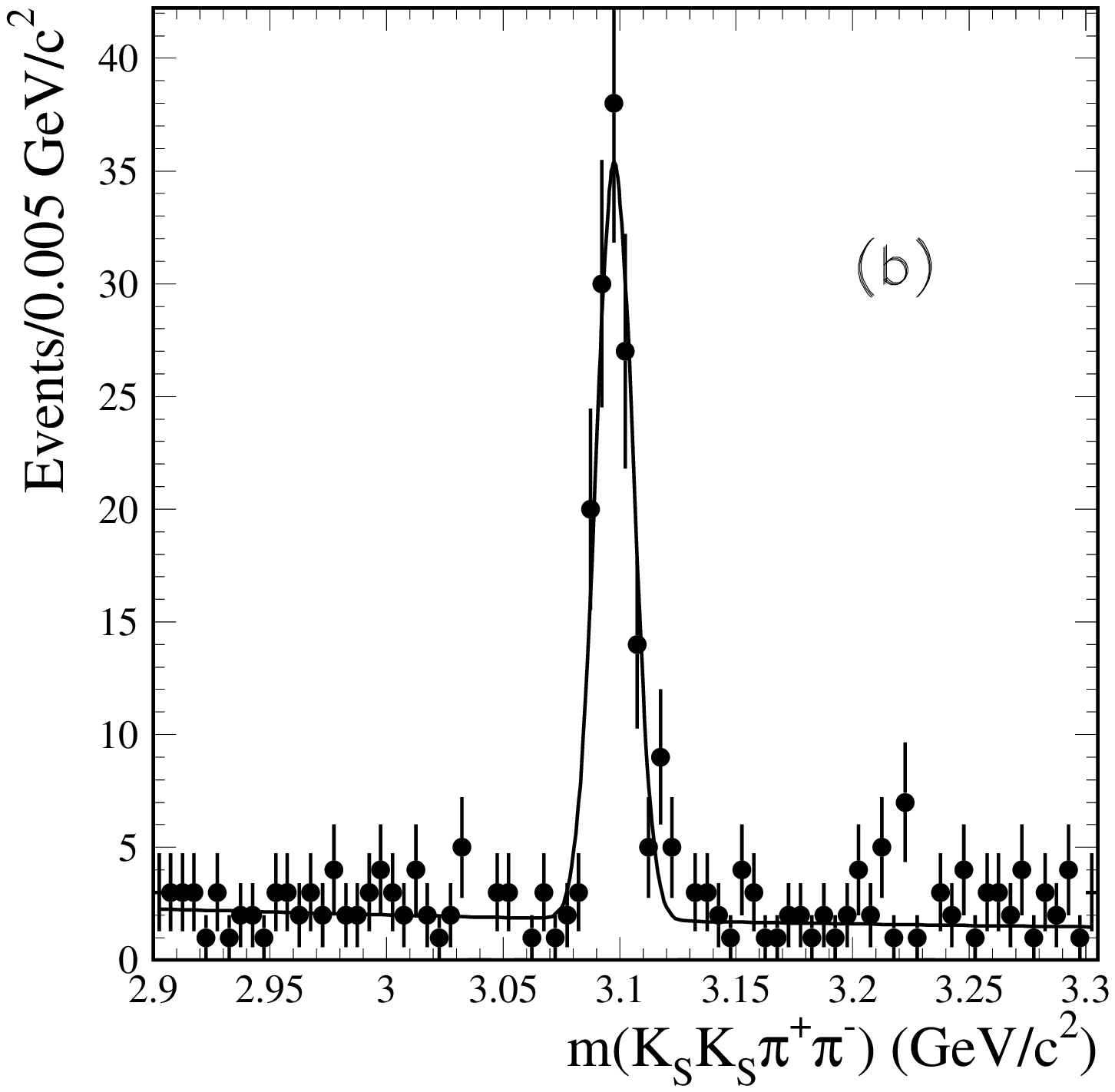}
\includegraphics[width=0.33\linewidth]{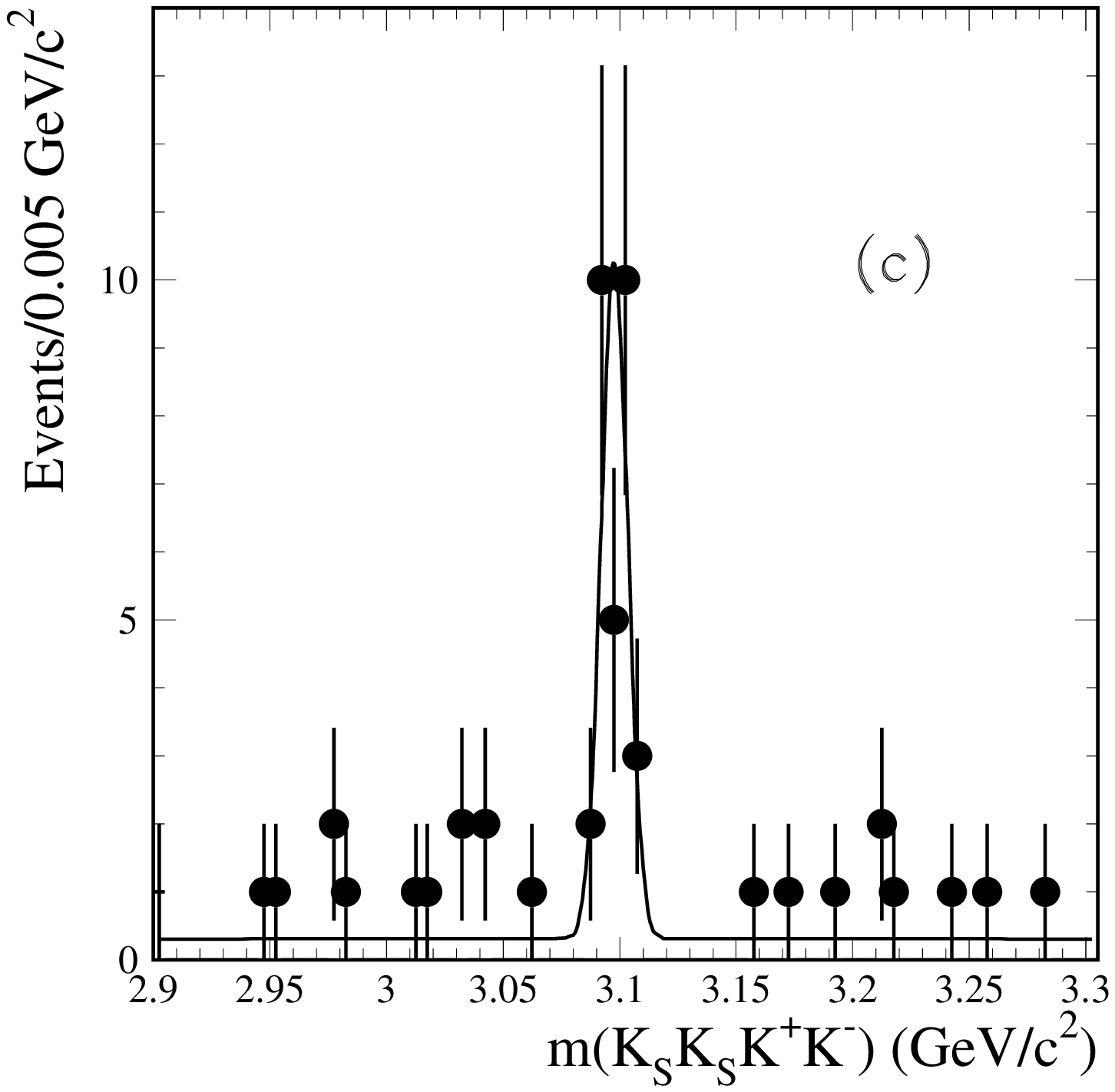}
\vspace{-0.4cm}
\caption{
Expanded views of the invariant mass distributions near the \jpsi mass
for the (a) $\KS\KL\pipi$,  (b) $\KS\KS\pipi$, 
and (c) $\KS\KS\Kp\Km$ final states.
The lines represent the results of the fits described in the text.
}
\label{kskljpsi}
\end{center}
\end{figure*}
\begin{table*}[tbh]
\caption{
  Summary of the $J/\psi$ parameters
  obtained in this analysis.
  }
\label{jpsitab}
\begin{ruledtabular}
\begin{tabular}{r@{~$\cdot$}l  r@{.}l@{$\pm$}l@{$\pm$}l 
                               r@{.}l@{$\pm$}l@{$\pm$}l
                               r@{.}l@{$$}l } 
\multicolumn{2}{c}{Measured} & \multicolumn{4}{c}{Measured}    &  
\multicolumn{7}{c}{$J/\psi$ Branching Fraction (10$^{-3}$)}\\
\multicolumn{2}{c}{Quantity} & \multicolumn{4}{c}{Value (\ev)} &
\multicolumn{4}{c}{This work}    & 
\multicolumn{3}{c}{PDG2012} \\
\hline
$\Gamma^{J/\psi}_{ee}$  &  $\BR_{J/\psi  \to \KS\KL\pipi}$     &
\hspace*{0.4cm}    
 20&8  & 2.3  & 2.1    & 3&7  & 0.6  & 0.4  &  \multicolumn{3}{c}{no entry} \\

$\Gamma^{J/\psi}_{ee}$  &  $\BR_{J/\psi  \to \KS\KS\pipi}$     &
  9&3  & 0.9  & 0.5    & 1&68 & 0.16 & 0.08 &  \multicolumn{3}{c}{no entry} \\

$\Gamma^{J/\psi}_{ee}$  &  $\BR_{J/\psi  \to \KS\KS K^+ K^-} $  &
  2&3  & 0.4  & 0.1    & 0&42 & 0.08 & 0.02 &  \multicolumn{3}{c}{no entry} \\

$\Gamma^{J/\psi}_{ee}$  &  $\BR_{J/\psi  \to K^*(892)\KS\pi} 
\cdot   \BR_{K^*(892)\to\KS\pi}$  &
  3&7  & 1.2  & 0.3    & 2&6 & 0.9 & 0.2 &  \multicolumn{3}{c}{no entry} \\

$\Gamma^{J/\psi}_{ee}$  &  $\BR_{J/\psi \!\to\! K_2^{*}(1430)\KS\pi}  
\cdot   \BR_{K_2^{*}(1430)\to\KS\pi}$  &
  2&5  & 1.2  & 0.2    & 3&6 & 1.7 & 0.3 &  \multicolumn{3}{c}{no entry} \\

$\Gamma^{J/\psi}_{ee}$  &  $\BR_{J/\psi  \to K^*(892)^+K^*(892)^-} 
\cdot  \BR^{2}_{K^*(892)\to\KS\pi}$  &
  0&05 & 0.03 & 0.02   & $<\,$1& \multicolumn{3}{l}{\hskip-2pt7~ 90\% ~C.L.} & 
  1&0  &  $^{+0.2}_{-0.4}$ \\

$\Gamma^{J/\psi}_{ee}$  &  $\BR_{J/\psi \!\to\! K_2^{*}(1430)K^*(892)}  
\cdot   \BR_{K_2^{*}(1430)\to\KS\pi} \cdot \BR_{K^*(892)\to\KS\pi} $  &
  0&58 & 0.50 & 0.02   & $<\,$7& \multicolumn{3}{l}{\hskip-2pt8~ 90\% ~C.L.} & 
\multicolumn{3}{c}{no entry} \\

$\Gamma^{J/\psi}_{ee}$  &  $\BR_{J/\psi\to\KS\KS\phi(1020) }
                      \cdot \BR_{\phi\to K^+ K^-}  $          &
  1&6  & 0.4  & 0.1    & 0&58 & 0.14 & 0.03 &  \multicolumn{3}{c}{no entry} \\

$\Gamma^{J/\psi}_{ee} $ & $\BR_{J/\psi\to f_2^{'}(1525)\phi(1020)}
                        \cdot\BR_{\phi\to K^+ K^-} 
                         \cdot\BR_{f_2^{'}(1525)\to \KS \KS}$\hspace*{0.5cm}  &
  0&88 & 0.34 & 0.04  & 0&45 & 0.17 & 0.02 &  0&8  & $\pm$0.4 (S=2.7)  \\

$\Gamma^{J/\psi}_{ee} $ & $\BR_{J/\psi\to f_2^{'}(1525)K^+ K^-}
                         \cdot\BR_{f_2^{'}(1525)\to \KS \KS}$  &
\hspace*{0.5cm} 1&28 & 0.42 & 0.05\hspace*{0.4cm} & 
\hspace*{0.5cm} 0&32 & 0.11 & 0.02\hspace*{0.4cm} & \multicolumn{3}{c}{no entry} \\

\end{tabular}
\end{ruledtabular}
\end{table*}
\section{ The Charmonium Region }
Figures~\ref{kskljpsi}(a), (b), and (c) show expanded views of the mass
distributions in Figs.~\ref{kskl2pi_chi2}(c), \ref{ksks2pi_raw},
and~\ref{ksks2k_raw}, respectively,
in the \jpsi mass region. 
Fitting with Gaussian plus second-order polynomial functions yields
$248\pm27$ $\jpsi\to\KS\KL\pipi$ decays,  
$133\pm13$ $\jpsi\to\KS\KS\pipi$ decays, and  
$28.5\pm5.5$ $\jpsi\to\KS\KS\Kp\Km$ decays.
Using the respective simulated efficiencies with all the corrections
described above, and the differential luminosity, 
we calculate the products of the \jpsi electronic width and
branching fractions to these modes, and list them in Table~\ref{jpsitab}.
Using the PDG value of $\Gamma_{ee}(\jpsi) = 5.55$~\kev~\cite{PDG}, 
we obtain the corresponding branching fractions, 
also presented in Table~\ref{jpsitab}. 
These are the first observations of these \jpsi decay modes and
measurements of their branching fractions.
They can be compared with 
$\BR(J/\psi\to K^+ K^-\pipi) = (6.8\pm0.3)\times 10^{-3}$~\cite{PDG},
which is dominated by the \babar~ measurement.
\begin{figure}[tbh]
\begin{center}
\vspace{-0.2cm}
\includegraphics[width=1.0\linewidth]{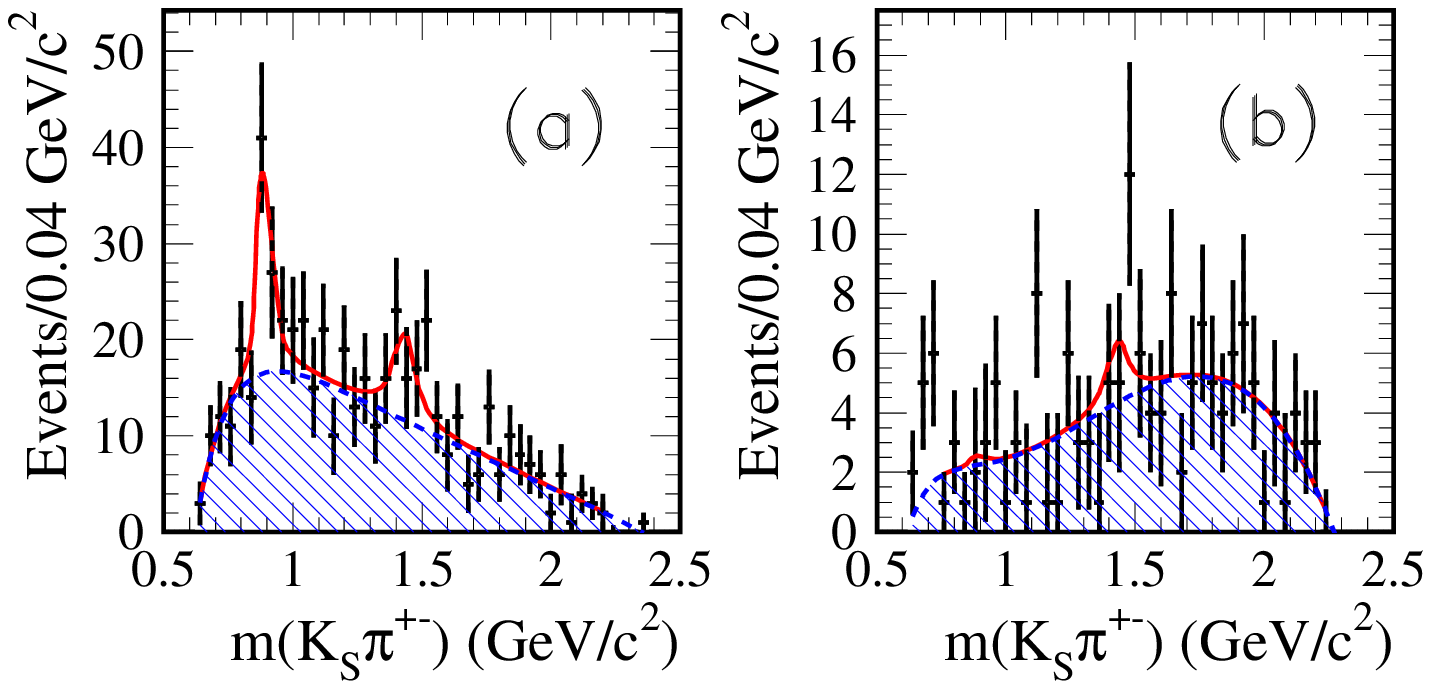}
\vspace{-0.4cm}
\caption{ (a)
The $\KS \pipm$ invariant mass distribution
(four entries per event) 
for the $\KS\KS\pipi$ events under the
\jpsi peak with the non-\jpsi contribution subtracted (see text). 
(b) The distribution of the other $m(\KS \pi^\mp)$ for those events
with one $m(\KS\pipm)$ value within 0.15~\gevcc of the $K^*(892)^\pm$ mass
(up to two entries per event, pairings in the overlap region taken once).
The lines represent the results of the fits described in the text, 
with the shaded areas representing the combinatorial components. 
}
\label{jpsi_kpi}
\end{center}
\end{figure}
\begin{figure}[tbh]
\begin{center}
\vspace{-0.2cm}
\includegraphics[width=1.0\linewidth]{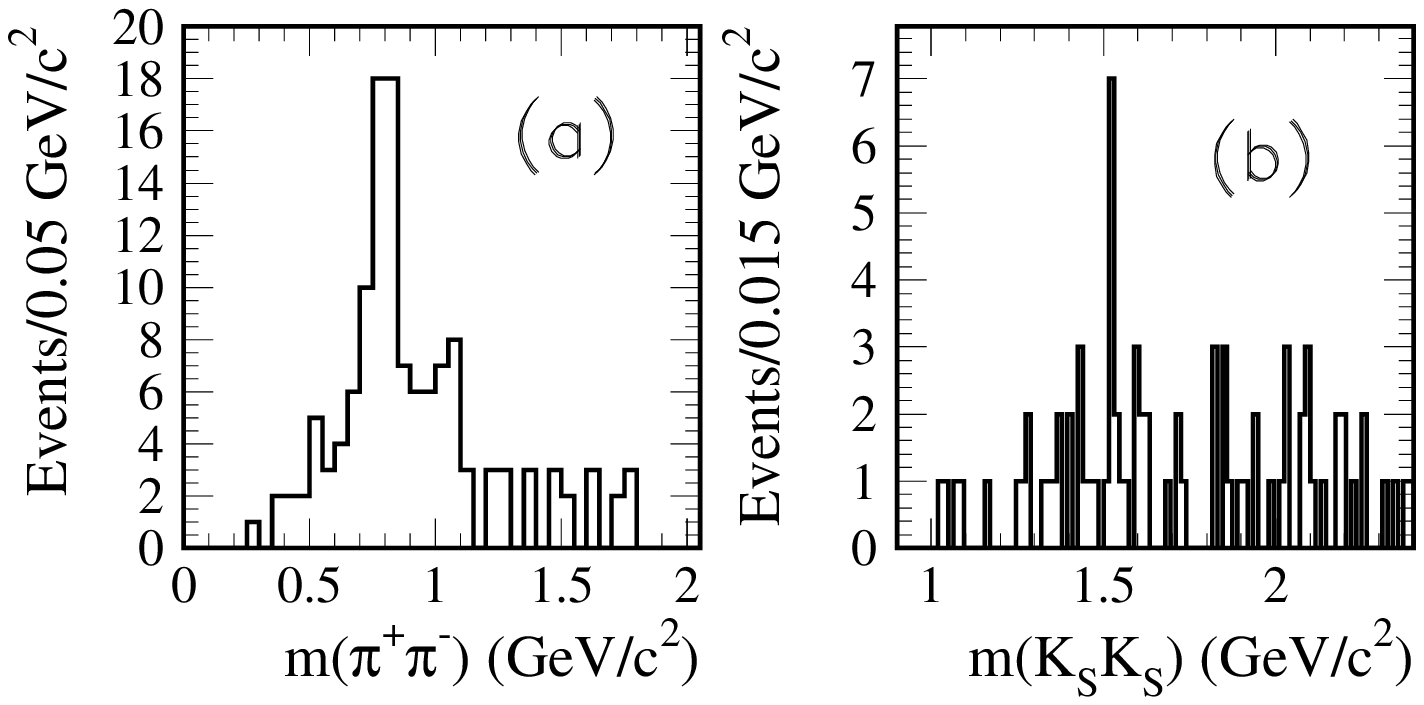}
\vspace{-0.4cm}
\caption{(a) The $m(\pipi)$ distribution for the 
 $\jpsi \to \KS\KS\pipi$ events.
 (b) The $m(\KS\KS)$ distribution for those events in the $\rho(770)$
  region, 0.6$<m(\pipi)<$1.0~\gevcc.
}
\label{jpsi_rho}
\end{center}
\vspace{-0.4cm}
\end{figure}
\subsection{\boldmath Internal structure of the
{\boldmath $\jpsi\to\KS\KS\pipi$ and $\KS\KS\Kp\Km$} decays}
The \jpsi signal in the $\KS\KL\pipi$ mode has a large non-resonant
background (see Fig.~\ref{kskljpsi}(a)),
and we are unable to quantify the contributions from the 
$K^*(892)\KS\pi$ and $\phi\pipi$ intermediate states with reasonable
accuracy.
The $\jpsi\to\phi\pipi$ decay rate is relatively well
measured~\cite{PDG}, dominated by \babar.

The $\KS\KS\pipi$ channel has much lower background 
(see Fig.~\ref{kskljpsi}(b)),
and we use the 157 events with invariant mass within 30~\mevcc
of the nominal \jpsi mass to study intermediate states. 
We use events in the 30~\mevcc intervals on each side of the
signal region to estimate a non-\jpsi contribution of 24 events, 
and to subtract the corresponding contributions from the histrograms
that follow.
The resulting $m(\KS\pi^{\pm})$ distribution (four entries per event)
is shown in Fig.~\ref{jpsi_kpi}(a).
Fitting with two Breit-Wigner (BW) functions plus a polynomial, 
we obtain $53\pm14$ events containing $K^*(892)\KS\pi$ and 
$35\pm15$ containing $K_2^{*}(1430)\KS\pi$.
To estimate decays to correlated $K^{*}(892)^+ K^{*}(892)^-$ or
$K_2^{*}(1430)^\mp K^{*}(892)^\pm$ pairs,
we consider events from the $K^{*}(892)^+$ and $K^{*}(892)^-$ bands
(see Fig.~\ref{kstarpi2dks})
defined by $|m(\KS\pi)-0.892|<0.15$~\gevcc;
a pairing in the overlap region gives only one entry, 
and there can be as many as two entries per event. 
Fitting the invariant mass distribution of the other $\KS\pi$ pair,
shown in Fig.~\ref{jpsi_kpi}(b), 
with two BW functions plus a polynomial,
we obtain $0.7\pm5.0$ and $8\pm8$ events for the
$K^{*}(892)^+ K^{*}(892)^-$ and $K_2^{*}(1430)^\mp K^{*}(892)^\pm$
combinations, respectively. 
Both are consistent with zero, i.e.,\ no correlated production. 

For each of these intermediate states we calculate the product of its
\jpsi branching fraction, $\Gamma^{J/\psi}_{ee}$, and the relevant
branching fractions for the intermediate resonances,
and list the values in Table~\ref{jpsitab}.
Using $\Gamma^{J/\psi}_{ee} = 5.55$~\ev, 
known branching fractions~\cite{PDG}, and
the assumptions that $K^*$ mesons decay equally to charged and neutral
kaons, and equally to \KS and \KL (e.g., $\BR_{K_2^{*}(1430)\to\KS\pi}=0.125$),
we calculate the corresponding branching fractions, 
also listed in Table~\ref{jpsitab}.
The only entry in the PDG tables for any of these channels is
$\BR_{J/\psi \!\to\! K^*(892)^+K^*(892)^-} =(1.0^{+0.2}_{-0.4}) \times 10^{-3} $.

Figure~\ref{jpsi_rho}(a)  
shows the $\pipi$ invariant mass distribution for the considered events.
A clear signal from the $\rho(770)$ resonance is seen, 
corresponding to $\jpsi \to \rho\KS\KS$ decays.
The $\KS\KS$ invariant mass distribution for those events
with $0.6<m(\pipi)<1.0~\gevcc$, shown in Fig.~\ref{jpsi_rho}(b),
features a narrow spike containing $9.4\pm 4.6$ events near 1.53~\gevcc.
We observe this same signal when no requirement is placed on the 
$\pipi$ invariant mass.  
Attributing this entirely to $\jpsi \to \rho(770) f_2^{'}(1525)$ decays,
we calculate the measured product and branching fraction,
using $\BR(f_2^{'}(1525)\to K \bar K) =0.71$~\cite{PDG},
and list them in Table~\ref{jpsitab}.
This channel also has no listing in the PDG tables~\cite{PDG}.
Due to uncertainties in the mass distributions for events without a
$\rho$ or $f_2^{'}$ meson, however,
we do not attempt to quantify the more inclusive $\rho\KS\KS$ or 
$\pipi f_2^{'}$ contributions. 
\begin{figure}[tbh]
\begin{center}
\vspace{-0.2cm}
\includegraphics[width=0.95\linewidth]{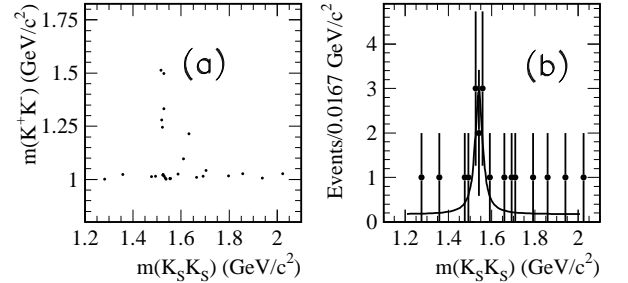}
\vspace{-0.4cm}
\caption{
(a) The $\Kp\Km$ versus $\KS\KS$ invariant mass for the $\KS\KS\Kp\Km$ 
events under the \jpsi peak (see text).
(b) The $\KS\KS$ invariant mass distribution for the events in (a) 
with $m(\Kp\Km)<1.04$~\gevcc.
The line represents the result of the fit described in the text.
}
\label{jpsi_ksksvs2k}
\end{center}
\vspace{-0.4cm}
\end{figure}

Figure~\ref{jpsi_ksksvs2k}(a) shows the $\Kp\Km$ versus $\KS\KS$
invariant mass for the 30 $\KS\KS\Kp\Km$ events with total invariant
mass within 30~\mevcc of the nominal \jpsi mass,
29$\pm$6 of which are \jpsi events.
Horizontal and vertical bands are visible, corresponding to the
$\phi(1020)$ and $f_2'(1525)$ resonances, respectively. 
We select 20 $\jpsi \to \phi(1020)\KS\KS$ candidate decays by
requiring $m(\Kp\Km)<1.04$~\gevcc, 
and plot their $m(\KS\KS)$ distribution in Fig.~\ref{jpsi_ksksvs2k}(b).
Fitting with a Breit-Wigner plus a constant function,
we obtain 11$\pm$4 $\jpsi \!\to\! f_2^{'}(1525)\phi(1020)$ decays;
including the five events with $m(\KS\KS)$ near 1525~\mevcc but higher
$m(\Kp\Km)$ values (see Fig.~\ref{jpsi_ksksvs2k}(b)) gives
16$\pm$5 $\jpsi \!\to\! f_2^{'}(1525)K^+K^-$ decays.

Using these numbers we calculate the products of
$\Gamma^{J/\psi}_{ee}$ and the relevant branching fractions,
and list them in Table~\ref{jpsitab}.
Using the PDG values of $\Gamma^{J/\psi}_{ee}$, 
$\BR(\phi\to K^+K^-) = 0.49$, 
and $\BR(f_2^{'}(1525)\to K \bar K) =0.71$~\cite{PDG},
we obtain the corresponding branching fractions, 
also shown in Table~\ref{jpsitab}.
Only one value can be compared with an existing PDG
listing~\cite{PDG}, 
namly $\BR(\jpsi \!\to\! f_2^{'}(1525)\phi(1020)) = (8 \pm 4) \times 10^{-4}$,
which has a scale factor of 2.7. 
Our result can be compared to the MarkII value $(4.8 \pm 1.8) \times 10^{-4}$,
and to the DM2 measurement $(12.3 \pm 0.26 \pm 2.0)\times 10^{-4}$~\cite{PDG}.

\section{Summary}
We have presented a study of the processes
$\epem\to \KS \KL$ and $\epem\to \KS \KL\pipi$ 
at low center-of-mass energies using 
using events with initial-state radiation (ISR)
collected with the \babar\ detector.
From the dominant $\epem\to\phi\gamma\to \KS\KL\gamma$ process near
$\KS\KL$ threshold,
we measure the probability of detecting the $\KL$ via its nuclear
interaction in the electromagnetic calorimeter with about 0.6\%
uncertainty,
as well as its angular resolution.
Using the positions of candidate $\KL$ clusters in the calorimeter as
input to kinematic fits,
we obtain clean samples of $\KS\KL\gamma$ and $\KS\KL\pipi\gamma$
events, 
and extract the $\epem\to\KS\KL$ and $\epem\to\KS\KL\pipi$ cross
sections from threshold to 2.2 and 4~\gev, respectively.

For the $\KS \KL$ final state,
we perform fits to the $\phi(1020)$ and $\phi(1680)$ resonances,
and report the resonance parameters and $\Gamma_{ee}\cdot\BR(\KS \KL)$
values.
The results are consistent with previous measurements, and much more
precise for c.m.\ energies above 1.2~\gev, especially for the
$\phi(1680)$ mass region. 
The $\epem\to \KS \KL\pipi$ cross section is measured for the first
time,
and is dominated by the $K^{*}(892)^+ K^{*}(892)^-$ intermediate
state. 
Additional contributions from the
$K^{*}(892)^\pm K_2^{*}(1430)^\mp$ and $\phi\pipi$
intermediate states are observed.

We also obtain the first measurements of the $\epem\to \KS \KS\pipi$
and $\epem\to\KS\KS\Kp\Km$  cross sections, and provide results from threshold to
4 and 4.5~\gev, respectively.
For the former process,
we again find the $K^{*}(892)^+K^{*}(892)^-$ intermediate state to be
dominant, and measure a contribution from $\rho(770) \KS \KS$.
However, no significant contribution from 
$K^{*}(892)^\pm K_2^{*}(1430)^\mp$ is observed.
For the latter process,
we observe contributions from the $\KS \KS\phi(1020)$ and
$f_2^{'}(1525)\phi(1020)$ intermediate states. 

We observe the $\jpsi\to \KS\KL\pipi$, $\KS\KS\pipi$, and
$\KS\KS\Kp\Km$ decays for the first time, 
and measure the product of the \jpsi electronic width and branching
fraction to each of these modes.
We study the substructure of these decays,
and obtain the first measurements of  
the $\jpsi\to K^*(892)^\pm\KS\pipm$,
$K_2^*(1430)^\pm\KS\pipm$,
$\rho(770)f_2^{'}(1525)$,
$\phi(1020)\KS\KS$, and
$f_2'(1525)\Kp\Km$ branching fractions.
In addition, we measure the  $\jpsi\to f_2'(1525)\phi(1020)$ branching fraction with improved
precision, and observe the  $\rho(770)\KS\KS$ and
$f_2^{'}(1525)\pipi$ decay modes.
We do not observe
$K^*(892)^+K^*(892)^-$ or
$K_2^*(1430)^\pm K^*(890)^\mp$ decays and set limits on their contributions.

\section*{Acknowledgements} 
We are grateful for the 
extraordinary contributions of our \pep2\ colleagues in
achieving the excellent luminosity and machine conditions
that have made this work possible.
The success of this project also relies critically on the 
expertise and dedication of the computing organizations that 
support \babar.
The collaborating institutions wish to thank 
SLAC for its support and the kind hospitality extended to them. 
This work is supported by the
US Department of Energy
and National Science Foundation, the
Natural Sciences and Engineering Research Council (Canada),
the Commissariat \`a l'Energie Atomique and
Institut National de Physique Nucl\'eaire et de Physique des Particules
(France), the
Bundesministerium f\"ur Bildung und Forschung and
Deutsche Forschungsgemeinschaft
(Germany), the
Istituto Nazionale di Fisica Nucleare (Italy),
the Foundation for Fundamental Research on Matter (The Netherlands),
the Research Council of Norway, the
Ministry of Education and Science of the Russian Federation, 
Ministerio de Econom\'{\i}a y Competitividad (Spain), the
Science and Technology Facilities Council (United Kingdom),
and the Binational Science Foundation (U.S.-Israel).
Individuals have received support from 
the Marie-Curie IEF program (European Union) and the A. P. Sloan Foundation (USA). 


\end{document}